\documentclass{./raa}
\usepackage{amssymb,amsmath}
\usepackage{graphicx,times}
\usepackage{./natbib}
\bibpunct[, ]{(}{)}{;}{a}{}{,}

\begin{document}

   \title{Multiwavelength study of low-luminosity 6.7-GHz methanol masers
 $^*$ \footnotetext{\small $*$ Supported by the National Natural
Science Foundation of China.} }

 \volnopage{ {\bf 2011} Vol.\ {\bf 11} No. {\bf XX}, 000--000}
   \setcounter{page}{1}

   \author{Yuan-Wei Wu,
      Ye Xu  \and
     Ji Yang
      }

   \institute{   Purple Mountain Observatory, Chinese Academy of Sciences,
             Nanjing 210008, China;
             {\it ywwu@pmo.ac.cn} \\
      \vs \no
   {\small Received 2010 April 11; accepted 2010 July 16 }
}

\abstract{ We present results of $^{13}$CO(1--0), C$^{18}$O(1--0),{ and}
HCO$^+$(1--0) map observations and N$_2$H$^+$(1--0) single point
observations{ directed} towards a sample of {nine} low-luminosity
6.7-GHz masers. N$_2$H$^+$ line emission has been detected from
{six} out of {nine} sources, C$^{18}$O line
emission {has} been detected from {eight} out of
{nine} sources,{ and} HCO$^+$ and $^{13}$CO emission
{has been} detected in all sources. In particular, a
``blue profile'' of{ the} HCO$^+$ spectrum, {a }signature of inflow, is found
towards one source. From integrated intensity emission maps, we
identified 17 cores in the sample. Among them{,} {nine}
cores are closely associated with low-luminosity methanol masers.
For these cores, we derive the column densities, core sizes, masses
and molecular abundances. Comparison of our results with
similar molecular line survey{s} towards {the }southern sky methanol masers
indicates that linewidths of our sample, including only the
low-luminosity masers, are smaller than the sample that includes
both low- and high-luminosity masers. For the maser associated
cores, their gas masses have the same order of magnitude as their
virial masses, indicating that these cores are gravitationally bound
systems. {In addition}, we have {found} from our
observations that the low-luminosity methanol masers tend to coexist
with H$_2$O masers and outflows rather than with OH masers.
   \keywords{masers --- ISM: abundances --- ISM: molecules --- stars: formation}
   }

   \authorrunning{Y. W. Wu, Y. Xu \& J. Yang}            
   \titlerunning{Multiwavelength Study of Faint 6.7-GHz Methanol Masers}  
   \maketitle

\section{Introduction}

Interstellar masers, such as the main-line hydroxy (OH) masers,
22-GHz water (H$_2$O) masers and Class II methanol (CH$_3$OH){ masers}, are
one of the most readily observed signposts of star formation. These
interstellar masers are powerful not only {for
}sign{aling} star formation regions, but also {for}
diagnos{ing} physical conditions
\citep{pavla96a,pavla96b}, prob{ing}
the kinematics of the{se} regions \citep{torr05,motogi08} and
measur{ing} trigonometric parallaxes
\citep{xu06a,reid09}. Recently, {with} the advances of
our understanding {of} these interstellar masers,
\citet{elling07a} even recommended us{ing} them to trace different
evolutionary phases of massive star formation, which may shed light
on our understanding of early evolutionary stages of massive stars.

Among these species, Class II methanol masers have some advantages over H$_2$O and OH
masers {since} they exclusively trace massive star-forming regions
\citep{minier03,xu08}, while H$_2$O and OH are known to also be associated with low
mass star formation and evolved stars. To date, extensive surveys have yielded more
than 800 6.7-GHz maser sites \citep{caswell95, pestal05, pandian07, elling07b, xu08,
caswell09,green09}. Recently, \citet{purcell09} found distinctions between
radio-quiet and radio-loud subgroups of 6.7-GHz methanol masers. \citet[hereafter
Paper I]{wu10} found remarkable physical and kinematic distinctions between faint and
bright subgroups of 6.7-GHz masers. Hence, there {is the} potential to use
elaborat{e} classifications of 6.7-GHz masers based on their associations, e.g., UCH
II regions, and inherent attribute{s}{. Such applications include using} maser
luminosities as a clock to indicate the evolutionary stage of ongoing massive star{
development}.

Observationally, molecular emission is a powerful tool {for}
investigat{ing} the physical and chemical
conditions in hot cores. Transitions requiring different
temperatures and densities for excitation constitute an excellent
probe of physical conditions. {It} is known that
$^{12}$CO is the most {useful} molecule tracer. When the
$^{12}$CO(1--0) lines {are} optically thick in most
opaque regions of molecular clouds, rarer CO isotopes, i.e.,
$^{13}$CO and C$^{18}$O, are usually used{ instead} to trace the cloud{'s} mass. In addition, HCO$^+$ and N$_2$H$^+$ are important ionic
molecules. {Determining the a}bundance of HCO$^+$ can reflect the
ionization rate of clouds. Saturated and self absorbed line profiles
of HCO$^+$ {are} also used to trace dynamics,
{like the} outflow and inflow of young protostellar objects
\citep{fuller05,klassen07, wu07,sun09}. In contrast, N$_2$H$^+$ is
{an} excellent tracer of quiescent high density gas
\citep{womack92}. Moreover, {since} the chemical
properties of hot cores vary with time, the relative molecular
abundances can also be used as indicators of evolution
\citep{bergin97,langer00}. In this paper, we report our 3-mm
spectral line observations, including $^{13}$CO(1--0),
C$^{18}$O(1--0), HCO$^+$(1--0) and N$_2$H$^+$(1--0) transitions,
towards a sample of {nine} faint 6.7-GHz
methanol masers (Table~\ref{Tab:faintmaser}). 
 All {nine} sources have also been mapped in transitions of NH$_3$(1,1),
(2,2), (3,3) and $^{12}$CO(1--0) in Paper~I. In Section~2{,} we describe the sample
and observations. In Section~3{,} we present the results {which }includ{e} molecular
line maps and individual descriptions. Analysis is given in Section~4. Finally{,}
conclusions are drawn in Section~5.

\begin{table}[h!!]
\centering
\begin{minipage}{60mm}
\caption{List of Faint 6.7-GHz Masers }\label{Tab:faintmaser}\end{minipage}

\vspace{-3mm} \fns\tabcolsep 1.2mm
\begin{tabular}
{lcccccccr} \hline \noalign{\smallskip}
Source    &RA (J2000)        &Dec (J2000)       & $S_{\rm peak}$& $V_{\rm LSR}$
&D         & $L_{\rm peak}$ 
& Other  & Ref.   \\
Name    &(h~~m~~s)        &($^\circ$~~$^\prime$
~~$^{\prime\prime}$)
& (Jy)& (km s$^{-1}$)  &(kpc) &($L_\odot$) & Name   &     \\
(1)& (2)& (3)& (4)& (5)& (6)& (7)& (8) & (9)\\
\noalign{\smallskip} \hline\noalign{\smallskip}
106.80+5.31   &   22:19:18.3    &   +63:18:48     &0.5    & --2.0  &0.9$^4$      &7.0E--10 &  S 140              &  13      \\
111.25--0.77   &   23:16:09.7    &   +59:55:29     &4.0    & --38.5 &3.5$^8$      &8.5E--08 &  IRAS 23139+5939    &  12      \\
121.24--0.34   &   00:36:47.358  &   +63:29:02.18  &10     & --22.8 &0.85$^{15}$  &1.3E--08 &  L 1287             &  9,14    \\
133.72+1.22   &   02:25:41.9    &   +62:06:05     &5      & --44.5 &2.3$^5$      &4.6E--08 &  W3 IRS5            &  11      \\
183.35--0.59   &   05:51:10.8    &   +25:46:14     &19     & --4.5  & 2.1$^6$     &1.5E--07 &  IRAS 05480+2545    &  11      \\
188.80+1.03   &   06:09:07.8    &   +21:50:39     &4.8    & --5.5  & 2$^1$       &3.3E--08 &  AFGL 5182          &  12      \\
189.03+0.78   &   06:08:40.671  &   +21:31:06.89  &17     & 8.8   & 1.5$^7$     &6.6E--08 &  AFGL 6466          &  2,14    \\
189.78+0.34   &   06:08:35.28   &   +20:39:06.7   &15     & 5.7   & 1.5$^\star$ &5.8E--08 &  S 252A             &  2,3     \\
206.54--16.36  &   05:41:44.15   &   --01:54:44.9   &1.48   & 12.4  & 0.415       &4.4E--10 &  NGC 2024           &  10      \\
\noalign{\smallskip}\hline\noalign{\smallskip}
\end{tabular}
\parbox{125mm}{\baselineskip 3.6mm
Col.~(1) is the source name which is named after Galactic coordinates, Cols.~(2) and
(3) are equatorial coordinates. Peak flux density, central velocity, distance and
maser luminosity are listed in Cols.~(4)$-$({7}), respectively. Other name{s} and
reference{s} are listed
in Cols.~({8}) and ({9}), respectively.\\
References for sources and distances: \\
{[1]} \citet{carp95}; [2] \citet{caswell95}; [3] \citet{caswell09}; {[4]}
\citet{cram74}; [5] \citet{georg76}; [6] \citet{Hugh93}; {[7]} \citet{Humph78}; [8]
\citet{lari99}; [9] \citet{macleod98}; {[10]} \citet{minier03}; [11]\citet{slysh99};
[12]\citet{szym00a}; {[13]} \citet{xu08};
[14] \citet{xu09}; [15] \citet{yang1991};\\
   $^{\star}$ Heliocentric kinematic distance.}
\end{table}

\section{Sample and Observations}

\subsection{Sample}

The low-luminosity 6.7-GHz methanol masers in this study were mostly selected
({seven} out of {nine}) from the catalog of \citet{xu03} according to their lowest
luminosities (the luminosities are calculated from the peak flux density assuming a
typical linewidth of 0.25~km~s$^{-1}$ and isotropic emission). Two other sources with
very low luminosities, 106.80+5.31 and 206.54--16.36, were selected from \citet{xu08}
and \citet{minier03} respectively. The 6.7-GHz line luminosities of these sources
range from 4.4$\times 10^{-10}$ to 1.5$\times 10^{-7}$ $L_\odot$. Their properties,
including galactic and equatorial coordinates, peak flux densities, central
velocities, distances, luminosities, {possible }other names and references{,} are
listed in Table~\ref{Tab:faintmaser}. 
All of the sources have already been mapp{ed} in transitions of
NH$_3$(1,1), (2,2), (3,3) and $^{12}$CO($1-0$) in Paper~{I}.

\subsection{Observations}

The observations were performed during {2008 }January and {2010 }March with the
13.7-m millimete{r} wave telescope in Delingha, China. We mapped the {nine} sources
in transitions of $^{13}$CO(1--0), C$^{18}$O(1--0) and HCO$^+$(1--0). {A t}ransition
of N$_2$H$^+$(1--0) was observed {with a} single pointing, targeted at the maser
sites. A cooled SIS receiver was employed \citep{zuo04}, and system temperatures were
200 $\sim$ 300 K during the observations. The Acousto-Optical Spectrometer (AOS) was
used to measure the transitions of $^{13}$CO(1--0) and C$^{18}$O(1--0) and the Fast
Fourier Transform Spectrometer (FFTS) was used to measure the transitions of
HCO$^+$(1--0) and N$_2$H$^+$(1--0). The HPBW was 60$^\prime$$^\prime$ at 110 GHz. The
observations were performed in a position switched mode. The grid spacing of the
mapping observations was 30$^\prime$$^\prime$,{ and} the average integration time was
5 min per point. The pointing accuracy was better than 10$^\prime$$^\prime$. Data
were calibrated using the standard chopper wheel method. S140 and NGC 2264 were used
for flux calibration and observed every two hours during the observations. Absolute
calibration is estimated to be accurate to about 15\%. Basic information of the
observations is summarized in
Table~\ref{Tab:obs}. 

\begin{table}[h!!]

\centering

\begin{minipage}[]{50mm}

 \caption[]{Observation Parameters}\label{Tab:obs}\end{minipage}

\vspace{-3mm}

\fns\tabcolsep 2.5mm
\begin{tabular}{lrcccc}
 \hline\noalign{\smallskip}
Translation   & $\nu_{\rm rest}$  &   HPBW  & Bandwidth
 &    $\Delta\nu_{\rm res}$      &  1$\sigma$ rsm$^a$ \\
              & (GHz) &($^\prime$$^\prime$)&(MHz)&(km s$^{-1}$)&(K) \\
\hline\noalign{\smallskip}
$^{13}$CO(1--0)  &110.201353 &60 &43 &0.11   &0.10 \\
C$^{18}$O(1--0)  &109.782182 &60 &43 &0.12   &0.10 \\
HCO$^+$(1--0)    &89.188521  &74 &200 &0.20  &0.10 \\
N$_2$H$^+$(1--0) &93.171880  &71 &1000&0.04  &0.10 \\
\noalign{\smallskip}\hline\noalign{\smallskip}
\end{tabular}
\parbox{95mm}{Notes: $^a$ typical value in the scale of brightness temperature
for the reduced sp{e}ctra.}
\end{table}

Data were processed using the CLASS and GREG packages of GILDAS\footnote{CLASS and
GREG are part of the Grenoble Image and Line Data Analysis Software (GILDAS) working
group's software. {\it http://www.iram.fr/IRAMFR/GILDAS/}} software, including
baseline subtraction, fitting Gaussian line profiles and hyperfine structure fitting
(HFS) of N$_2$H$^+$(1--0) lines.

\section{Results}

N$_2$H$^+$ emission is detected in {six} sources,
C$^{18}$O emission is detected in {eight} sources,{ and}
$^{13}$CO and HCO$^+$ are detected in all sources. Detected spectral
lines are at least 3-$\sigma$ above the baseline and in most cases
have a signal to noise ratio greater than five. The detailed
detections for individual sources
are presented in Table~\ref{Tab:det}. 
Recently, similar 3-mm spectral line survey{s} towards southern 6.7-GHz methanol
masers presented high detection rates (99\%) of HCO$^+$(1--0), N$_2$H$^+$(1--0) and
$^{13}$CO(1--0) \citep{purcell09}. Our detection rates of $^{13}$CO(1--0) and
HCO$^+$(1--0) are 100\%, similar to{ the} 99\%{ detection rates} of
\citet{purcell09}{, w}hile our detection rate of N$_2$H$^+$(1--0) is 66\%, lower
than{ the} 99\%{ rate} of \citet{purcell09}. The lower detection rate of
N$_2$H$^+$(1--0) may be due to the relatively lower sensitivity of our observations.

\begin{table}

\centering

\begin{minipage}[]{63mm}

\vs\vs

 \caption{ Detection Rates of Molecular
Lines}\label{Tab:det}\end{minipage}

\vspace{-3mm} \fns
\begin{tabular}{lcccc}
\hline\noalign{\smallskip}
Source Name   & $^{13}$CO(1--0) & C$^{18}$O(1--0) & HCO$^+$(1--0) & N$_2$H$^+$(1--0) \\
\hline\noalign{\smallskip}
106.80+5.31        &Y  &Y   &Y   &Y  \\
111.25--0.77        &Y  &Y   &Y   &N   \\
121.24--0.34        &Y  &Y   &Y   &Y  \\
133.72+1.22        &Y  &N    &Y   &N   \\
183.35--0.59        &Y  &Y   &Y   &Y  \\
188.80+1.03        &Y  &Y   &Y   &N   \\
189.03+0.78        &Y  &Y   &Y   &Y  \\
189.78+0.34        &Y  &Y   &Y   &Y  \\
206.54--16.3        &Y  &Y   &Y   &Y  \\
\hline\noalign{\smallskip}
\scriptsize{Detection Rate}       &9/9    &8/9  &9/9  &6/9  \\
\noalign{\smallskip}\hline\noalign{\smallskip}
\end{tabular}
\parbox{85mm}{
Notes: ``Y'' indicates detection, ``N'' indicate{s} no detection.  }
\end{table}

\subsection{Spectra}

In Figure~\ref{Fig:spectra}, 
 we present the
$^{13}$CO(1--0), C$^{18}$O(1--0), HCO$^+$(1--0) and N$_2$H$^+$(1--0)
spectra at the positions of the peak of the maser associated cores.
$^{13}$CO(1--0), C$^{18}$O(1--0),{ and} HCO$^+$(1--0) lines are fitted
with Gaussian profiles. For HCO$^+$(1--0) spectra with evidence of
self absorbtion, though exhibiting {a }double-peak profile, are
fitted with {a }single Gaussian{ function}. Spectral
parameters, i.e., bright temperatures, line width, velocity of the
local standard of rest ($V_{\rm LSR}$), and integral intensities of
$^{13}$CO(1--0), C$^{18}$O(1--0),{ and} HCO$^+$(1--0){,} are listed in
Table~\ref{Tab:linepara}. 
For N$_2$H$^+$(1--0),
following \cite{purcell09}, we fit the spectra with two methods:
(1) the hyperfine structure fitting routine in the CLASS software,
considering{ the} seven-component structure; (2) three-Gaussian fitting
{of} the three blended groups. The fitted parameters of both methods
are presented in Table~\ref{Tab:ngauss}. 

\begin{figure}[h!]

\centering

\includegraphics[width=39mm,angle=0]{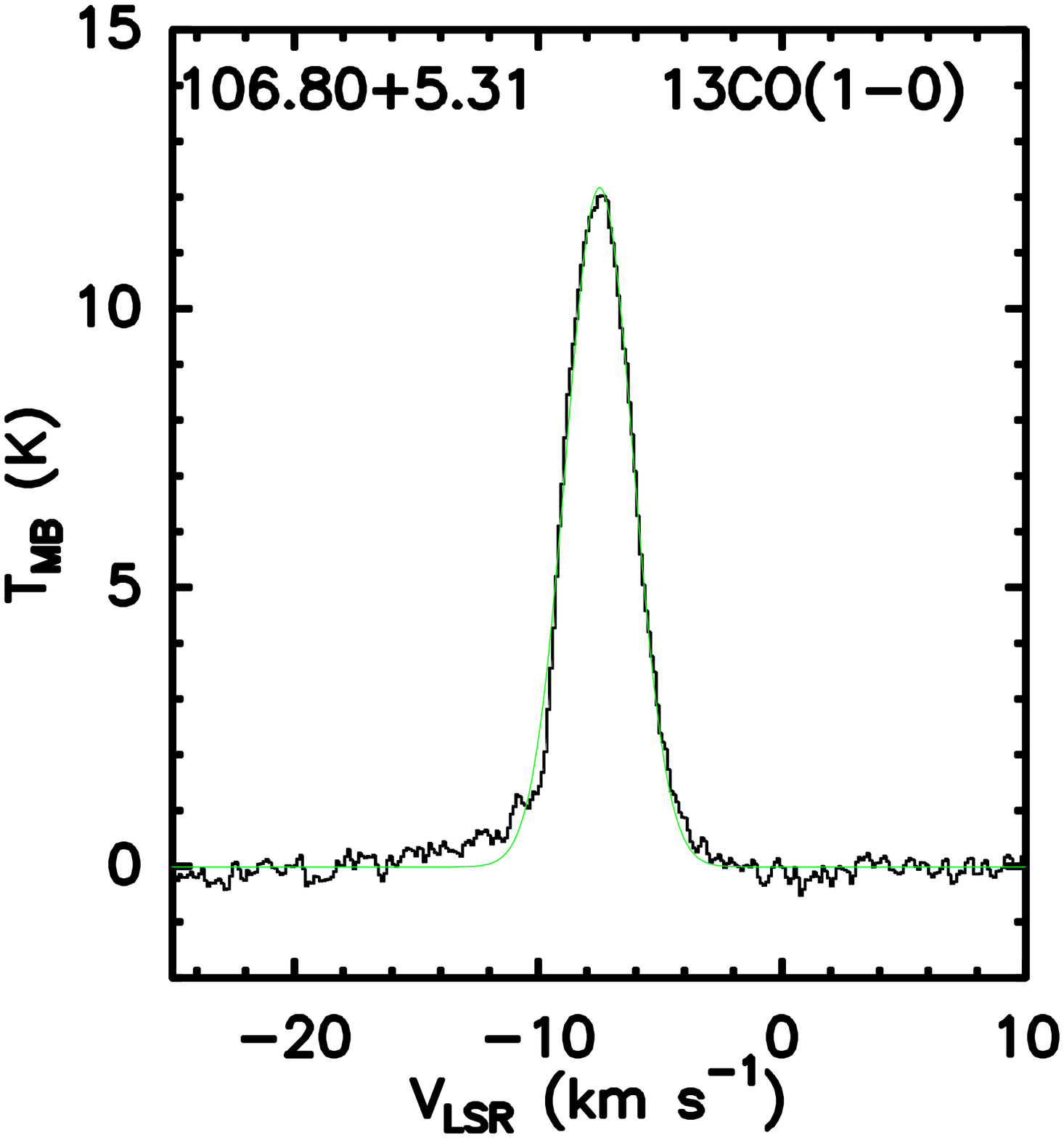}~~~ 
\includegraphics[width=39mm,angle=0]{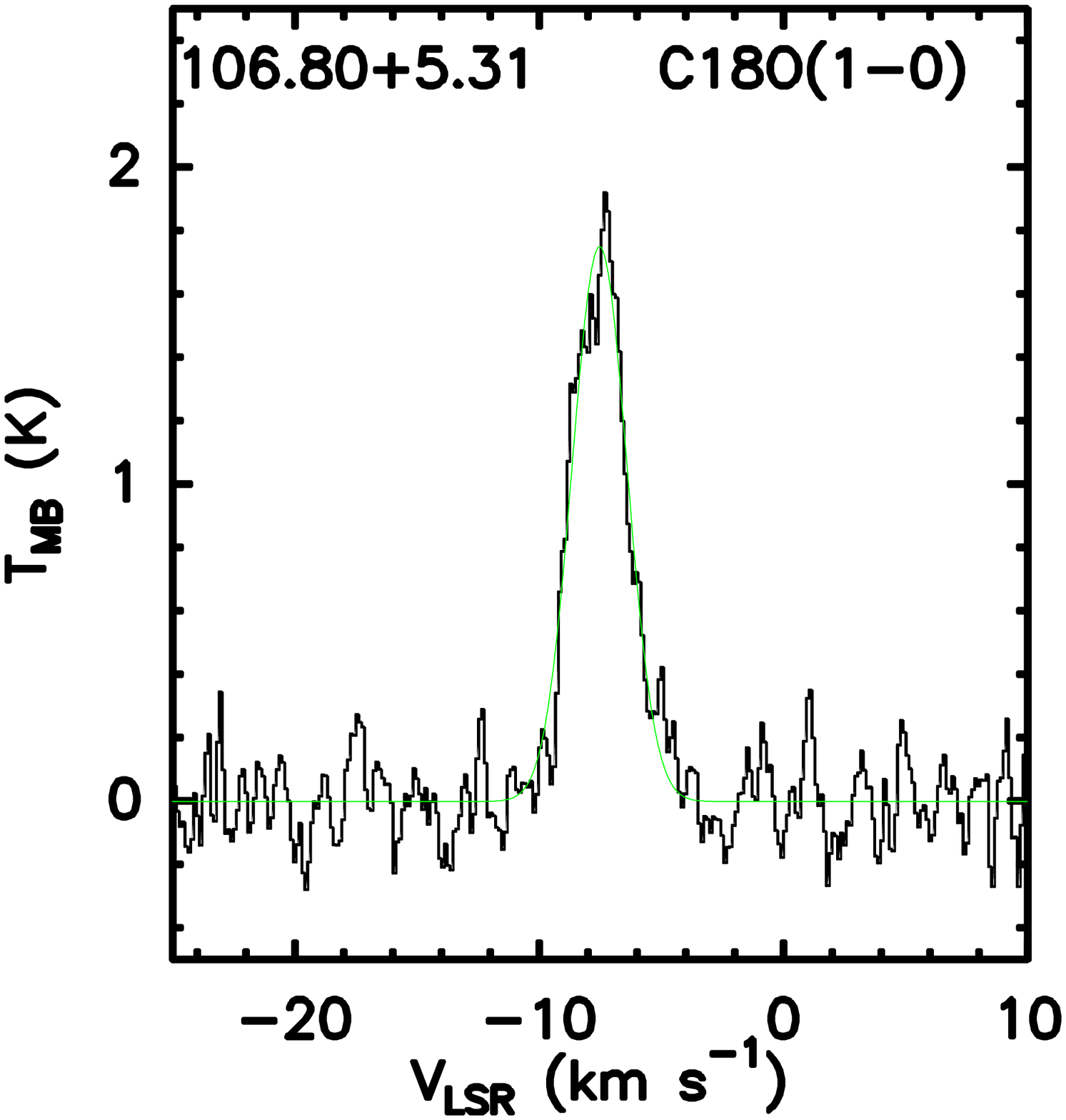}~~~
\includegraphics[width=40mm,angle=0]{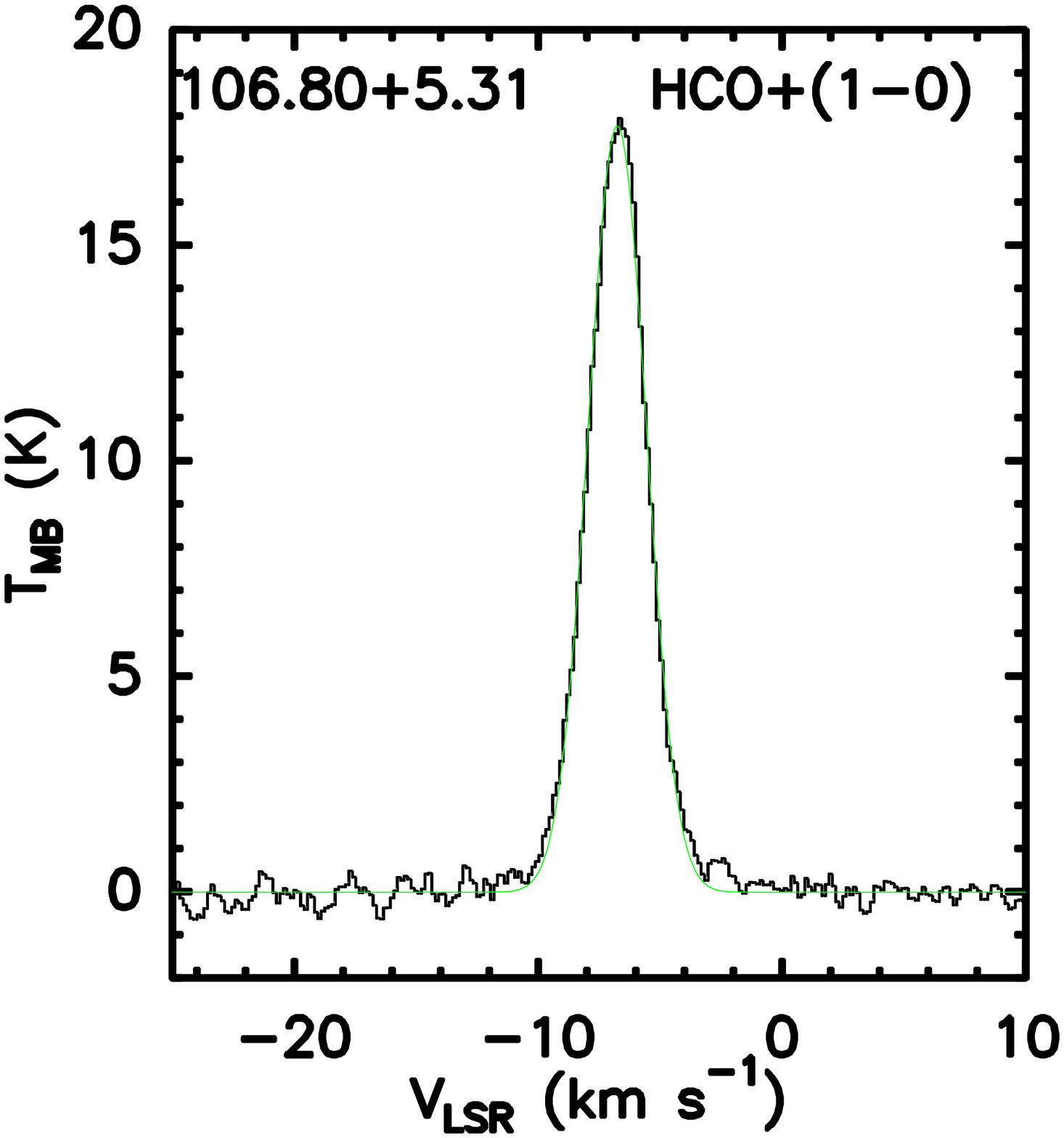}

\vspace{1mm}
\includegraphics[width=38mm,angle=0]{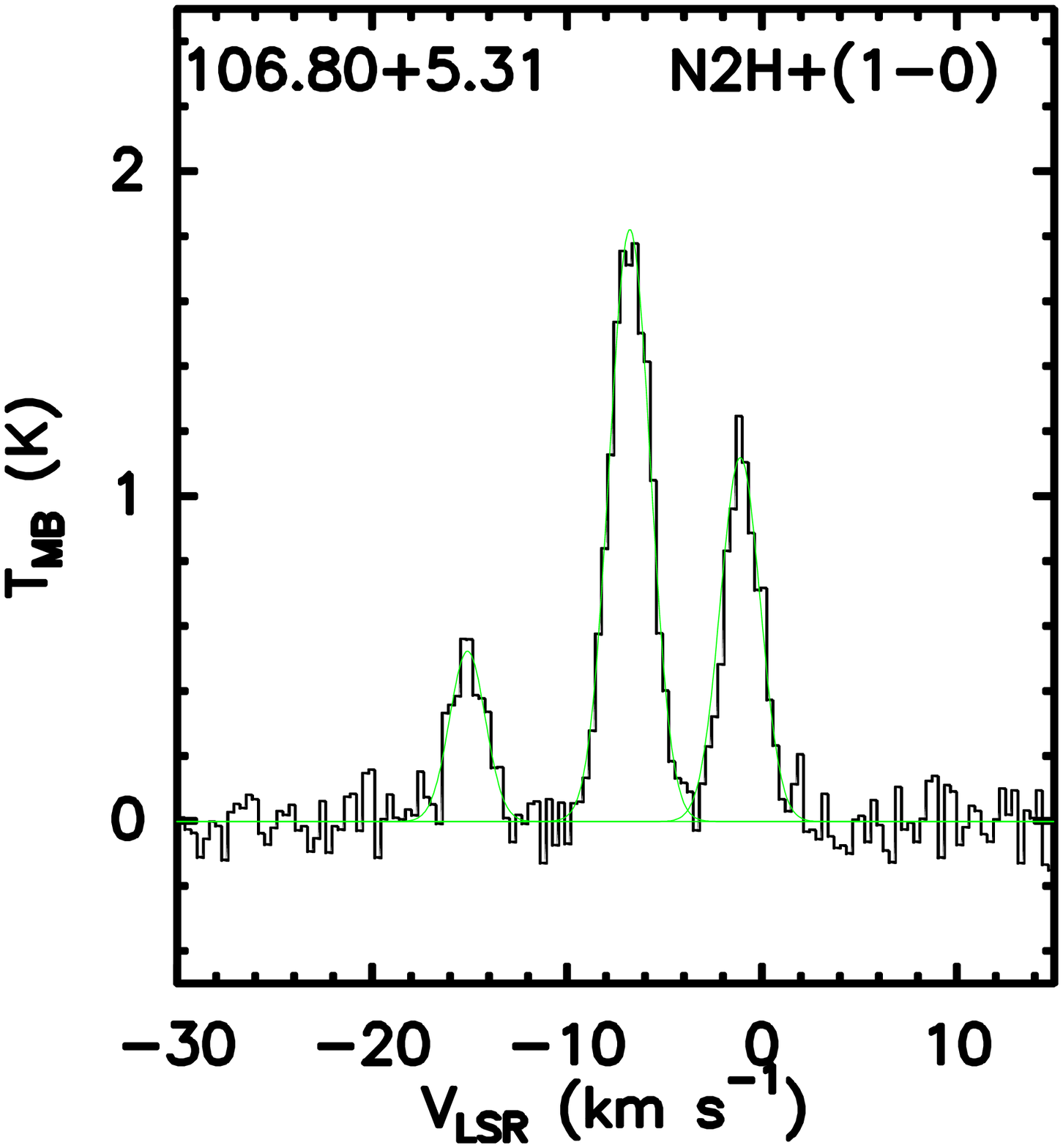}~~~ 
\includegraphics[width=39mm,angle=0]{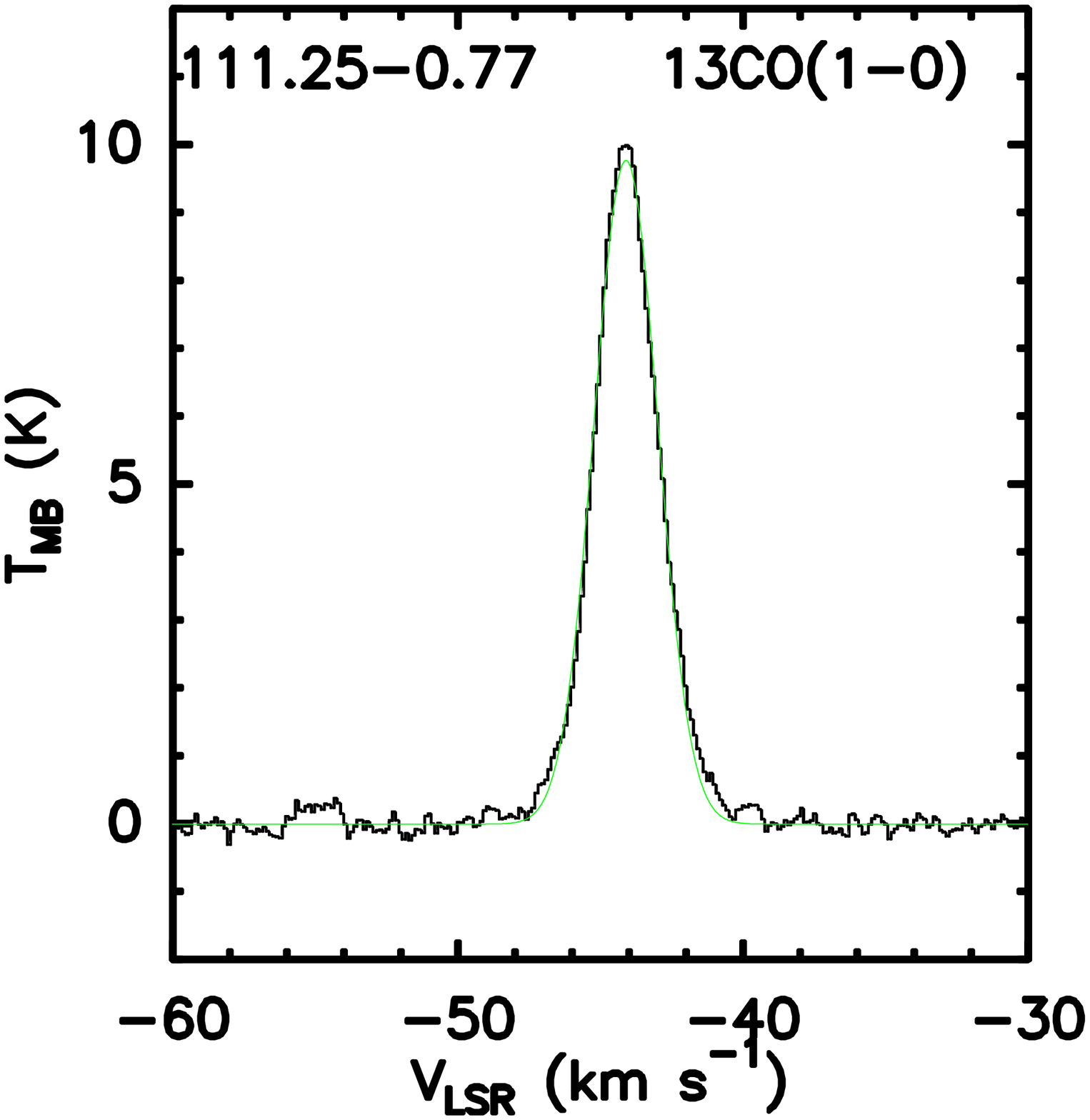}~~~
\includegraphics[width=40mm,angle=0]{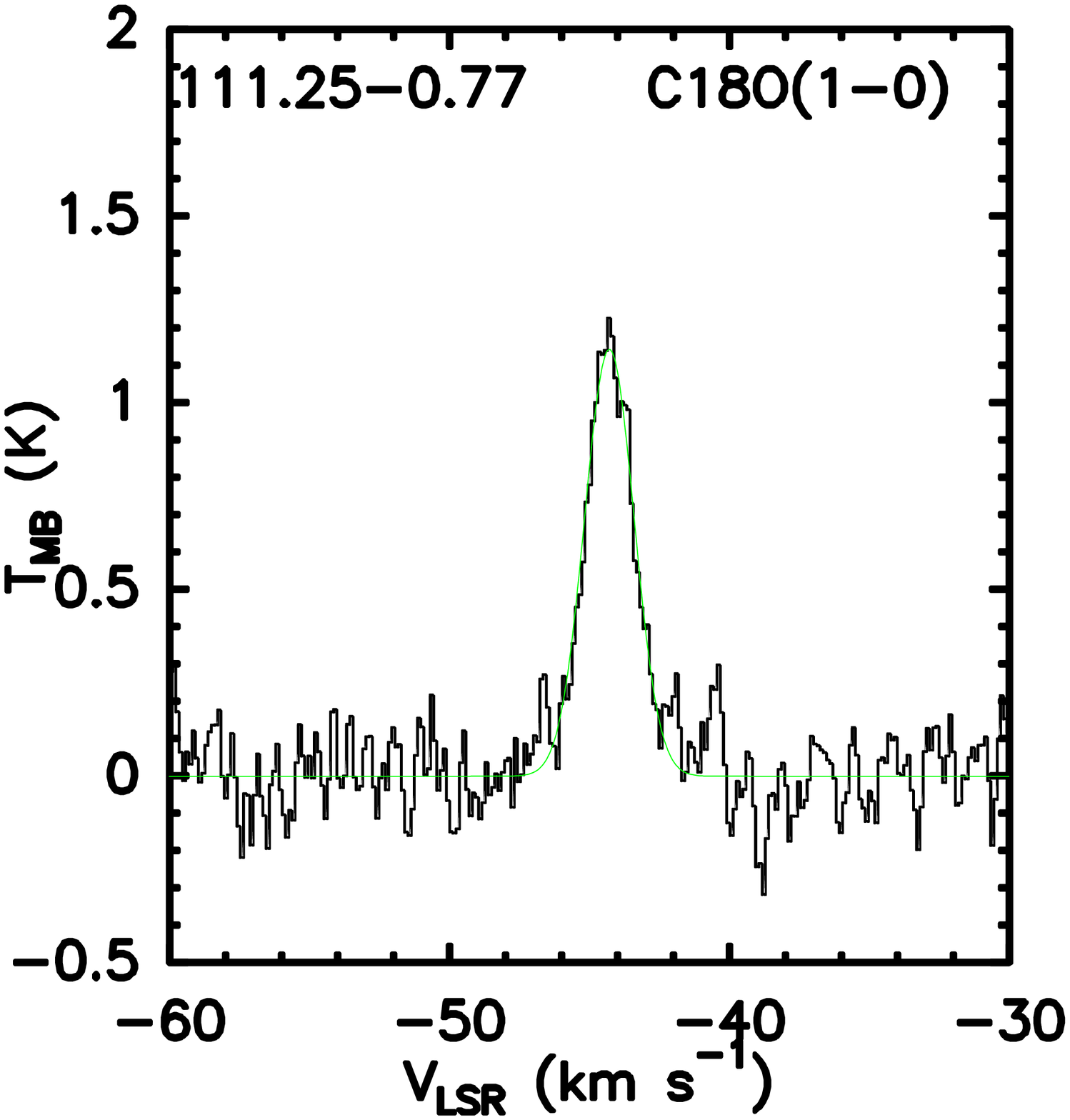}

\includegraphics[width=39mm,angle=0]{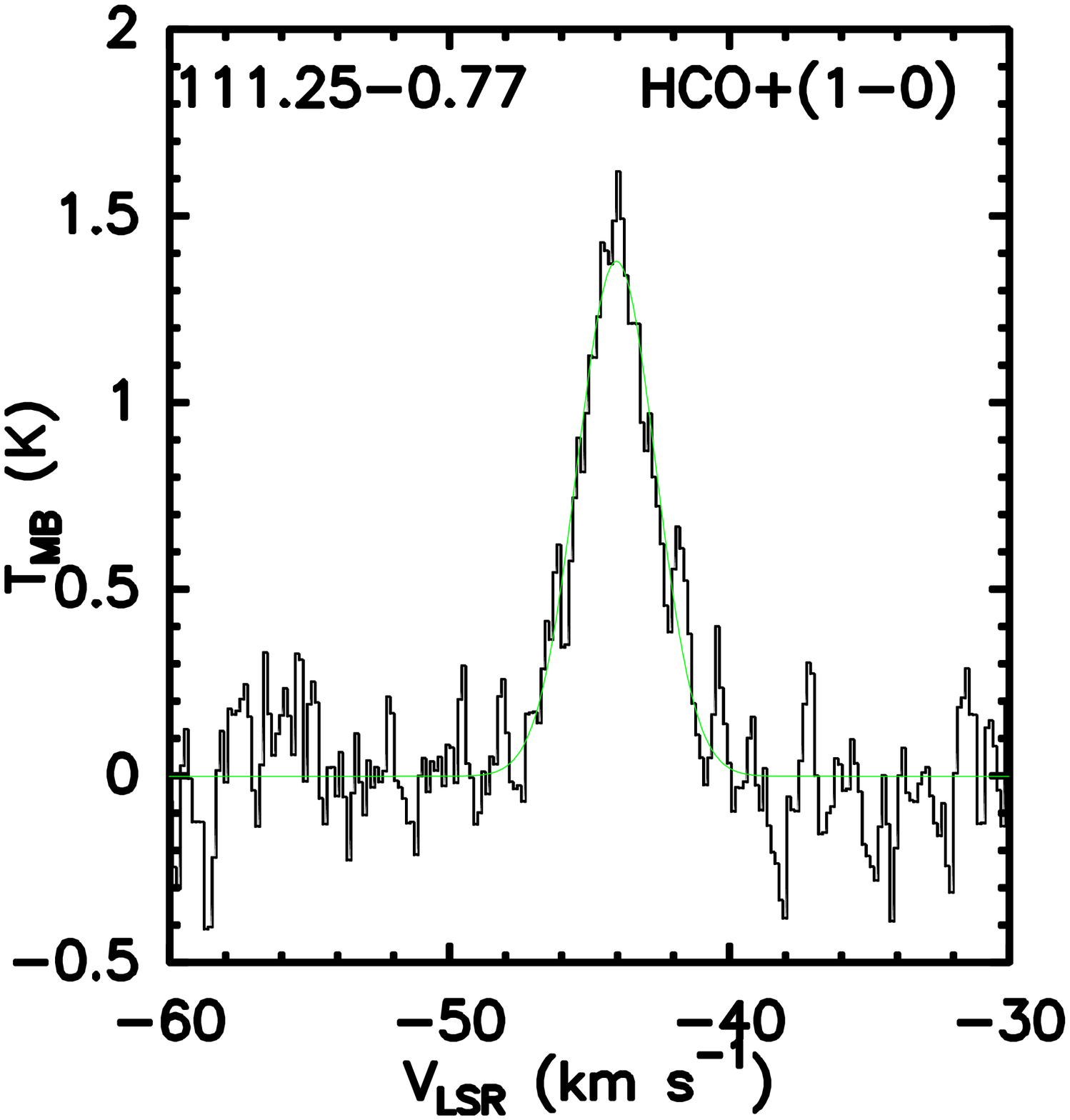}~~ 
\includegraphics[width=39mm,angle=0]{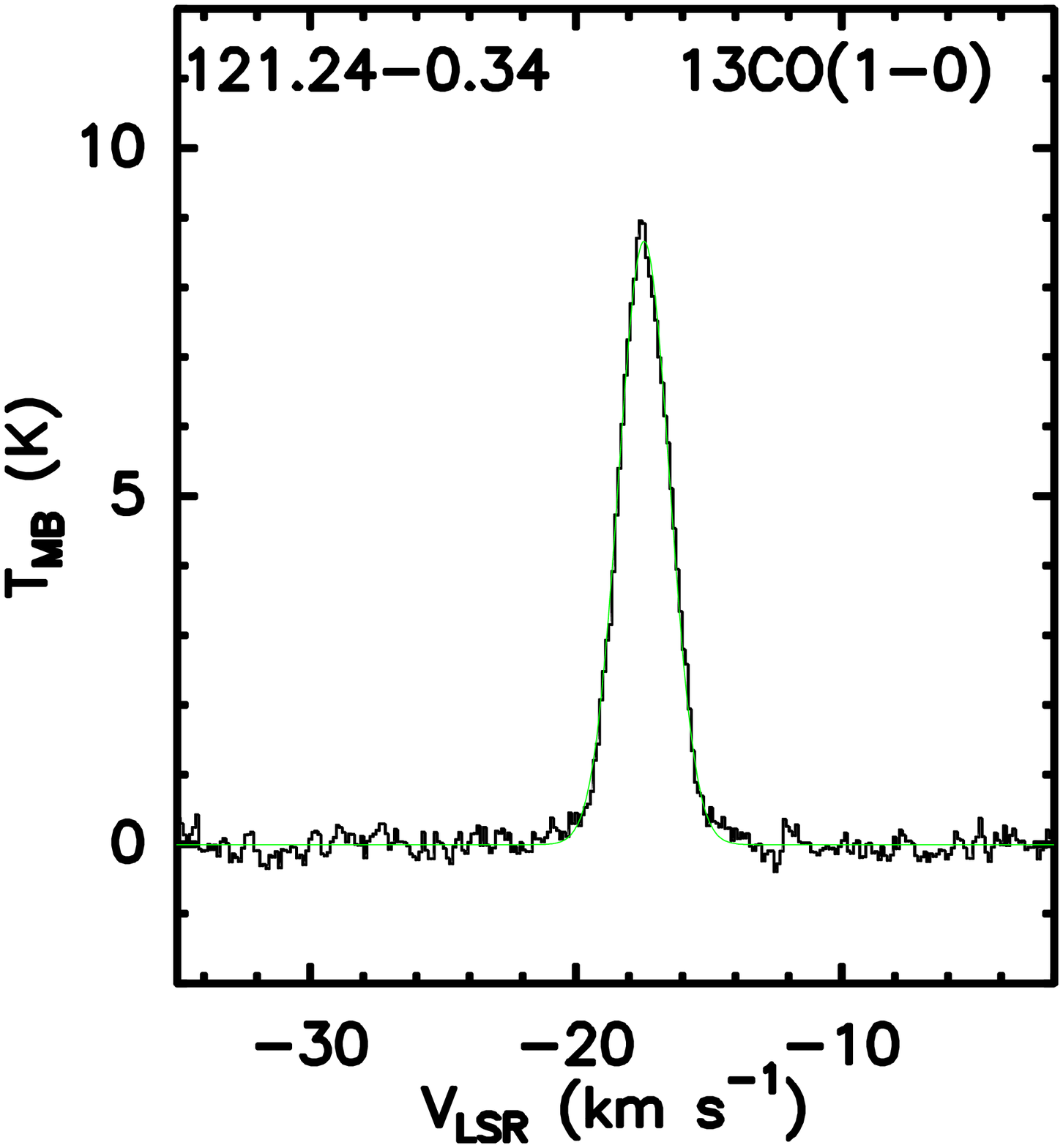}~~~
\includegraphics[width=40mm,angle=0]{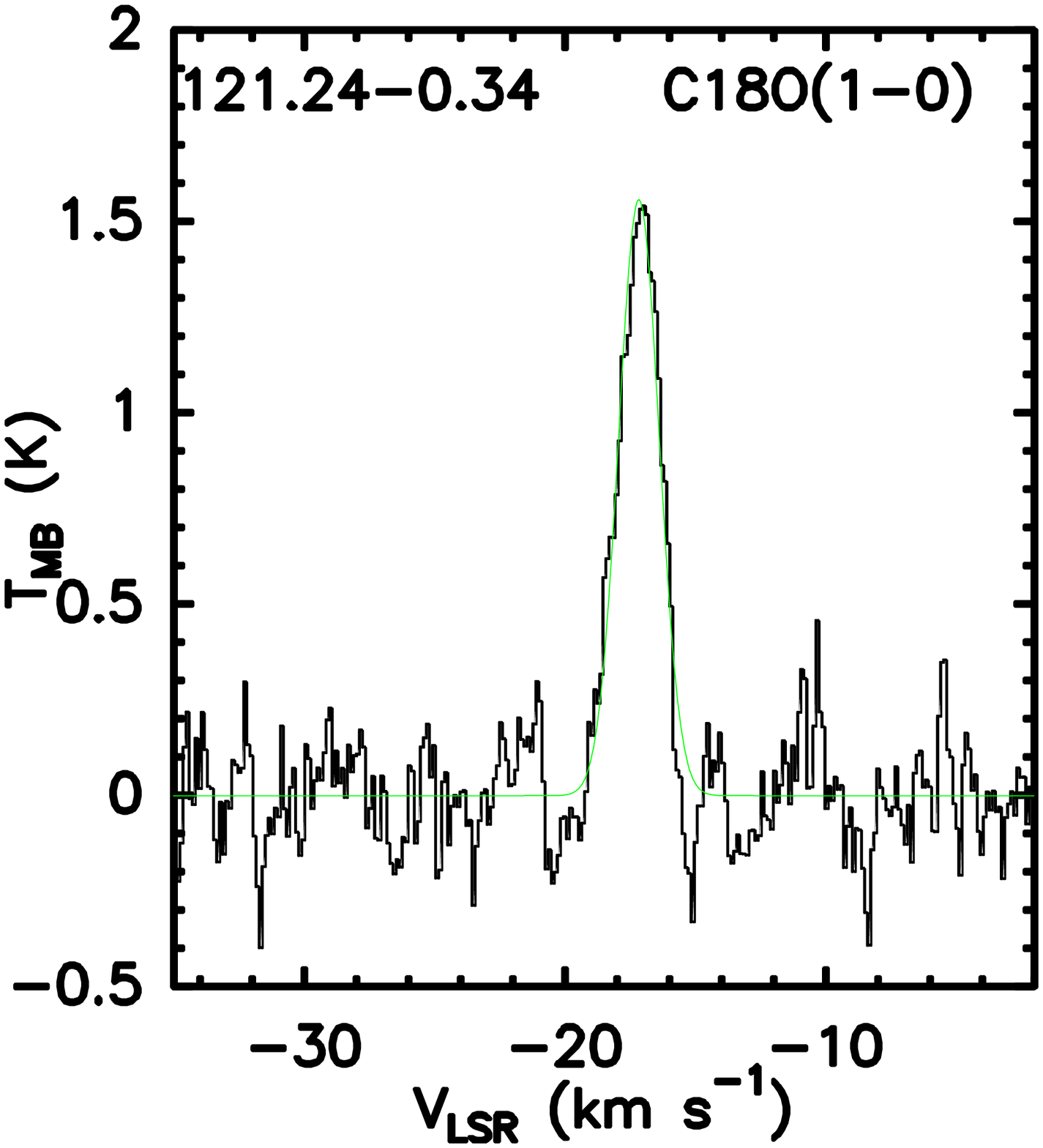}

\vs
\includegraphics[width=39mm,angle=0]{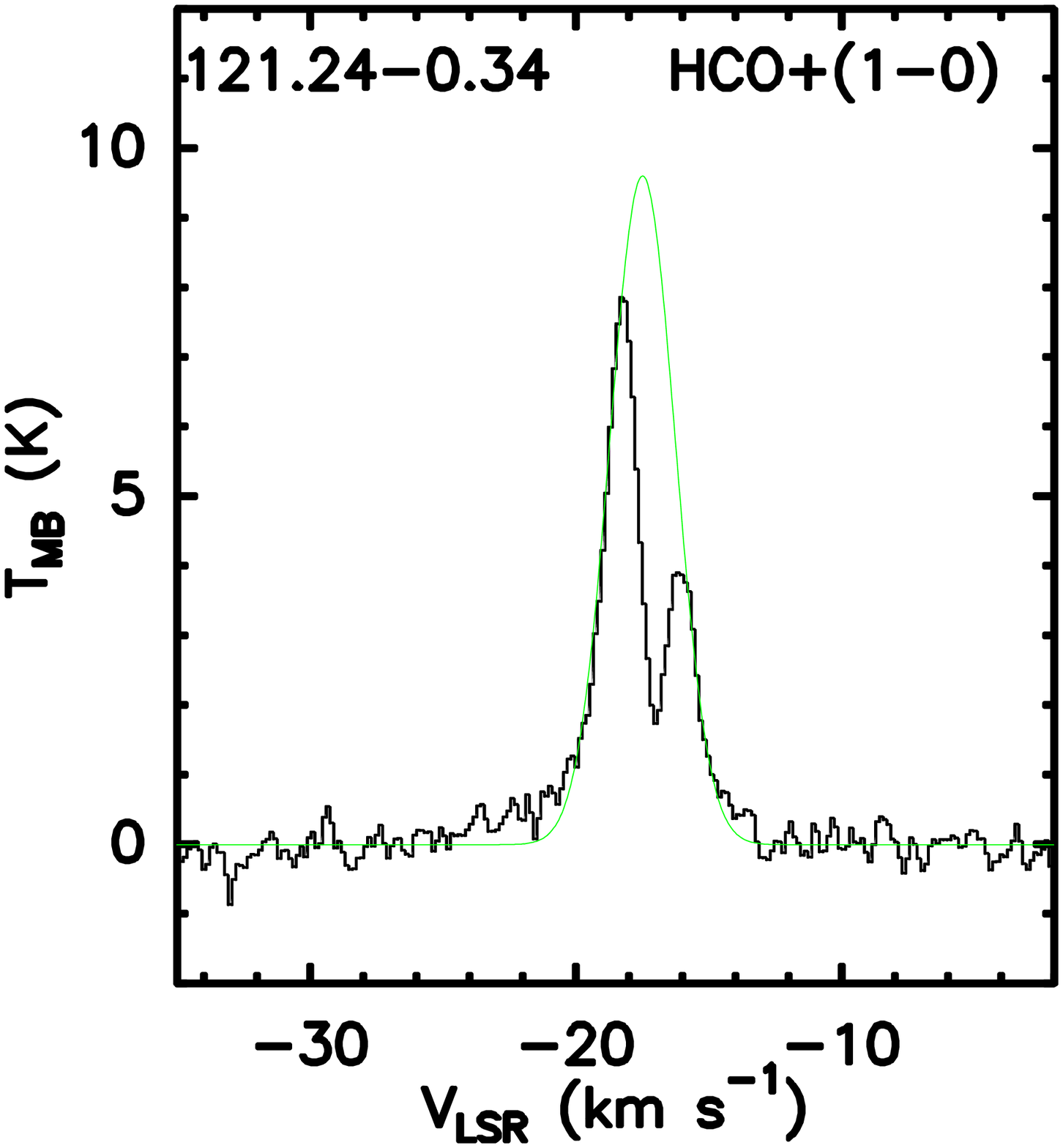}~~ 
\includegraphics[width=39mm,angle=0]{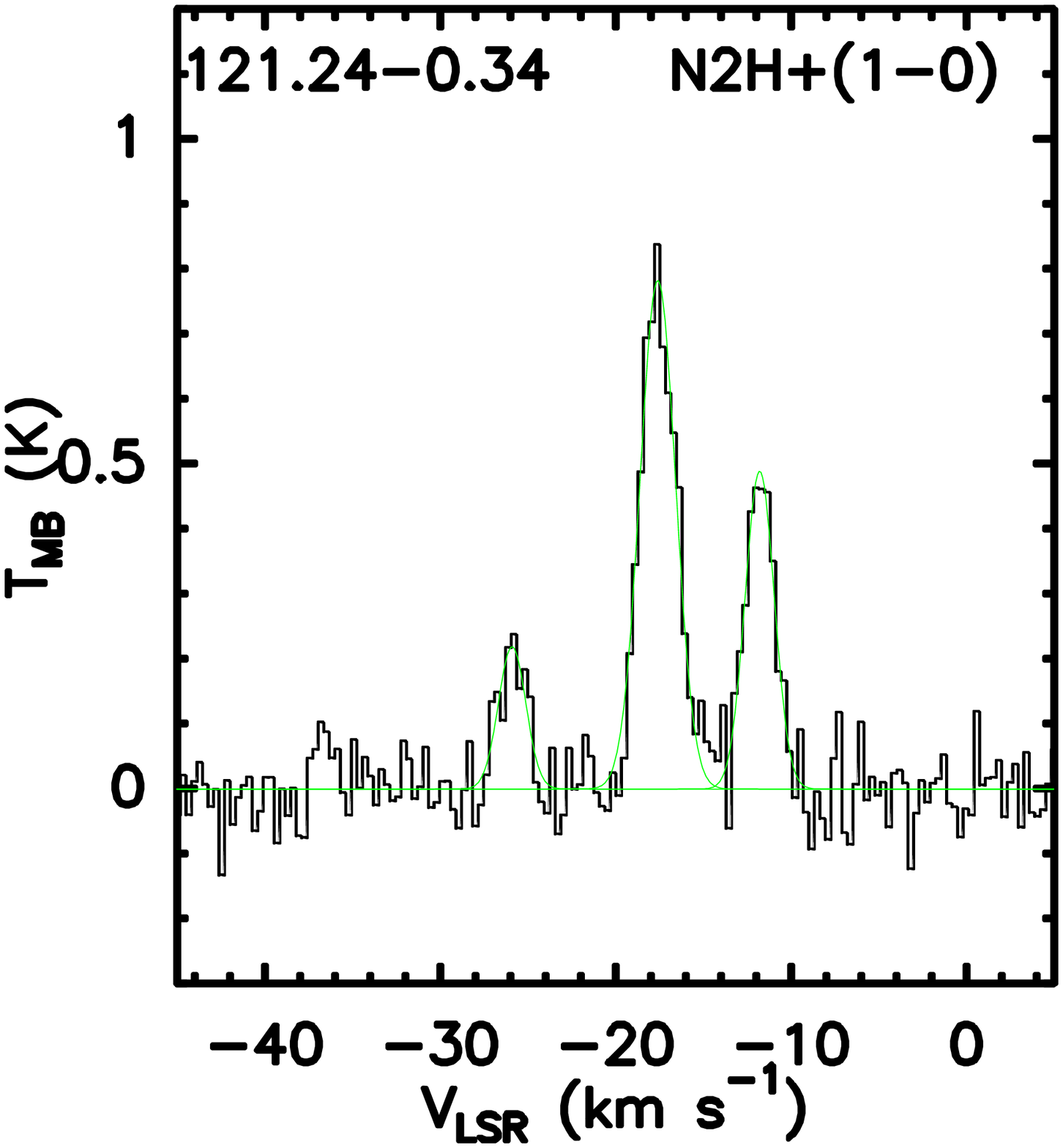}~~~
\includegraphics[width=40mm,angle=0]{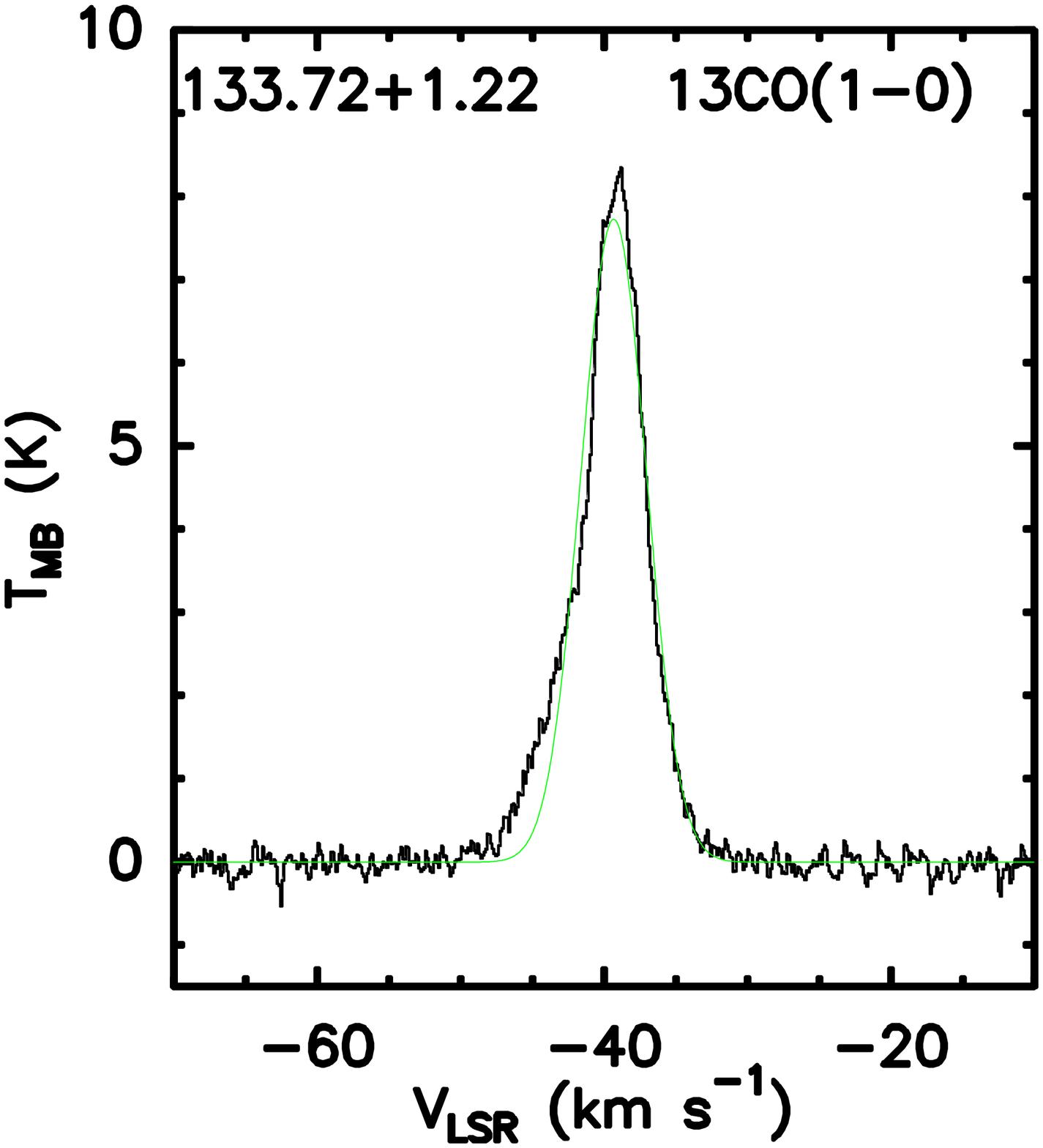}

\caption{\baselineskip 3.6mm $^{13}$CO(1--0), C$^{18}$O(1--0),
HCO$^+$(1--0) and N$_2$H$^+$(1--0) spectra at the positions of the
peak of the maser associated cores. The velocities are the radial
velocity with respect to the local standard of rest. The $y$-axis
is the main beam temperature (green lines are fitted profiles,
color online). }\label{Fig:spectra}
\end{figure}
\begin{figure}[h!]\setcounter{figure}{0}

\vs \centering
\includegraphics[width=38mm,angle=0]{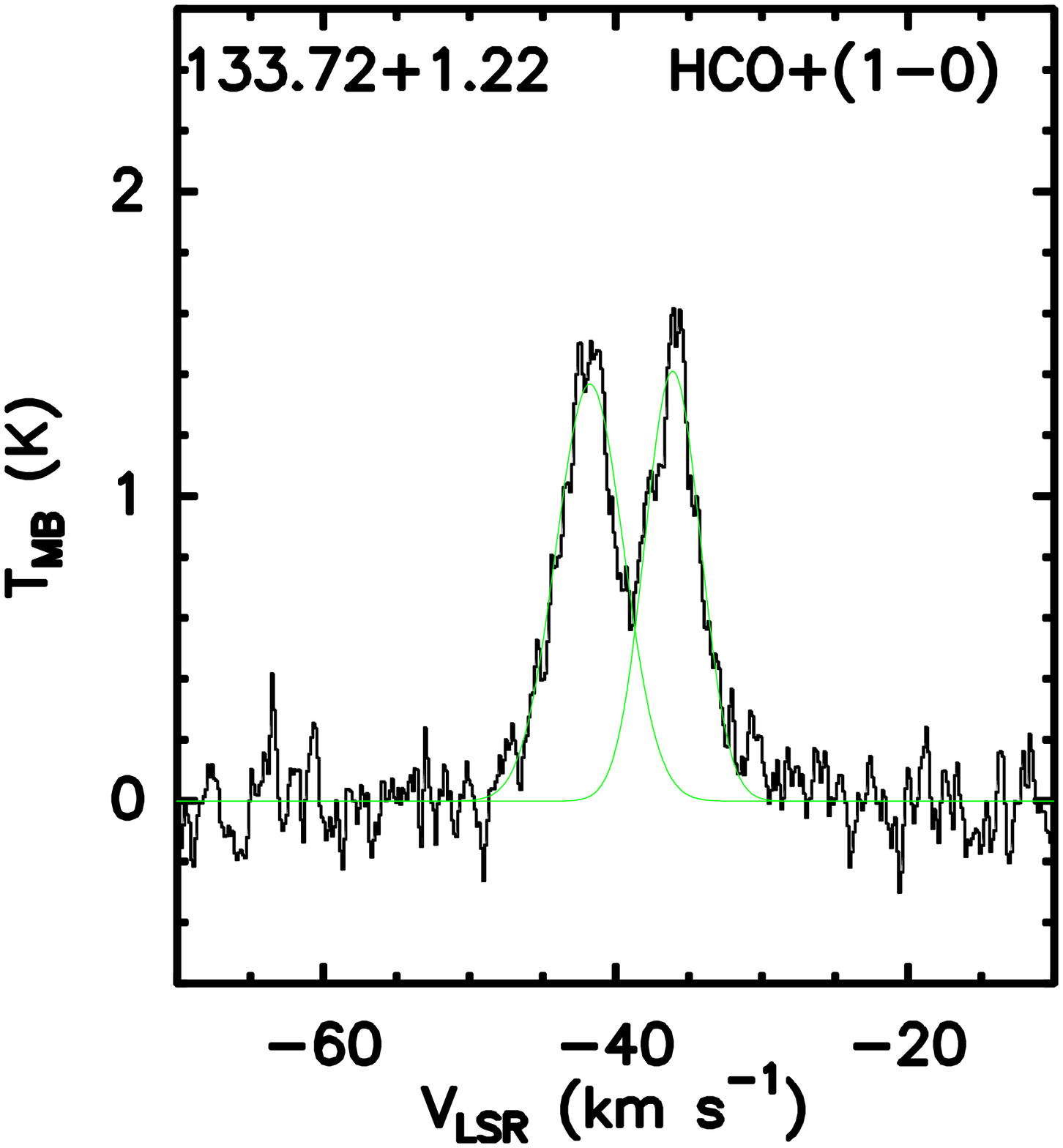}~~ 
\includegraphics[width=39mm,angle=0]{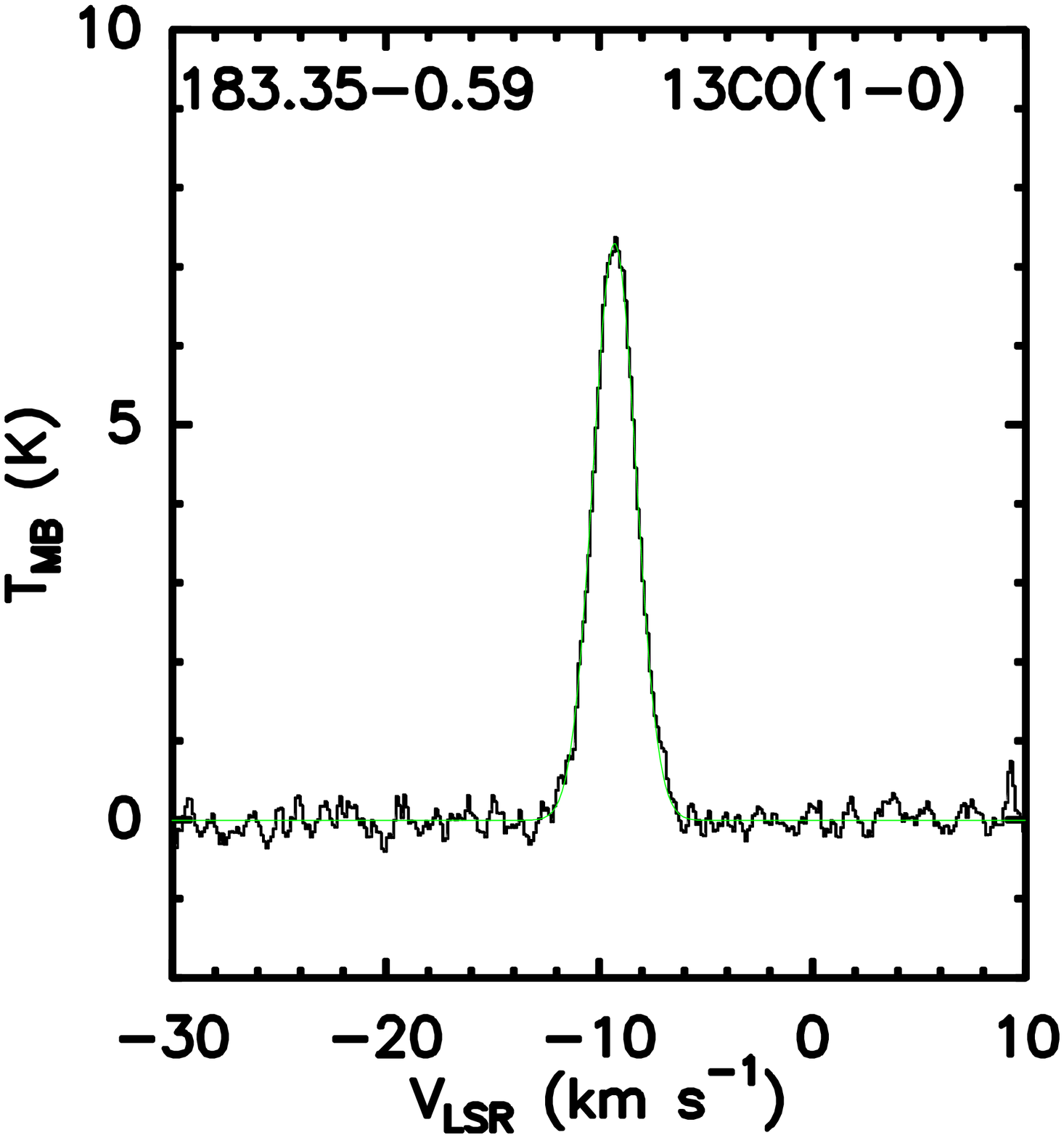}~~~
\includegraphics[width=38mm,angle=0]{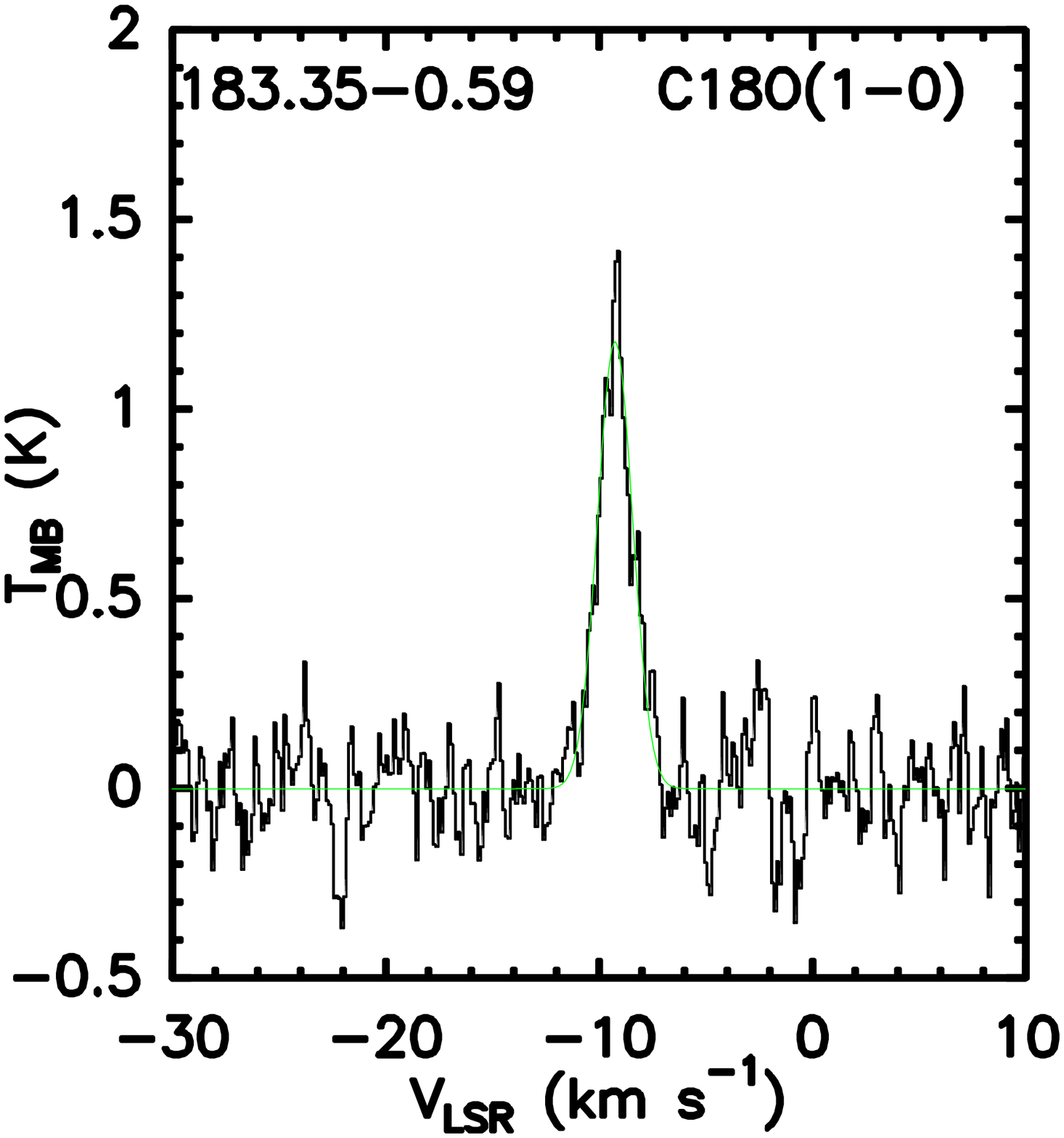}

\vs\vs
\includegraphics[width=38mm,angle=0]{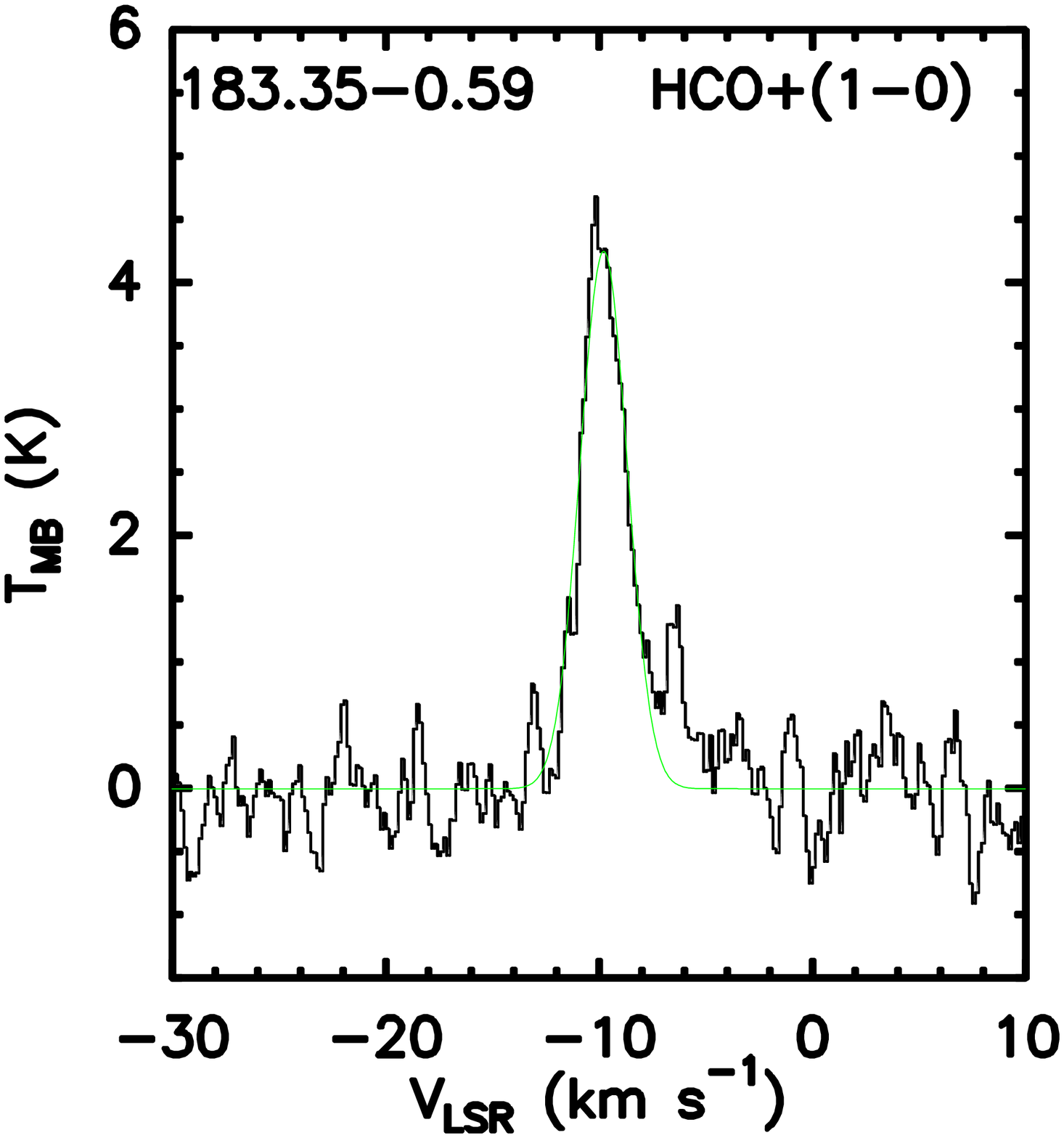}~~ 
\includegraphics[width=38mm,angle=0]{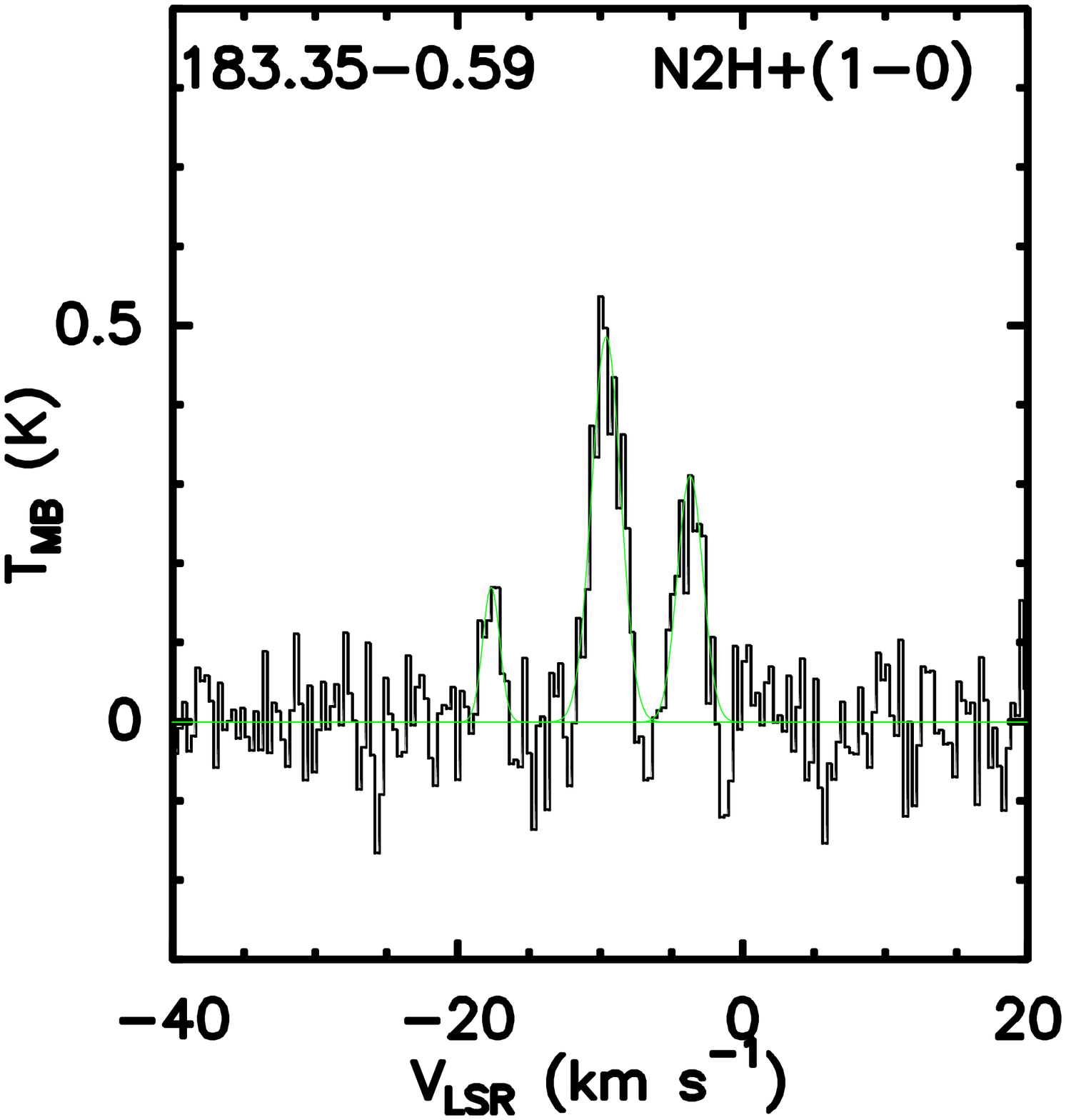}~~~
\includegraphics[width=38mm,angle=0]{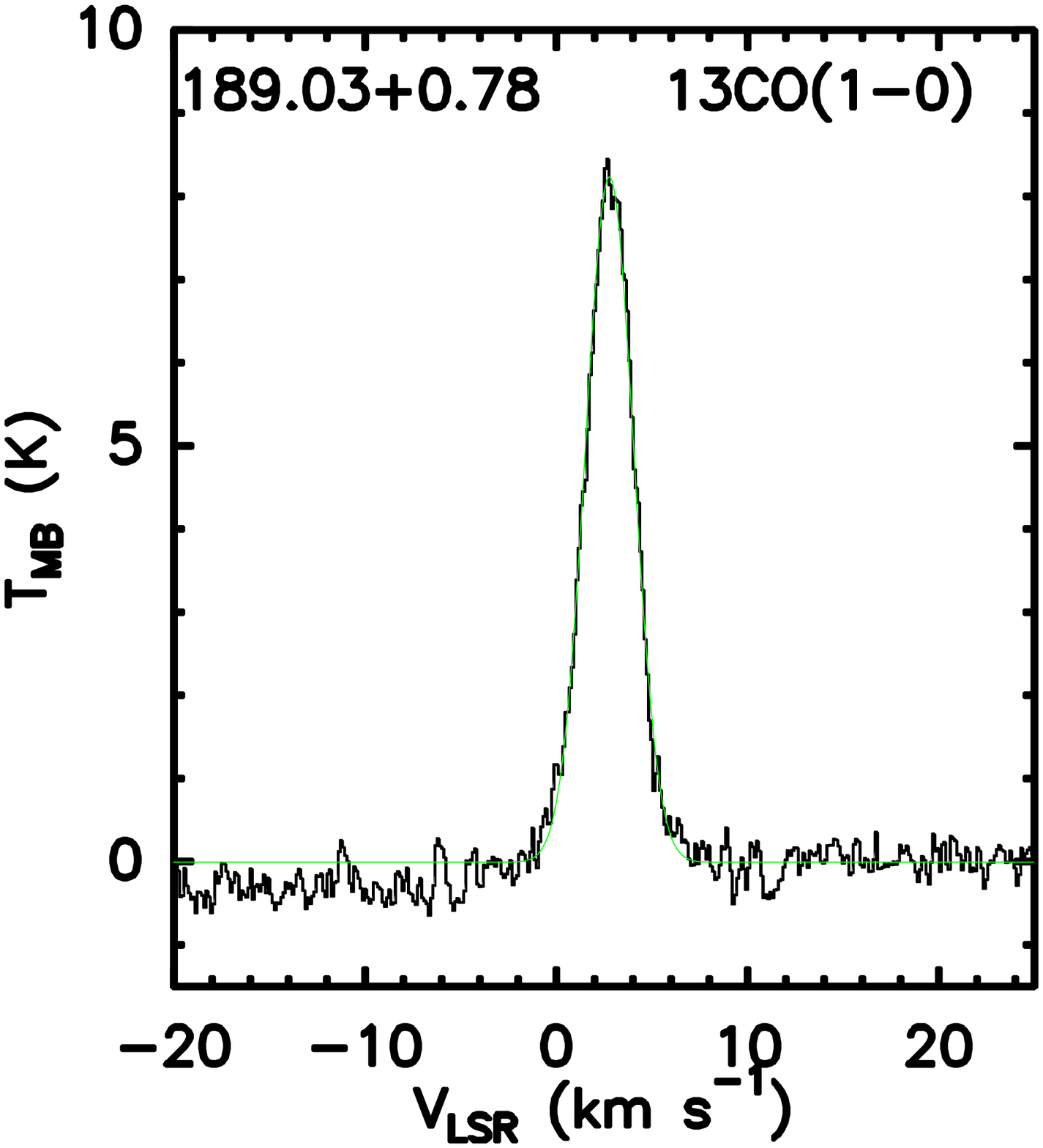}

\vs\vs
\includegraphics[width=38mm,angle=0]{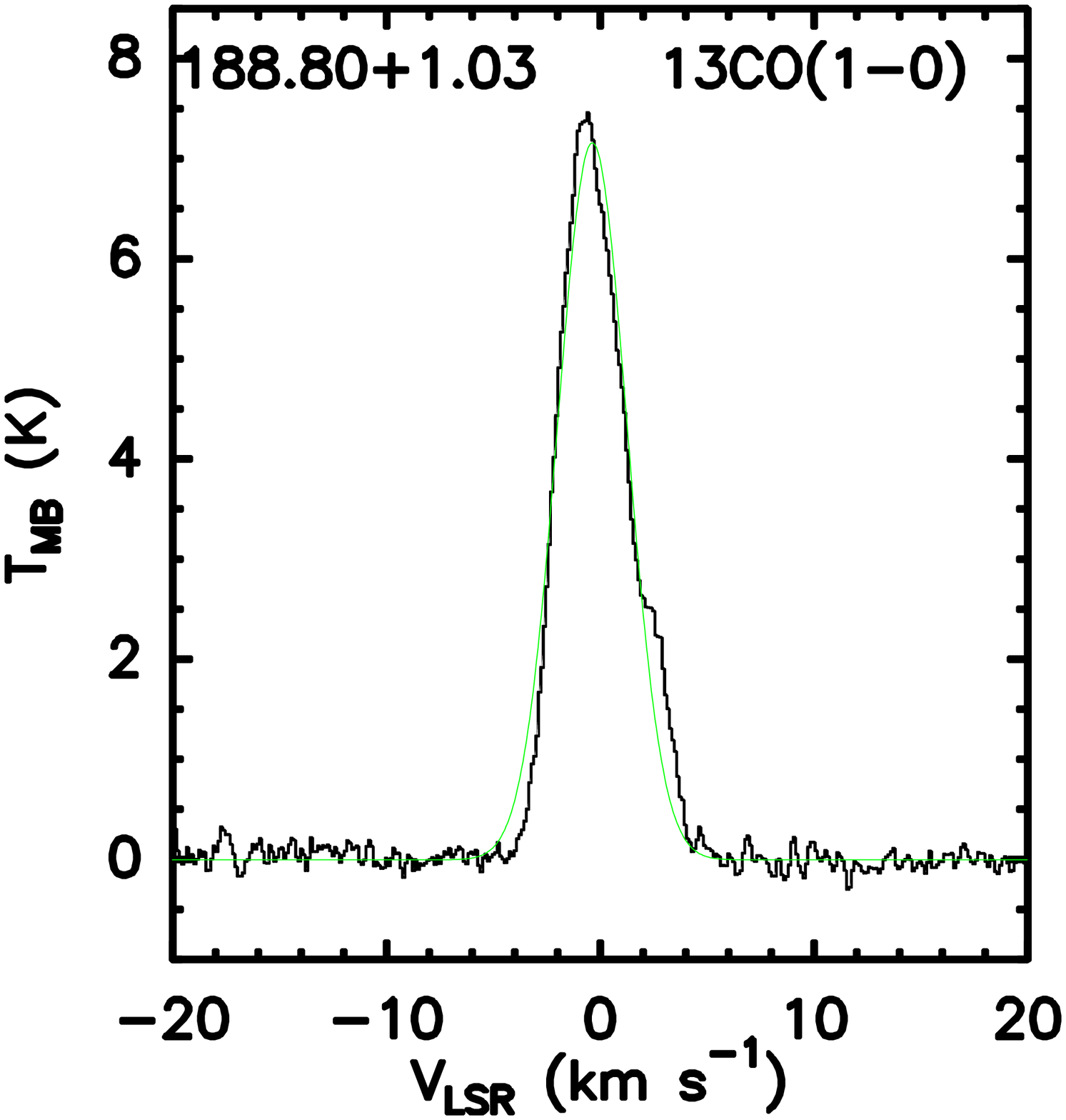}~~ 
\includegraphics[width=38mm,angle=0]{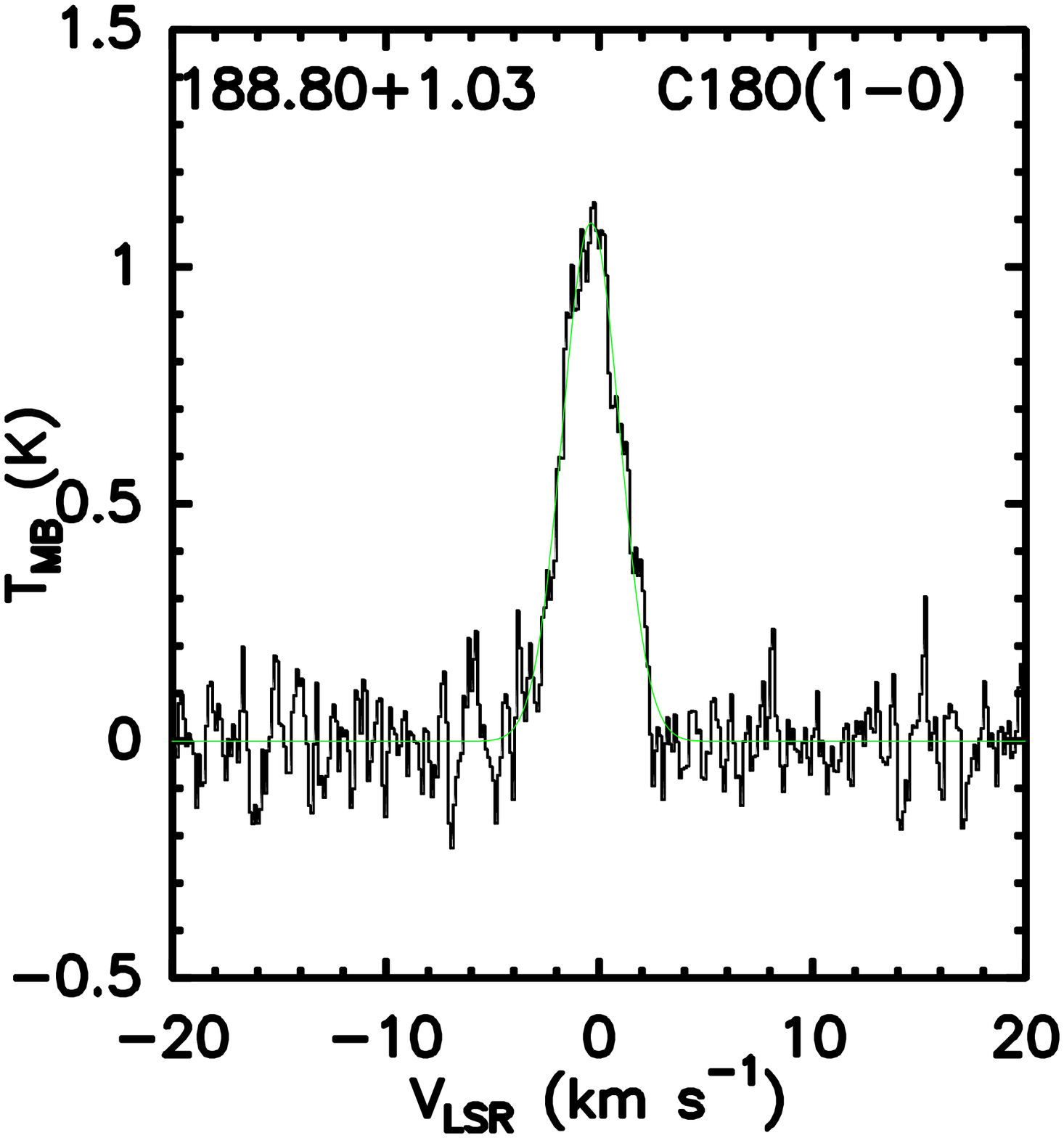}~~~
\includegraphics[width=38mm,angle=0]{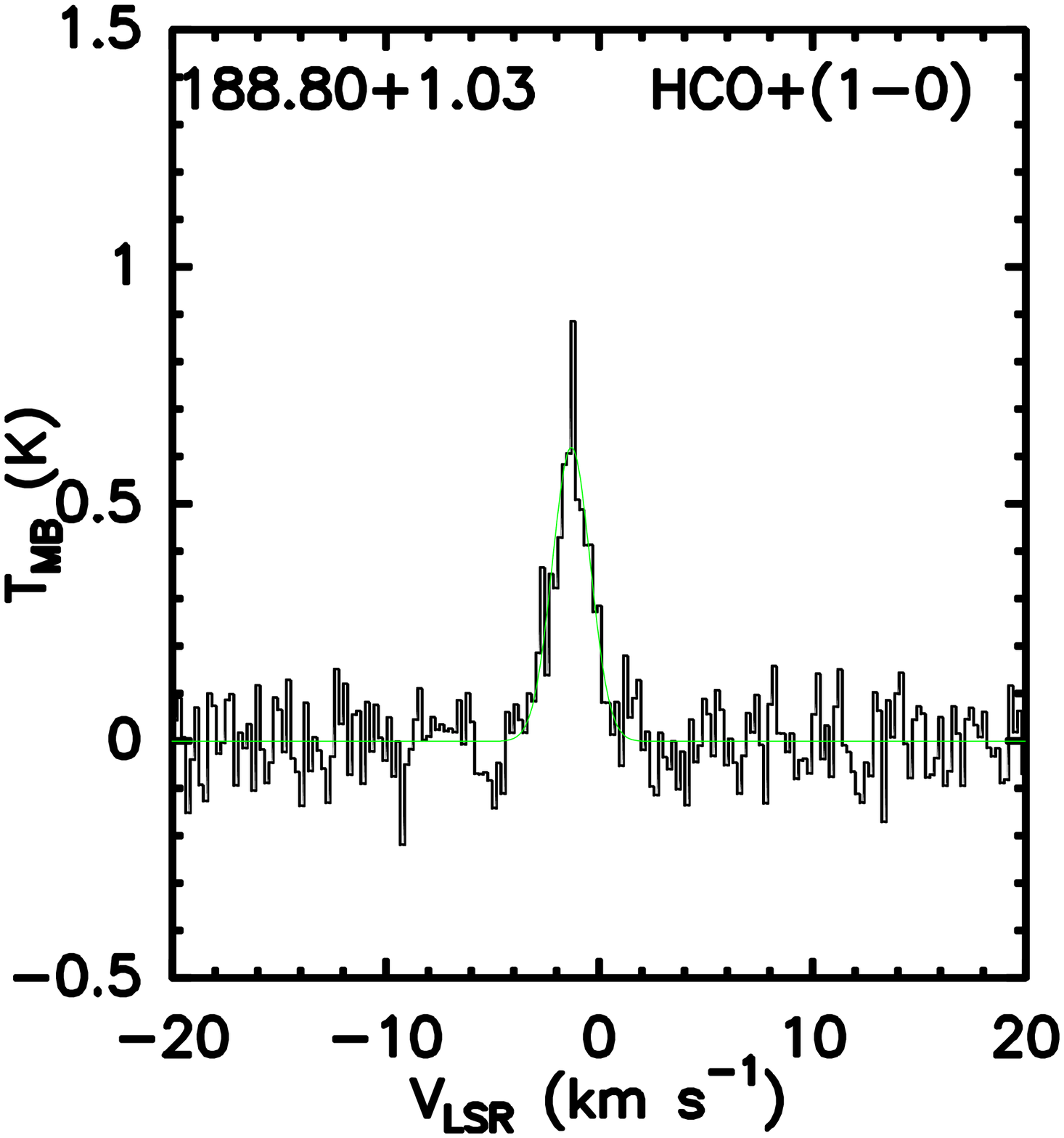}

\vs\vs
\includegraphics[width=38mm,angle=0]{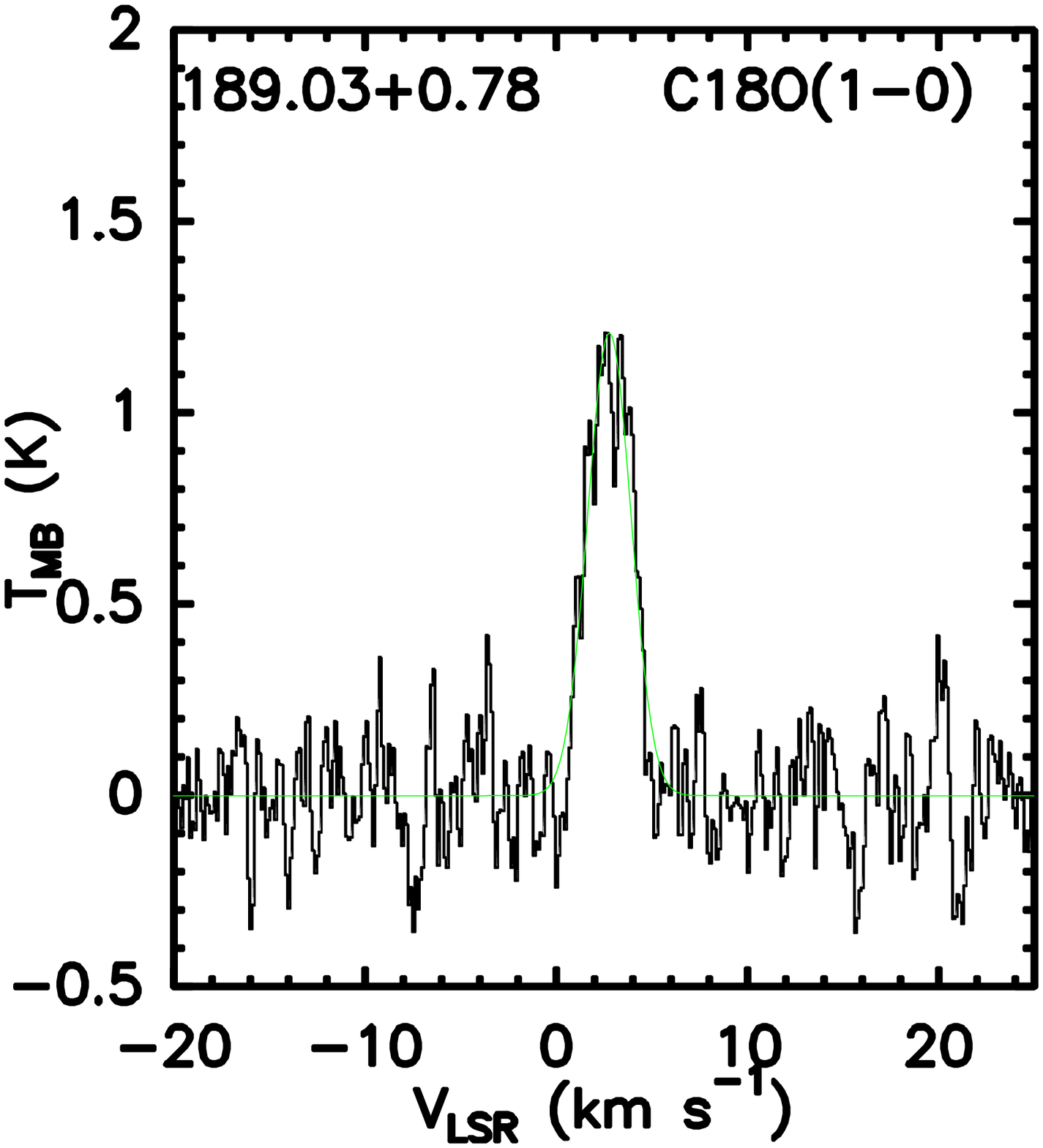}~~ 
\includegraphics[width=38mm,angle=0]{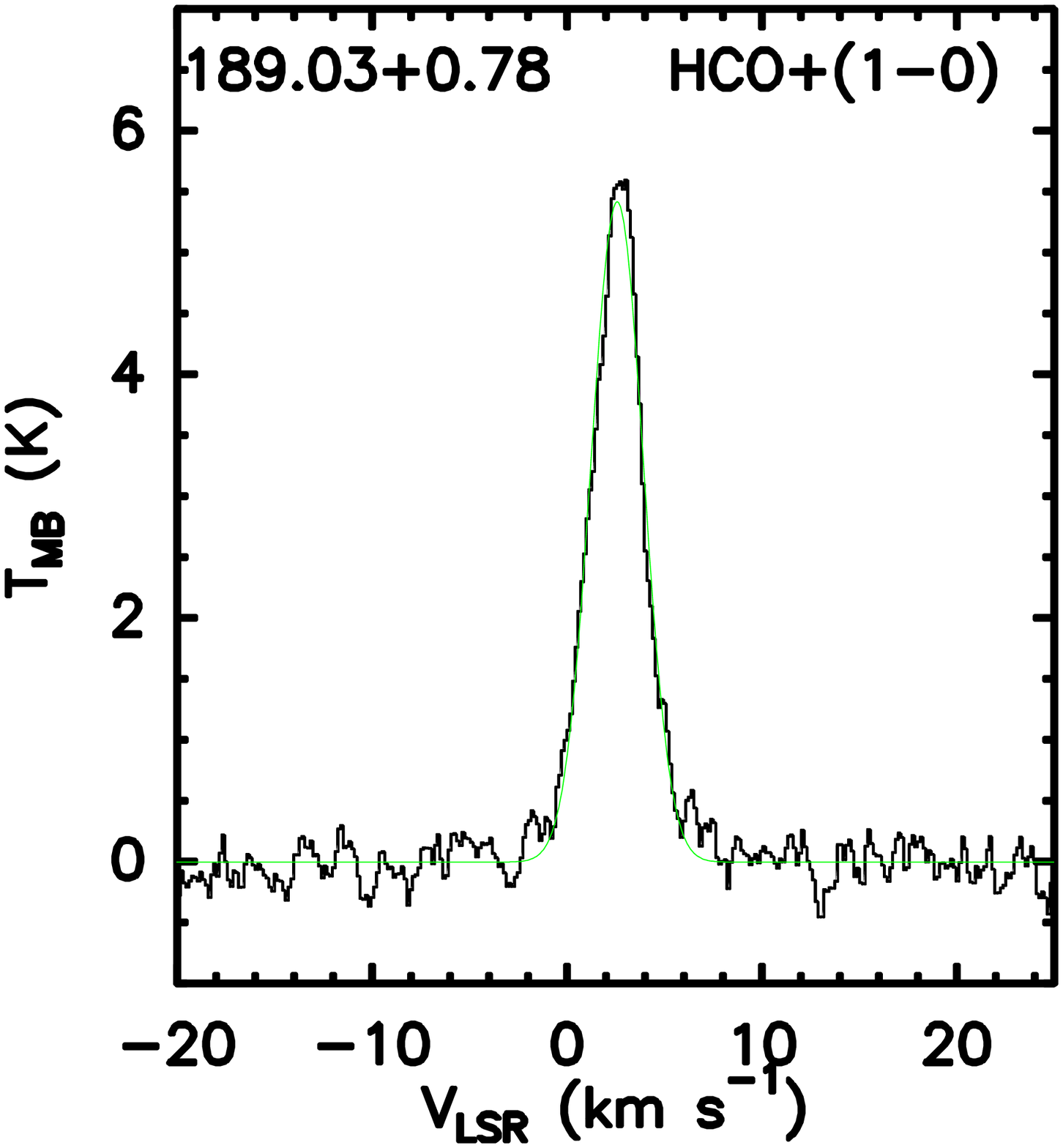}~~~
\includegraphics[width=39mm,angle=0]{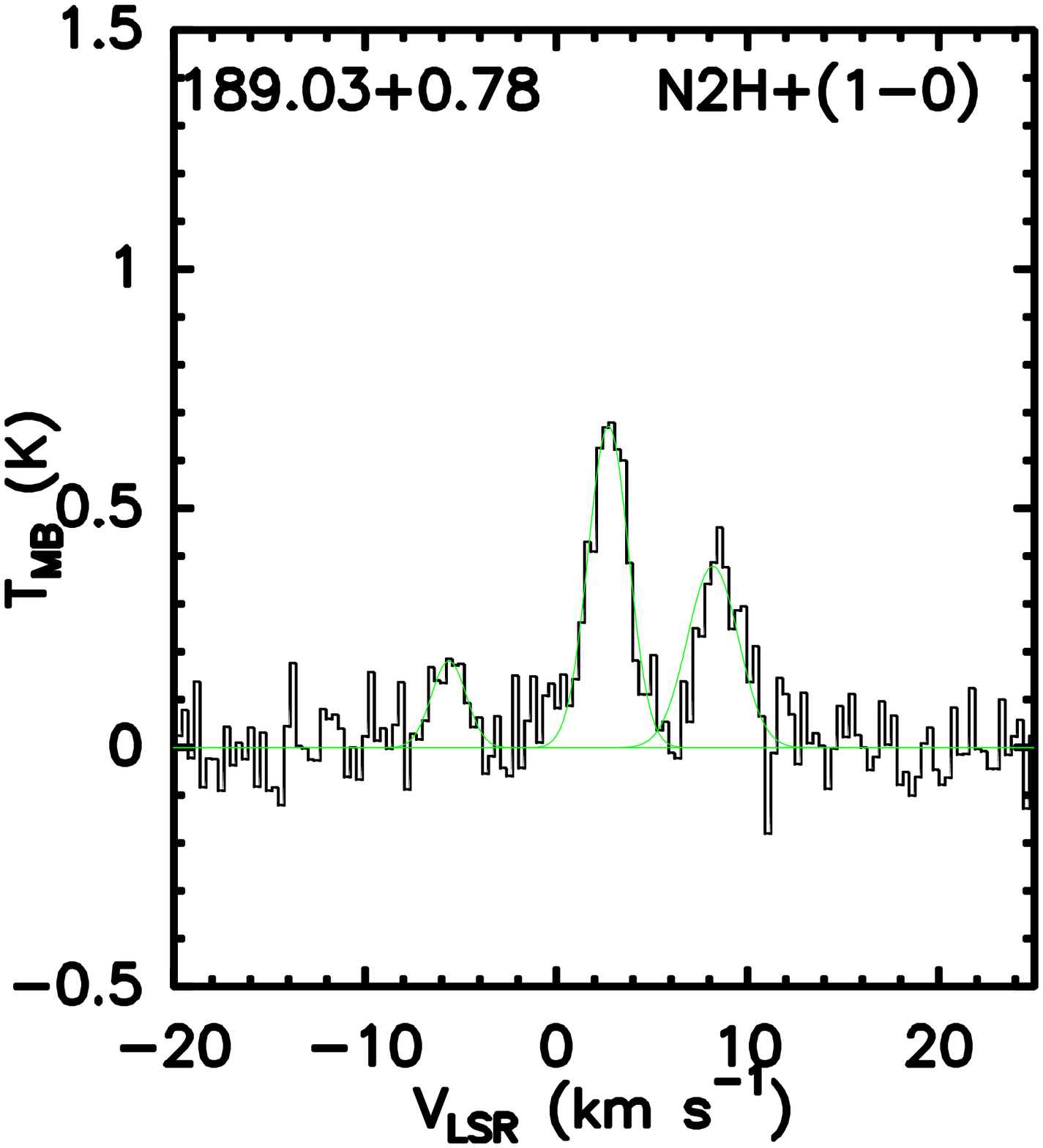}

\vs
\begin{minipage}[]{35mm}
\caption{ \it ---  Continued.} \end{minipage}
\vs\vs 
\end{figure}
\begin{figure}[h!!]\setcounter{figure}{0}

\centering \vs
\includegraphics[width=40mm,angle=0]{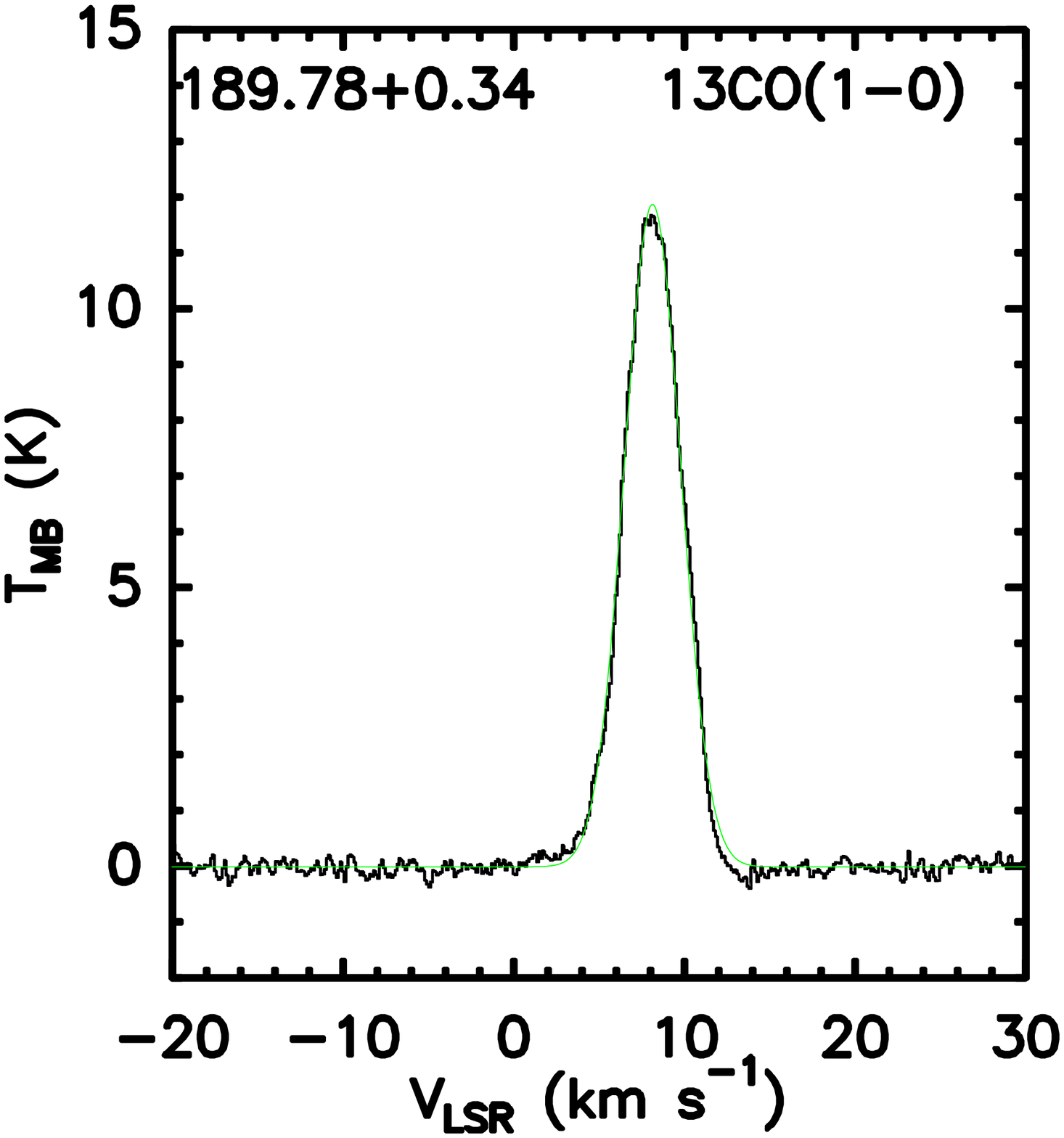}~~ 
\includegraphics[width=40mm,angle=0]{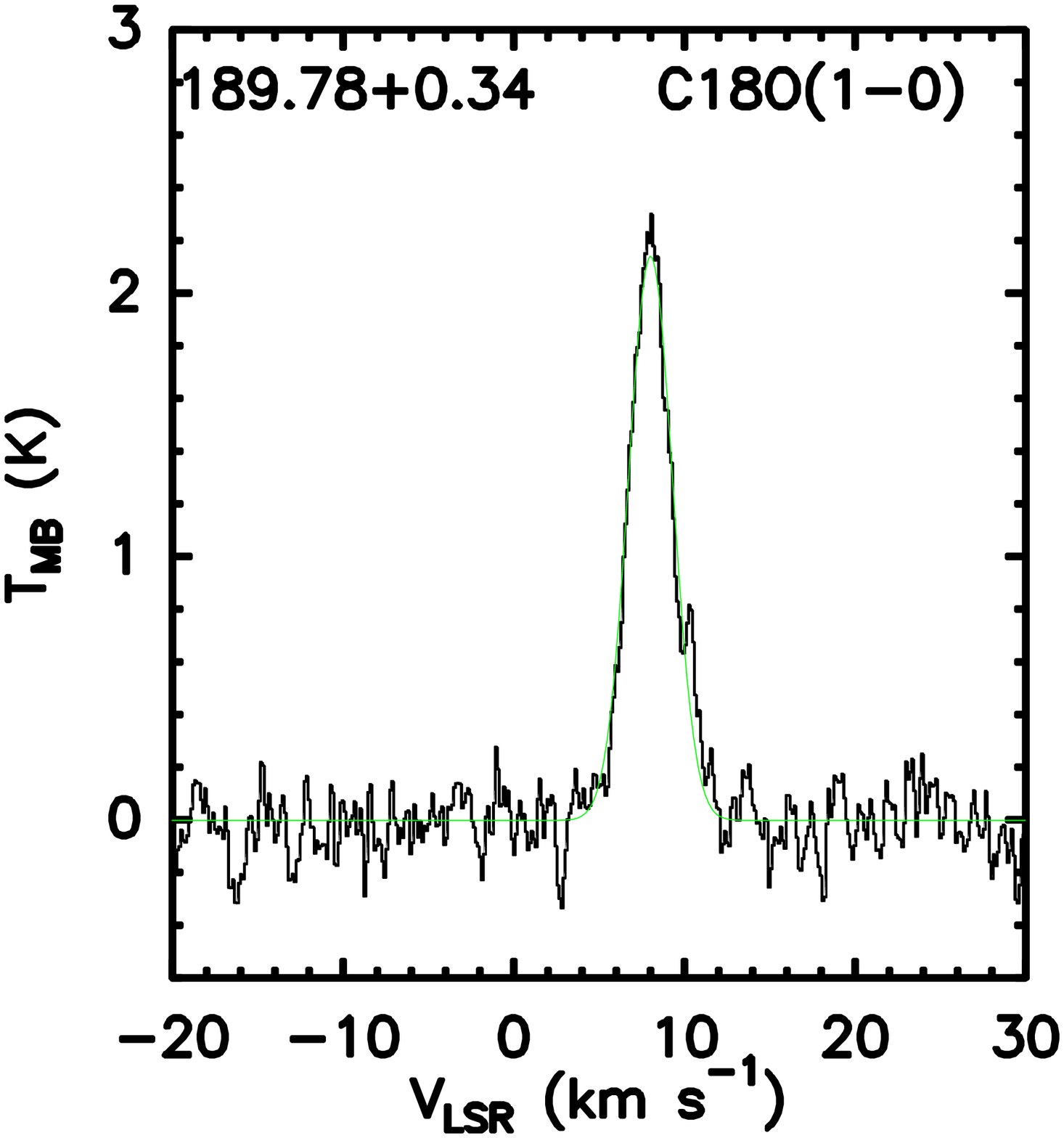}~~~
\includegraphics[width=40mm,angle=0]{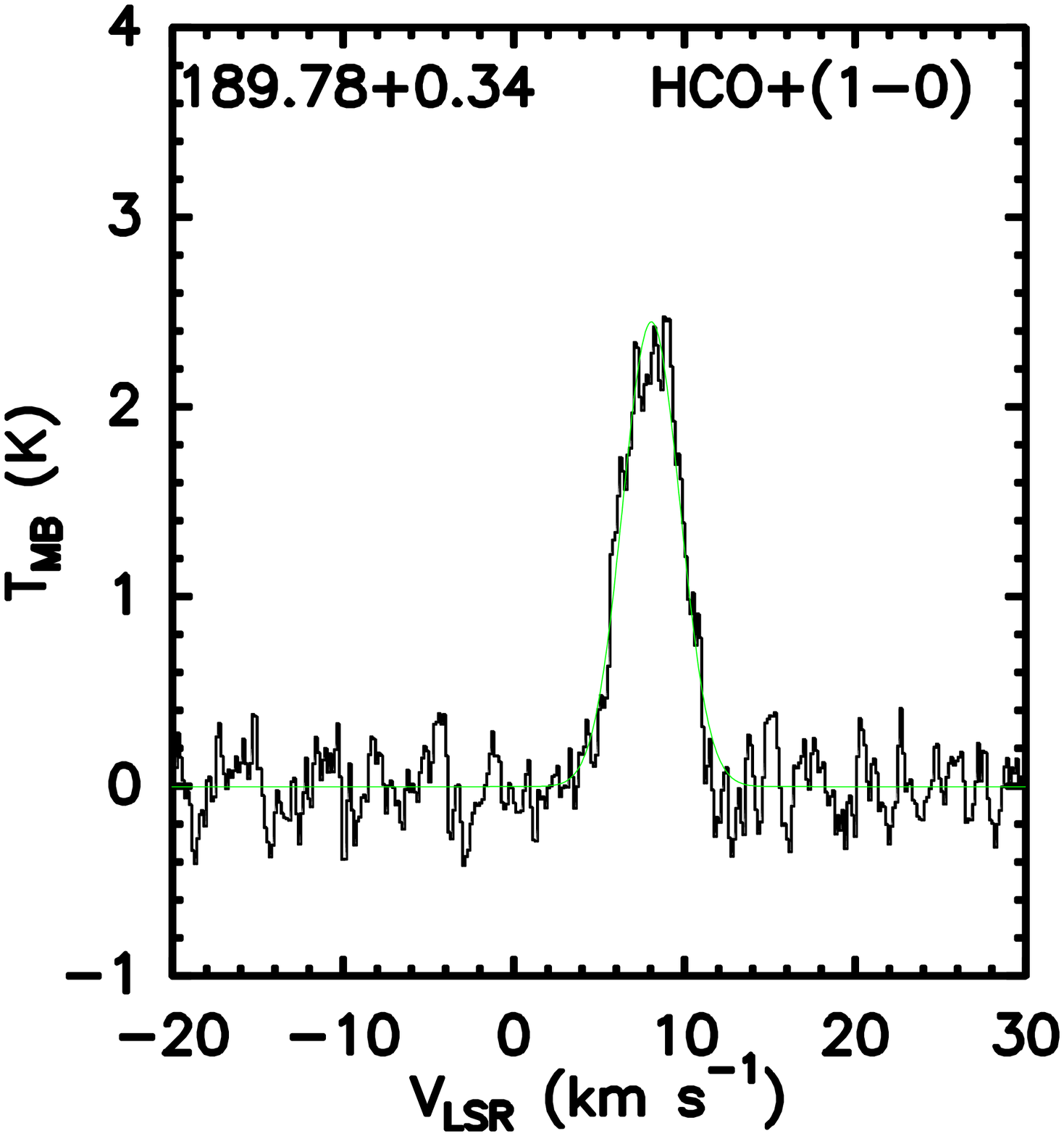}

\vs
\includegraphics[width=38mm,angle=0]{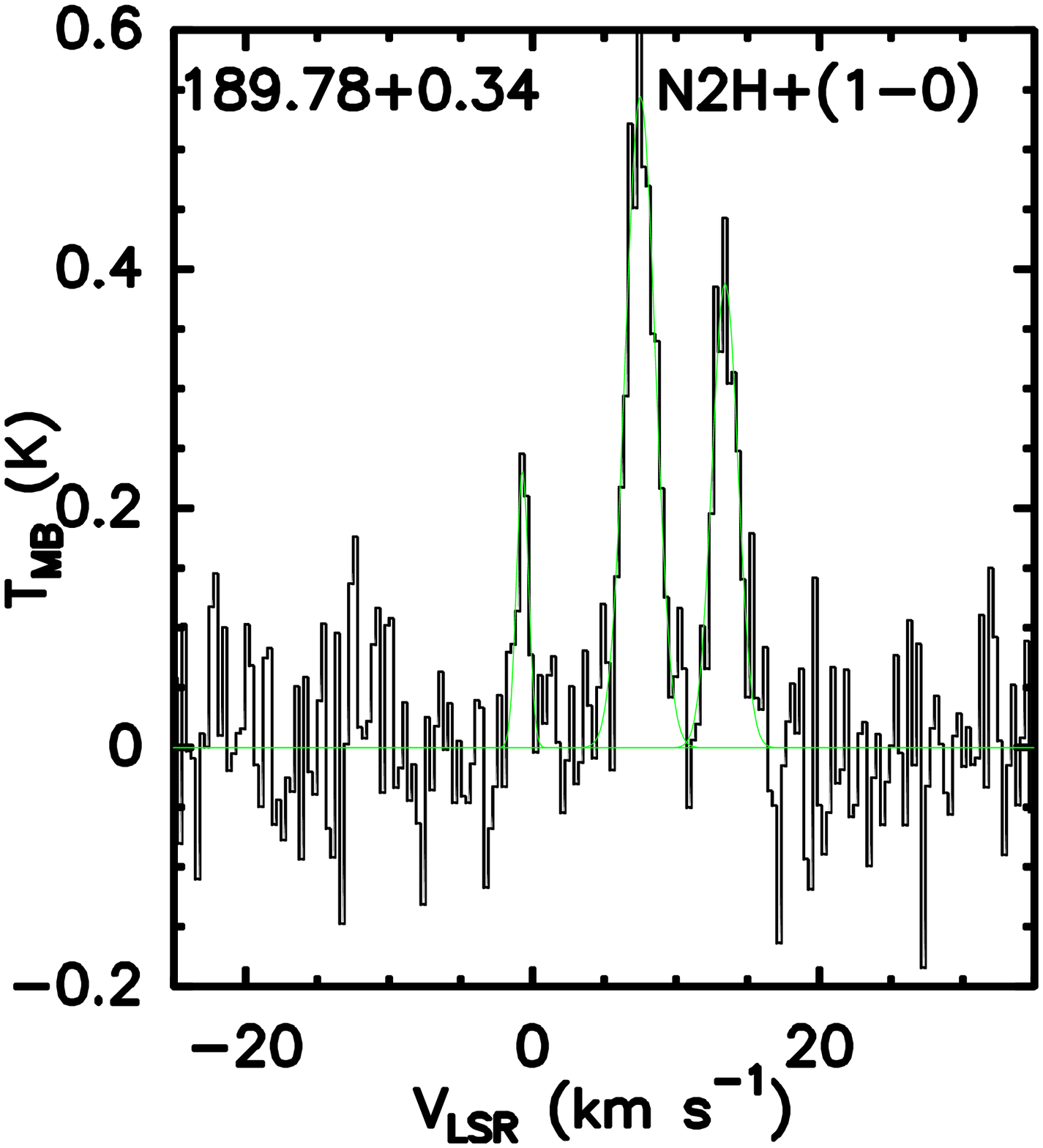}~~ 
\includegraphics[width=38mm,angle=0]{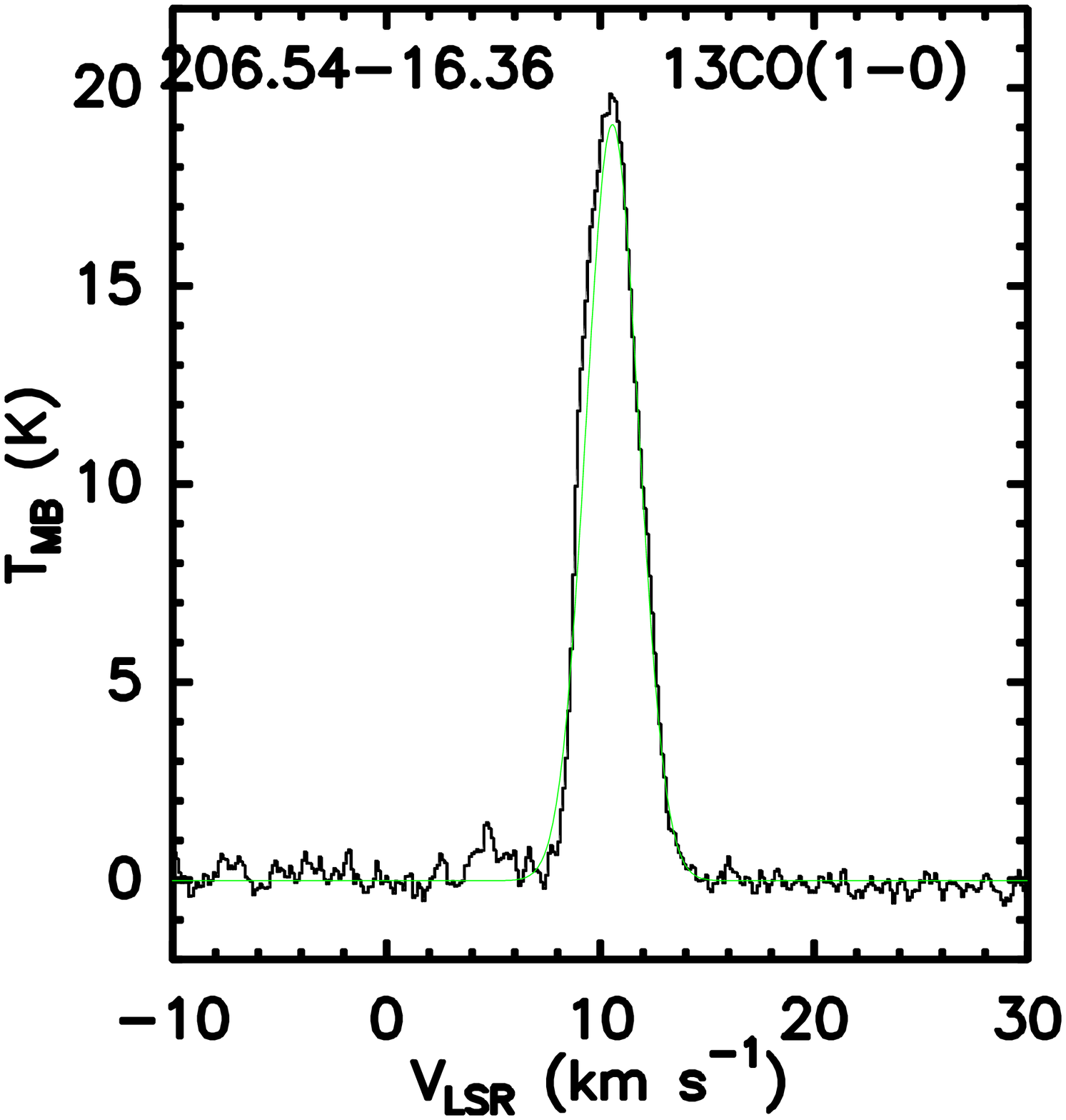}~~~
\includegraphics[width=38mm,angle=0]{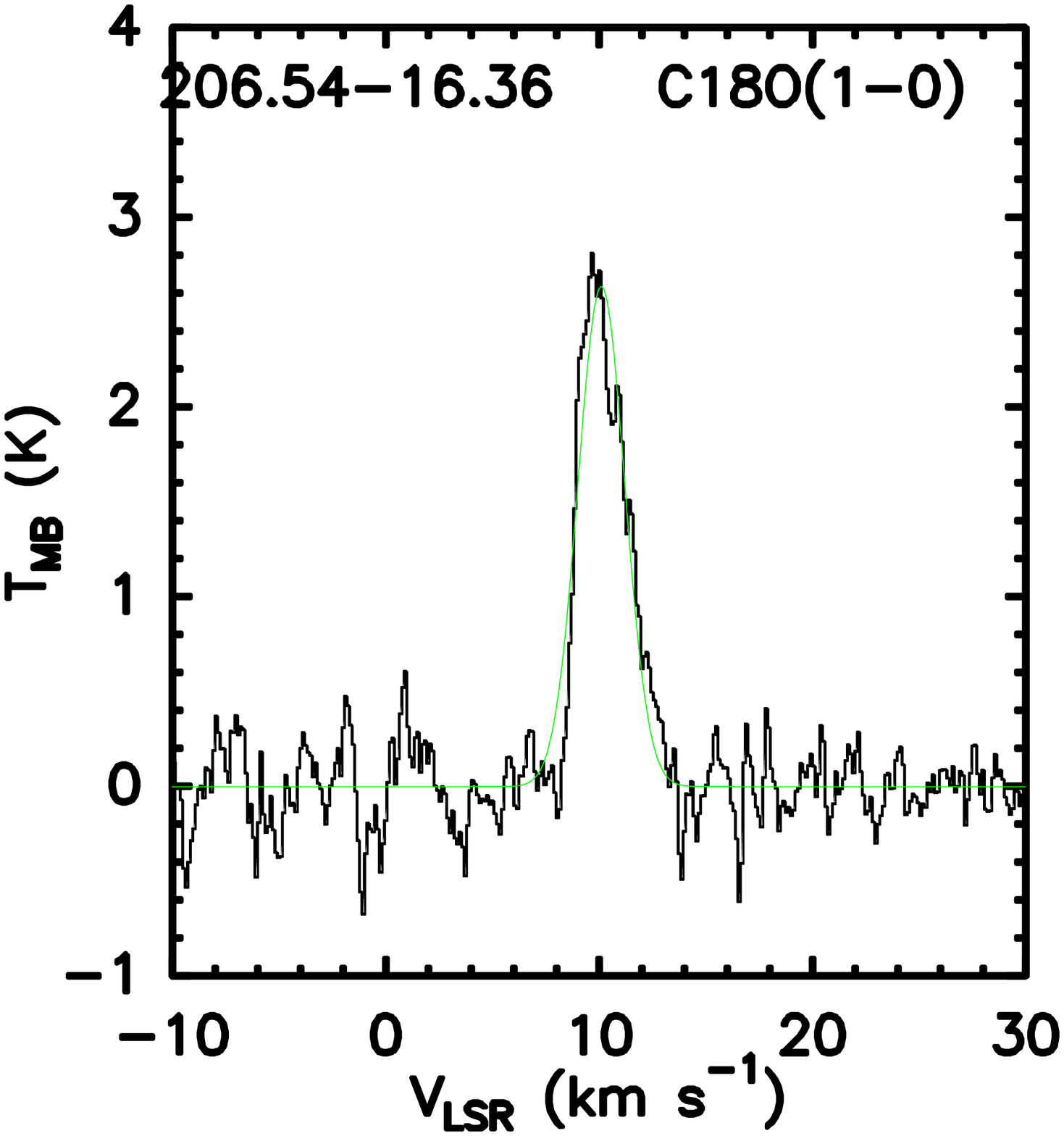}

\vs\vs
\includegraphics[width=40mm,angle=0]{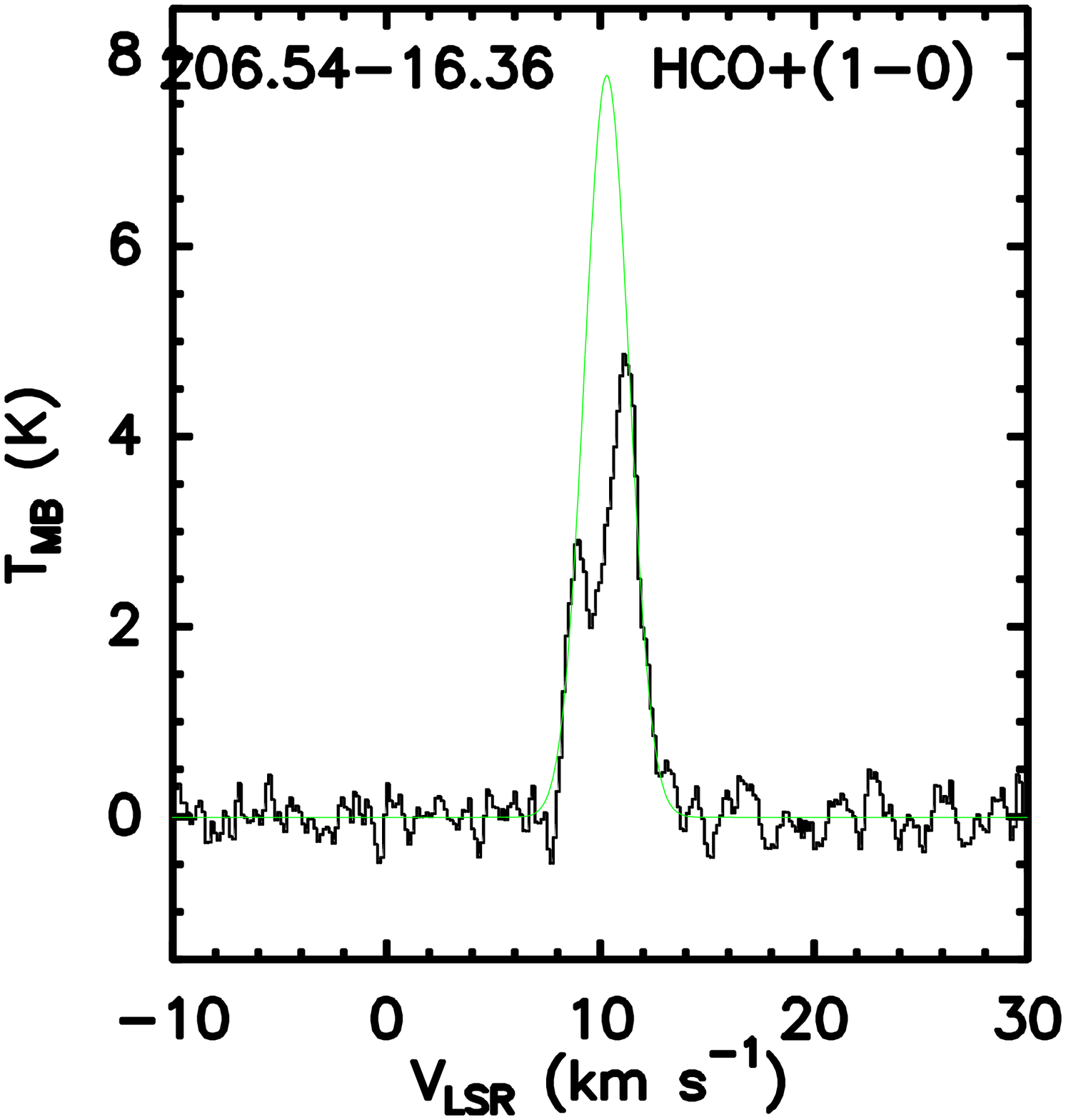}~~~~ 
\includegraphics[width=40mm,angle=0]{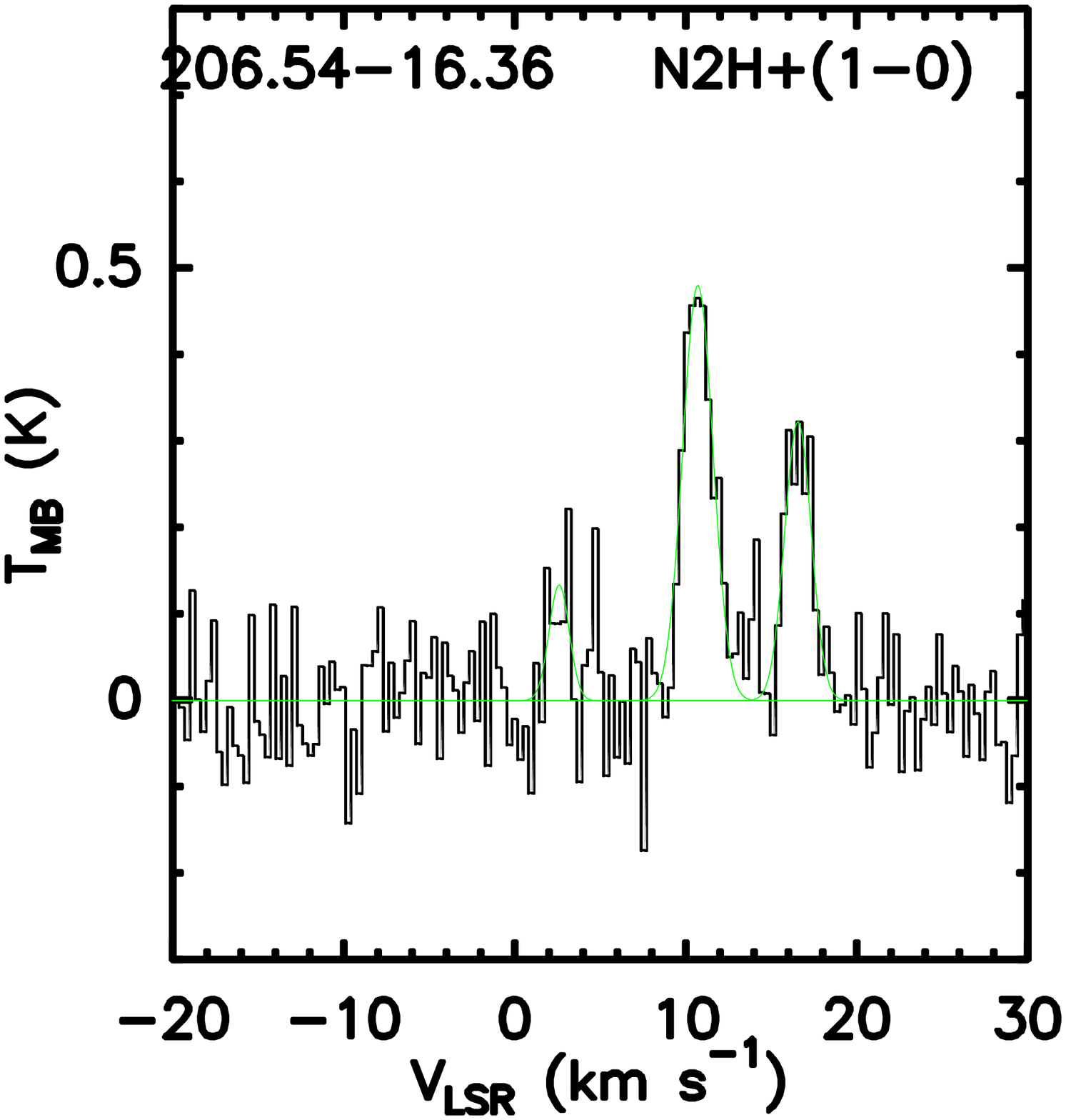}

\begin{minipage}[]{35mm}

\caption{ \it --- Continued.}\end{minipage} 

\pagebreak
\end{figure}

\subsection{Maps}

Contours of $^{13}$CO(1--0), C$^{18}$O(1--0) and HCO$^+$(1--0)
integrated intensities are presented in Figure~\ref{Fig:maps}.
 We use infrared dust emission, {from }the Midcourse
Space Experiment (MSX) E band (21\,$\mu$m) images, {as the }background {for} the
contours. For 106.80+5.31, where no MSX data are available, the
Multiband Imaging Photometer for Spitzer (MIPS) 24\,$\mu$m image is
used. The central blank pixels in the 24\,$\mu$m image {are}
due to saturation. The squares, triangles, pluses and
ellipses in Figure~\ref{Fig:maps} denote the 6.7-GHz CH$_3$OH
masers, H$_2$O masers, OH masers and error ellipses of the IRAS
point sources in the fields, respectively. We use C$^{18}$O(1--0)
(optical thin line) contours to define a concentrated structure,
i.e., a core, and determine the size and position of the core. {Although}
for 133.72+1.22 where there is no detection of C$^{18}$O, the cores
are identified from{ the} $^{13}$CO map and core 1 of 189.78+0.34, where
there is no C$^{18}$O emission peak, is identified from{ the} HCO$^+$ map.
Cores were identified using visual inspection. The methodology used
to define cores is similar to the operation of the ``CLUMPFIND''
algorithm \citep{william94}. Totally, we find 17 cores from the
{nine} fields, with {eight} cores {showing no}
maser associations.

\begin{table}[b!!]

\centering

\renewcommand\arraystretch{1}

\caption{$^{13}$CO(1--0), C$^{18}$O(1--0),{ and} HCO$^+$(1--0) spectral parameters,
including bright temperatures, integrated intensities, central velocities and fitted
line widths.}\label{Tab:linepara}

\tiny
\tabcolsep 0.5mm

\begin{tabular}{lccccccccccccccccc}
\hline \noalign{\smallskip}
 & & & \multicolumn{4}{c}{$^{13}$CO}&  & &
 \multicolumn{4}{c}{C$^{18}$O}& & \multicolumn{4}{c}{HCO$^+$}
 \tabularnewline\noalign{\smallskip}
\cline{4-7} \cline{9-12} \cline{14-18} \noalign{\smallskip} {Region}  &
{Core$^\alpha$} & & $T_{R}^{\ast}$ & {\small $\int$}$T_{R}^{\ast}dV$ & $V_{\rm LSR}$
& $\Delta V$ & & $T_{R}^{\ast}$ & {\small $\int$} $ T_{R}^{\ast}dV$ & $V_{\rm LSR}$ &
$\Delta V$ & & $T_{R}^{\ast}$ & {\small $\int$}$ T_{R}^{\ast}dV$ & $V_{\rm LSR}$ &
$\Delta V$ \tabularnewline & & & (K) & (K km s$^{-1}$) & (km s$^{-1}$) & ( km
s$^{-1}$) &   & (K) & (K km s$^{-1}$) & (km s$^{-1}$) & (km s$^{-1}$) &  &(K)  & (K
km s$^{-1}$) & ( km s$^{-1}$) & ( km s$^{-1}$) & \tabularnewline \hline 106.80+5.31 &
1 & &12.22 &42.34 &--7.46(01)  &3.26(02) & &1.85 &5.73
&--7.52(04) & 2.91(11)  & &17.82 &53.71 & --6.74(01)  &2.83(02)   \\
111.25--0.77  & 1 & &9.78   &27.43  &--44.09(01) &2.64(03) & &1.14  &2.49
 &--44.28(04)& 2.05(11)  & &1.37  &4.79  & --44.02(05) &3.27(14)   \\
121.24--0.34  & 1 & &9.06   &21.46  &--17.39(01) &2.22(03) & &1.57
 &3.11  &--17.15(04)& 1.87(08)  & &8.70  &27.75 & --17.53(02) &3.00(06)   \\
133.72+1.22  & 1 & &8.35   &29.18  &---        &---      & &---   &---   &---
 & ---       & &1.81  &9.52  & --40.57(05) &4.91(14)   \\
133.72+1.22  & 2 & &5.67   &29.75  &--43.37(01) &4.93(03) & &---   &---   &---
  & ---       & &1.41  &7.39  & --41.89(05) &4.91(14)   \\
133.72+1.22  & 3 & &8.17   &46.05  &--38.88(03) &5.40(02) & &---   &---   &---
    & ---       & &1.40  &7.00  & --36.14(05) &4.71(14)   \\
133.72+1.22  & 4 & &9.06   &28.15  &--38.88(02) &2.92(02) & &---   &---   &---
     & ---       & &1.38  &4.14  & --39.50(05) &2.81(15)   \\
183.35--0.59  & 1 & &7.30   &18.44  &--9.25(01)  &2.37(02) & &1.16  &2.34
 &--9.24(04) & 1.90(10)  & &4.25  &12.17 & --9.76(05)  &2.69(15)   \\
188.80+1.03  & 1 & &6.99   &20.63  &--0.58(01)  &2.77(02) & &1.00  &2.83
&--0.39(06) & 2.65(15)  & &0.68  &1.81  & --1.25(06)  &2.54(16)   \\
188.80+1.03  & 2 & &6.96   &21.90  &--0.17(01)  &2.96(02) & &1.35
 &2.97  &--0.41(02) & 2.08(05)  & &0.66  &1.99  & --1.15(07)  &2.83(16)   \\
189.03+0.78  & 1 & &8.31   &26.70  &2.79(01)   &3.02(02) & &1.26  &3.66  &2.81(06)
 & 2.74(13)  & &5.44  &18.47 & 2.59(01)   &3.19(04)   \\
189.03+0.78  & 2 & &8.56   &24.27  &2.32(01)   &2.66(03)
& &1.63  &3.82  &2.13(03)  & 2.20(09)  & &3.06  &10.19 & 2.28(35)   &3.12(09)   \\
189.78+0.34  & 1 & &12.26  &41.28  &8.64(01)   &3.16(02) & &2.21  &4.81  &8.55(02)
 & 2.24(05)  & &2.77  &13.52 & 7.99(04)   &4.58(09)   \\
189.78+0.34  & 2 & &11.89  &49.23  &8.15(01)   &3.89(01) & &2.39  &7.18  &8.04(02)  & 3.03(05)
 & &2.05  &8.32  & 7.70(05)   &3.83(13)   \\
189.78+0.34  & 3 & &10.92  &45.76  &8.31(01)   &3.94(01) & &2.68  &5.09  &7.67(02)  & 1.96(07)
 & &1.39  &5.83  & 7.78(08)   &3.94(19)   \\
206.54--16.36 & 1 & &20.43  &57.08  &10.33(01)  &2.68(01) & &3.81  &7.49  &10.04(02)
& 1.94(06)
& &3.47  &10.97 & 10.06(03)  &2.66(09)   \\
206.54--16.36 & 2 & &21.27  &66.54  &10.61(01)  &2.94(02) & &3.17  &9.48  &10.28(02)
& 2.81(04)
& &4.45  &17.54 & 10.72(04)  &3.70(07)   \\
 \hline\noalign{\smallskip}
\end{tabular}
\parbox{130mm}
{\fns Notes:
 $\alpha$: 1, 2, 3,{ and} 4 denote the different components in the same
 field.}
\end{table}

\begin{figure}[h!!]

\vs\vs

\centering

\hspace{0.5mm}\includegraphics[height=53mm,angle=90]{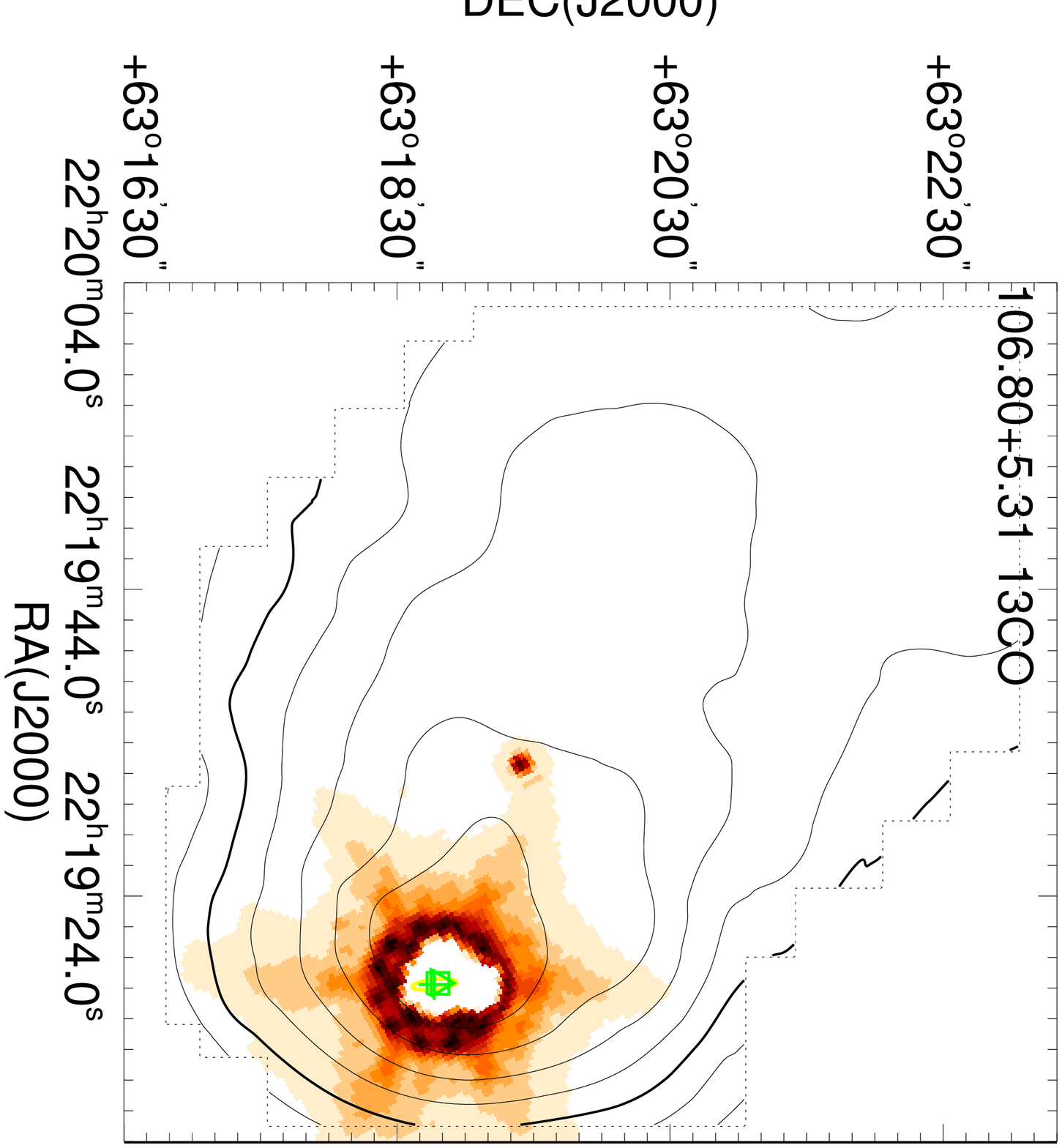} %
\hspace{-11mm}\includegraphics[height=53mm,angle=90]{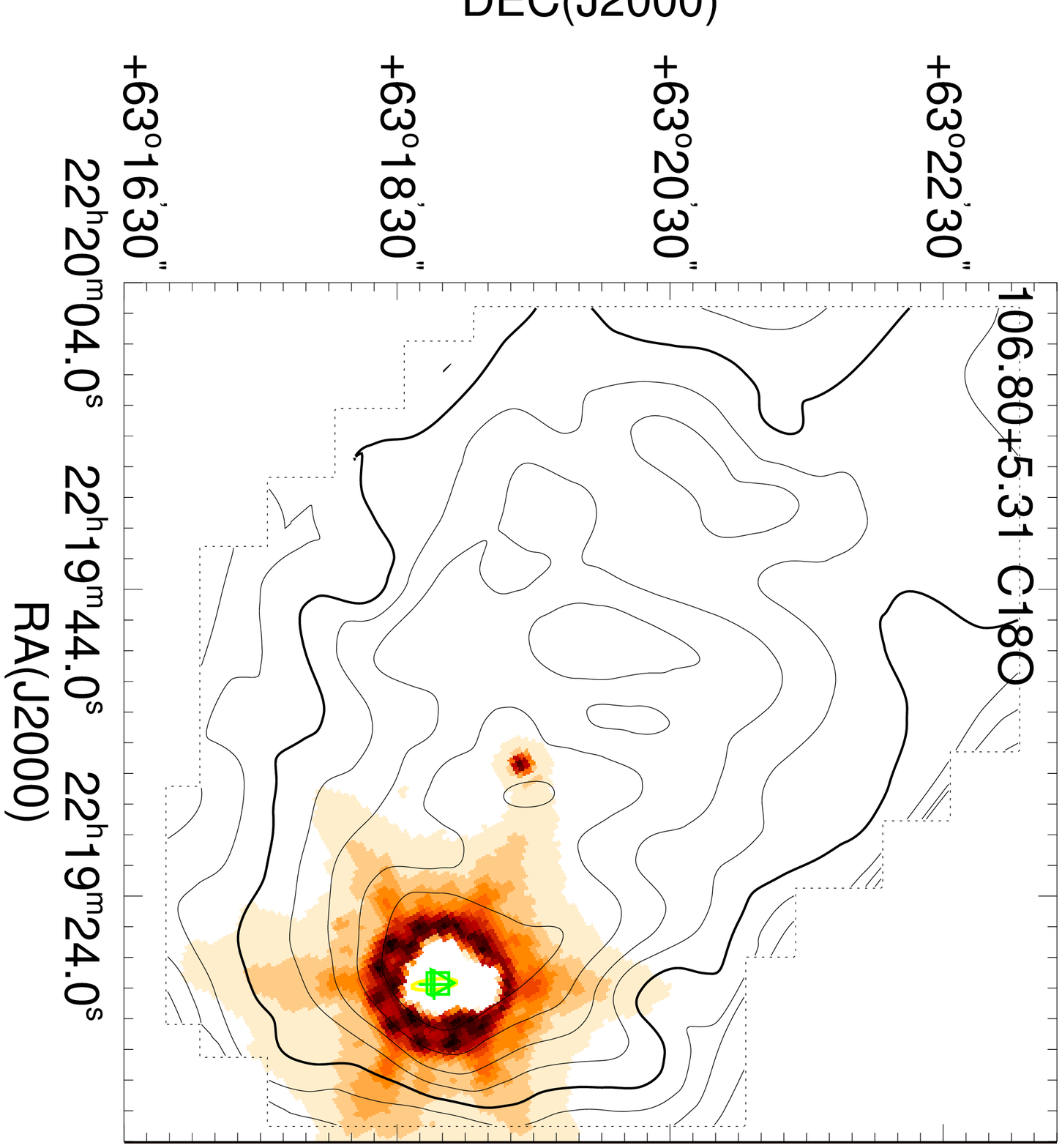}
\hspace{-11mm}\includegraphics[
height=53mm, angle=90]{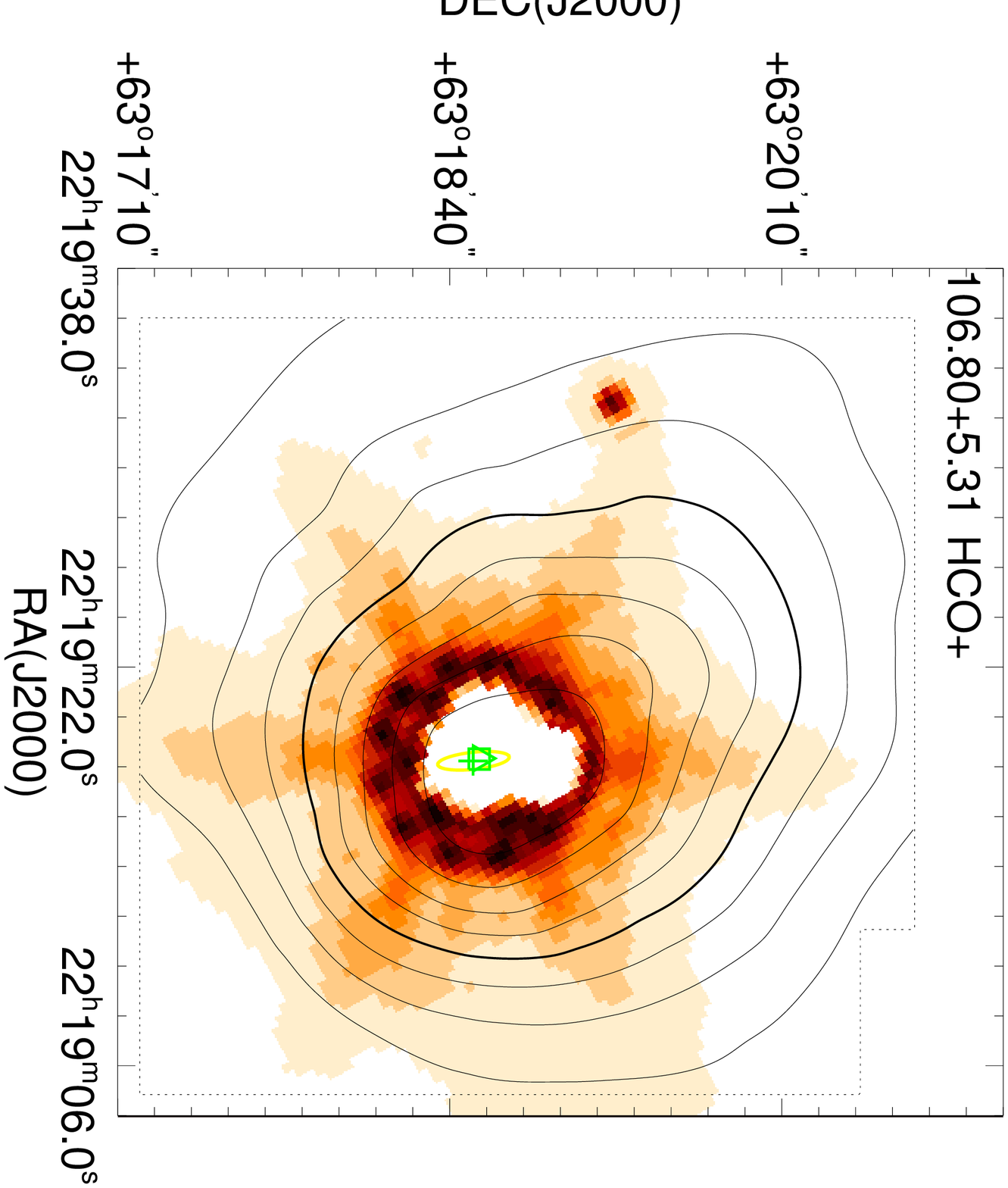}%

\vs
\hspace{0.5mm}\includegraphics[height=53mm,angle=90]{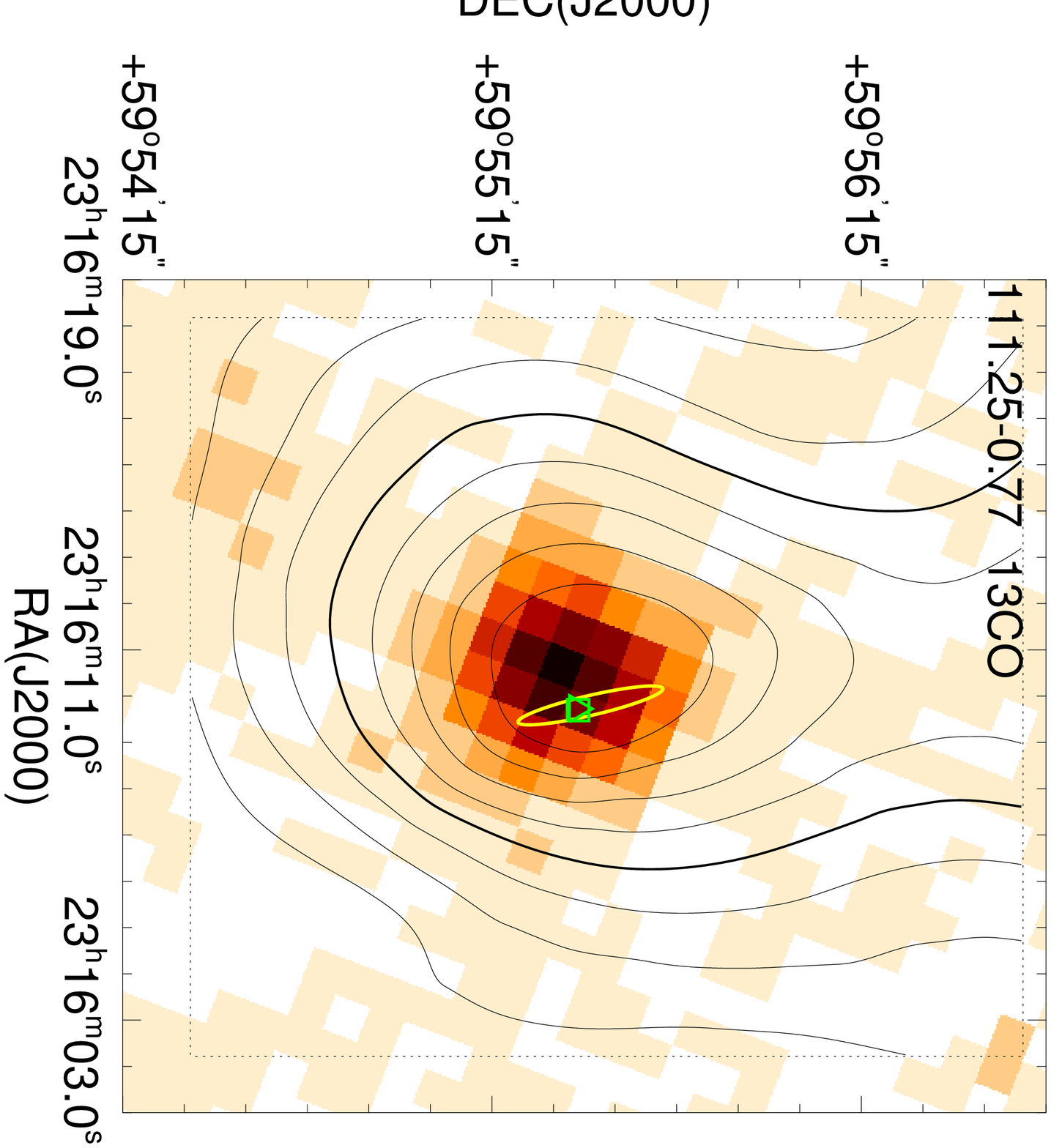}
\hspace{-11mm}\includegraphics[height=53mm,angle=90]{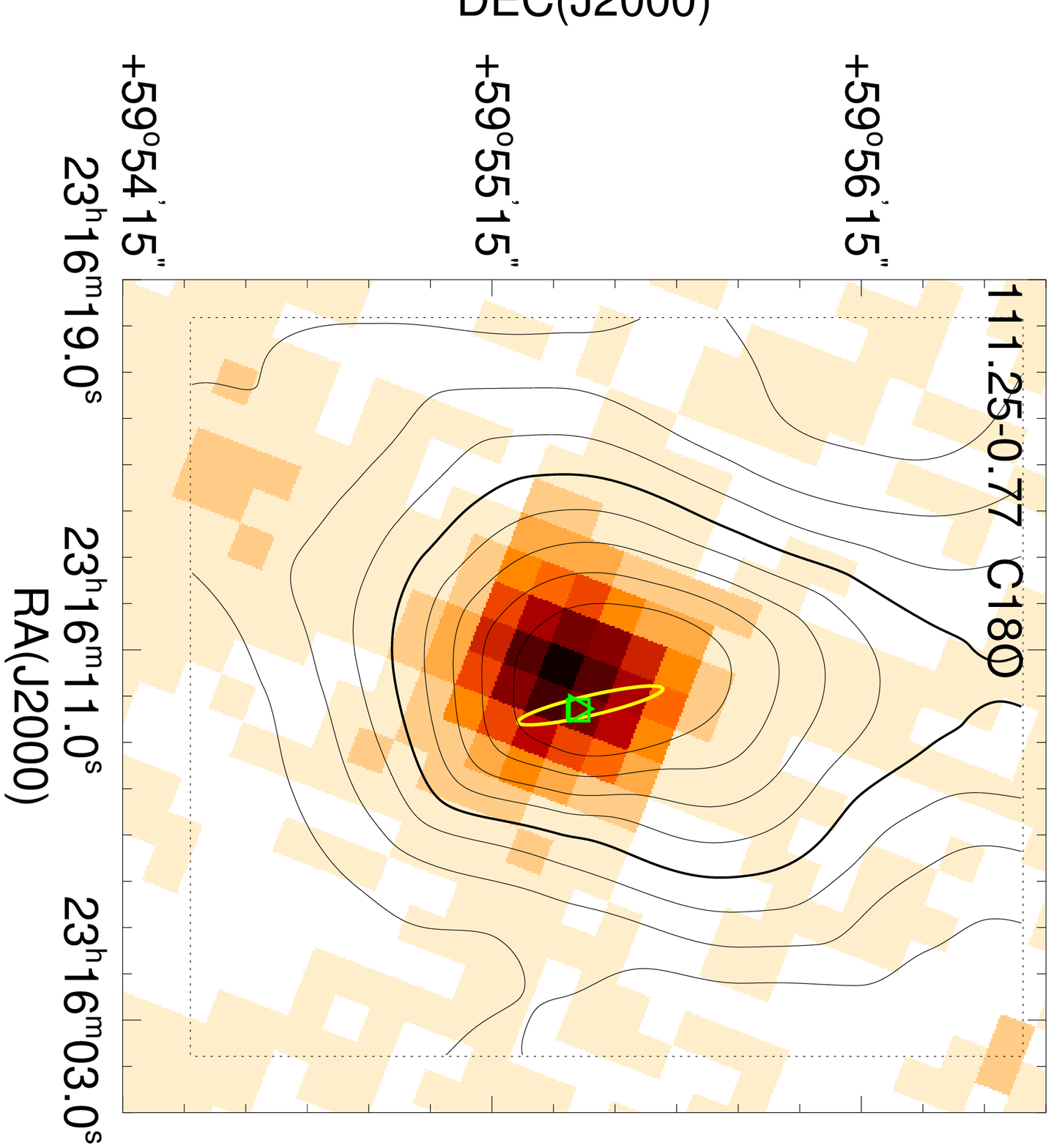}%
\hspace{-11mm}\includegraphics[height=53mm,angle=90]{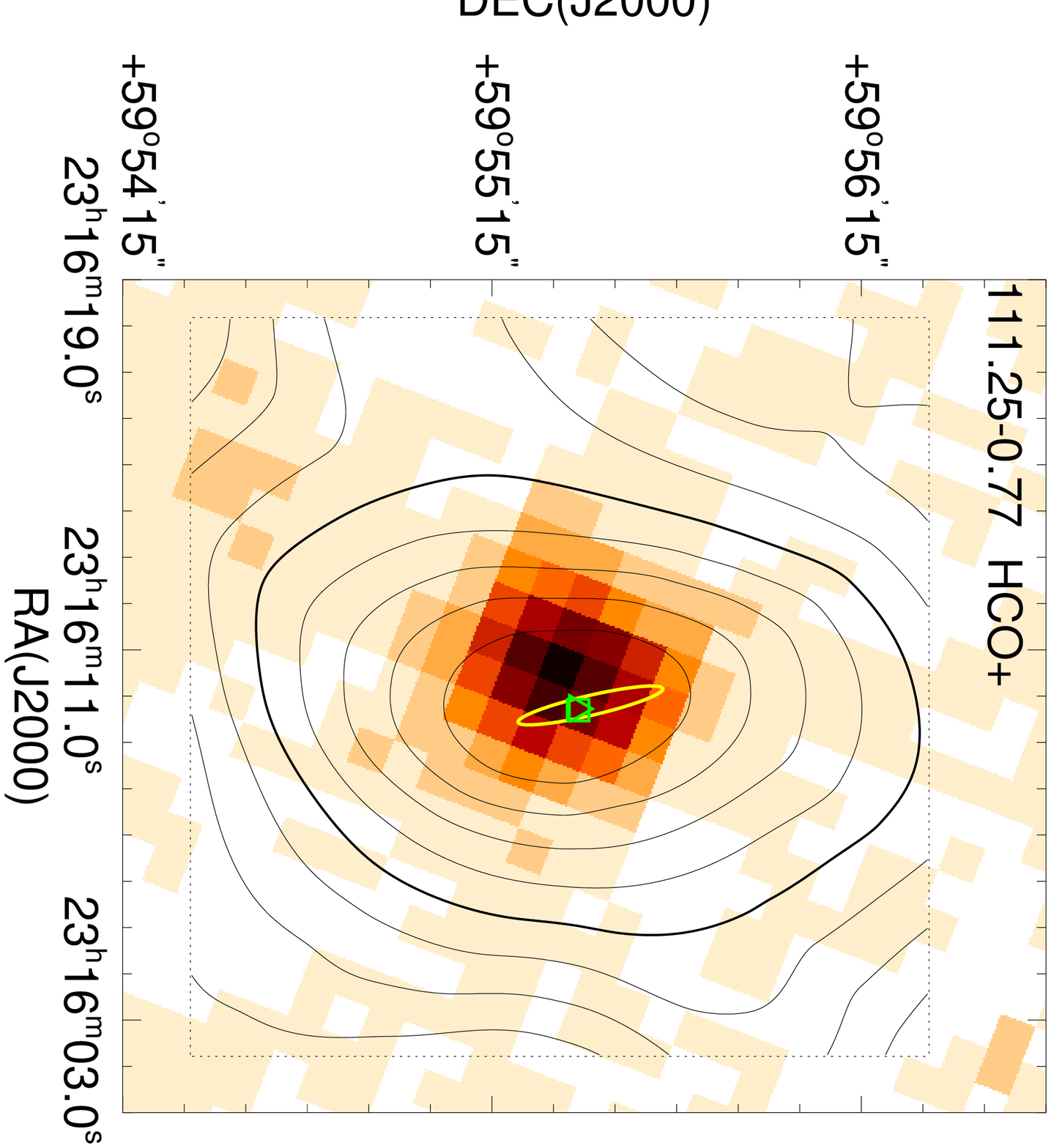}

\vs
\hspace{0.5mm}\includegraphics[height=53.5mm,angle=90]{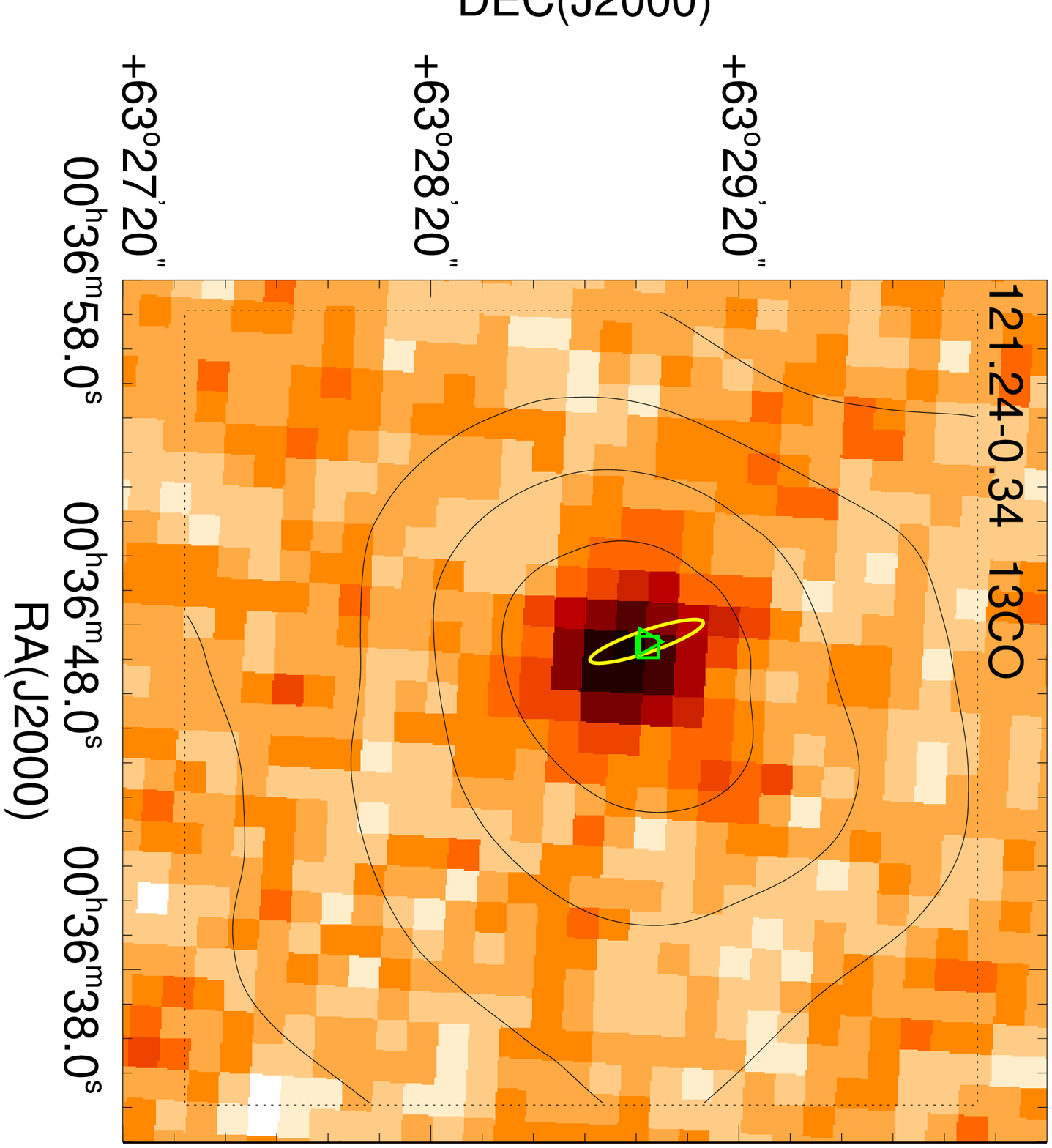}%
\hspace{-12mm}\includegraphics[height=53.5mm,angle=90]{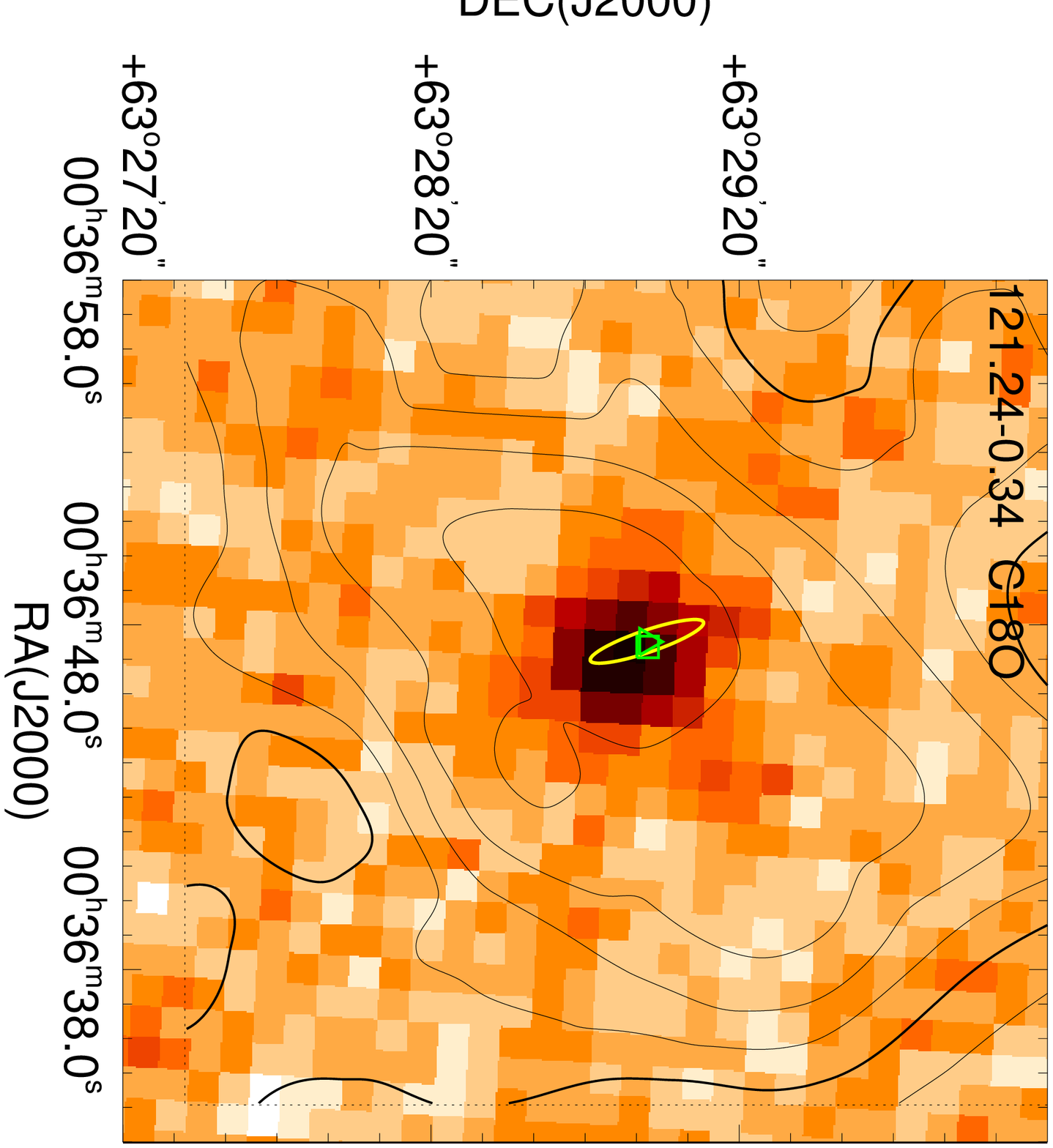}
\hspace{-12mm}\includegraphics[height=53.5mm,angle=90]{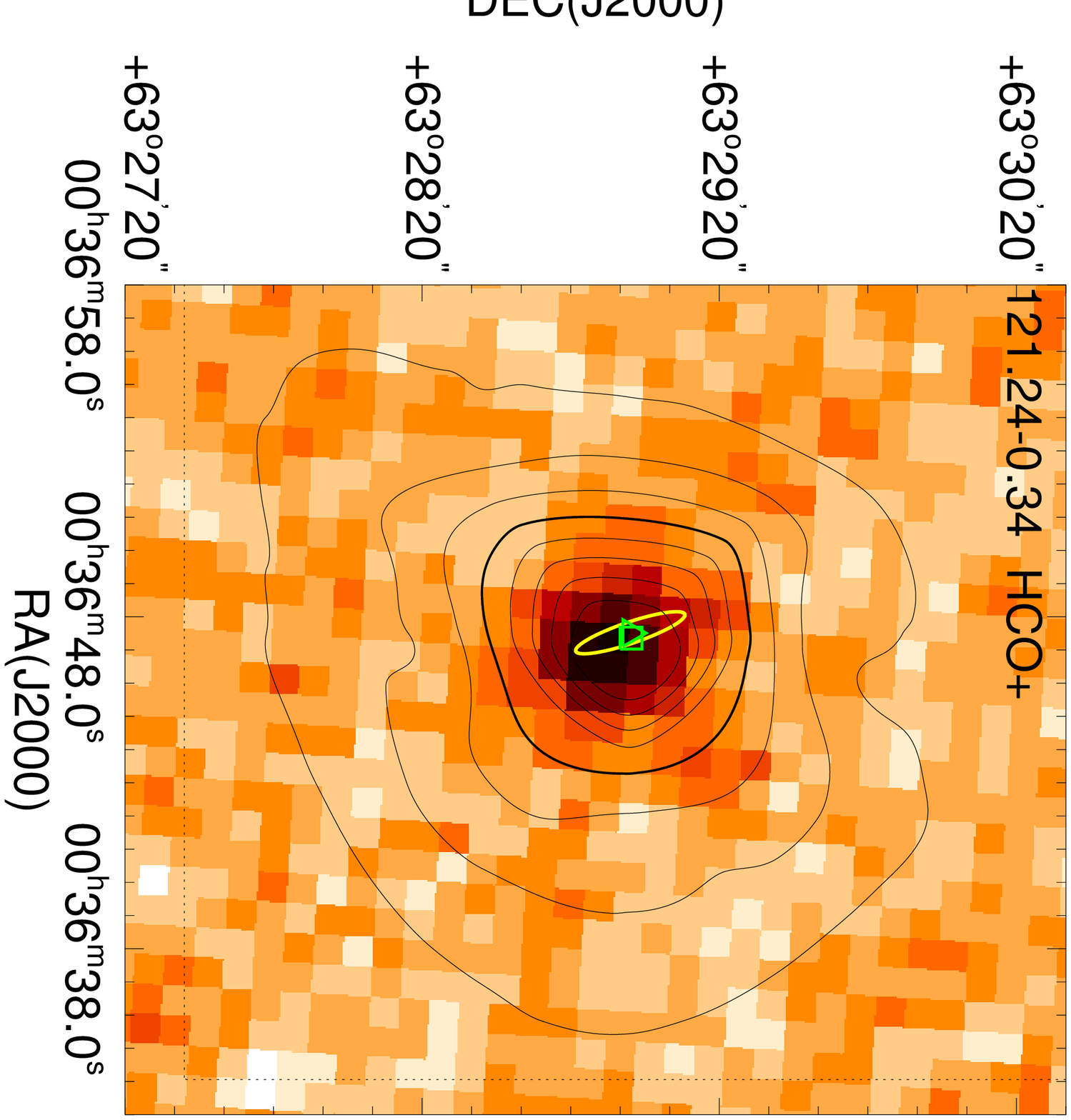}%

\vs\vs
\hspace{0.5mm}\includegraphics[height=45mm,angle=90]{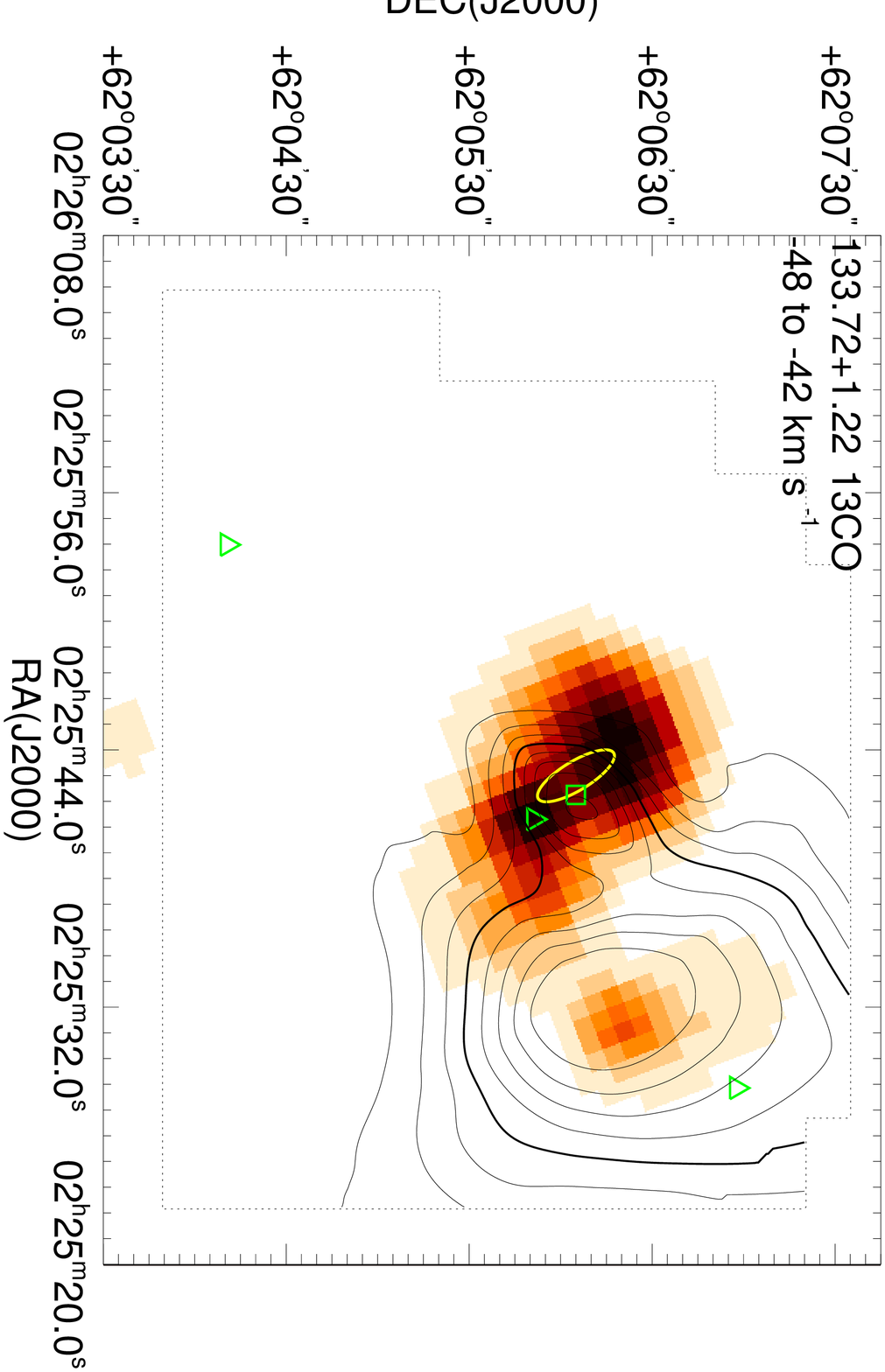}
\hspace{-2mm}\includegraphics[height=45mm,angle=90]{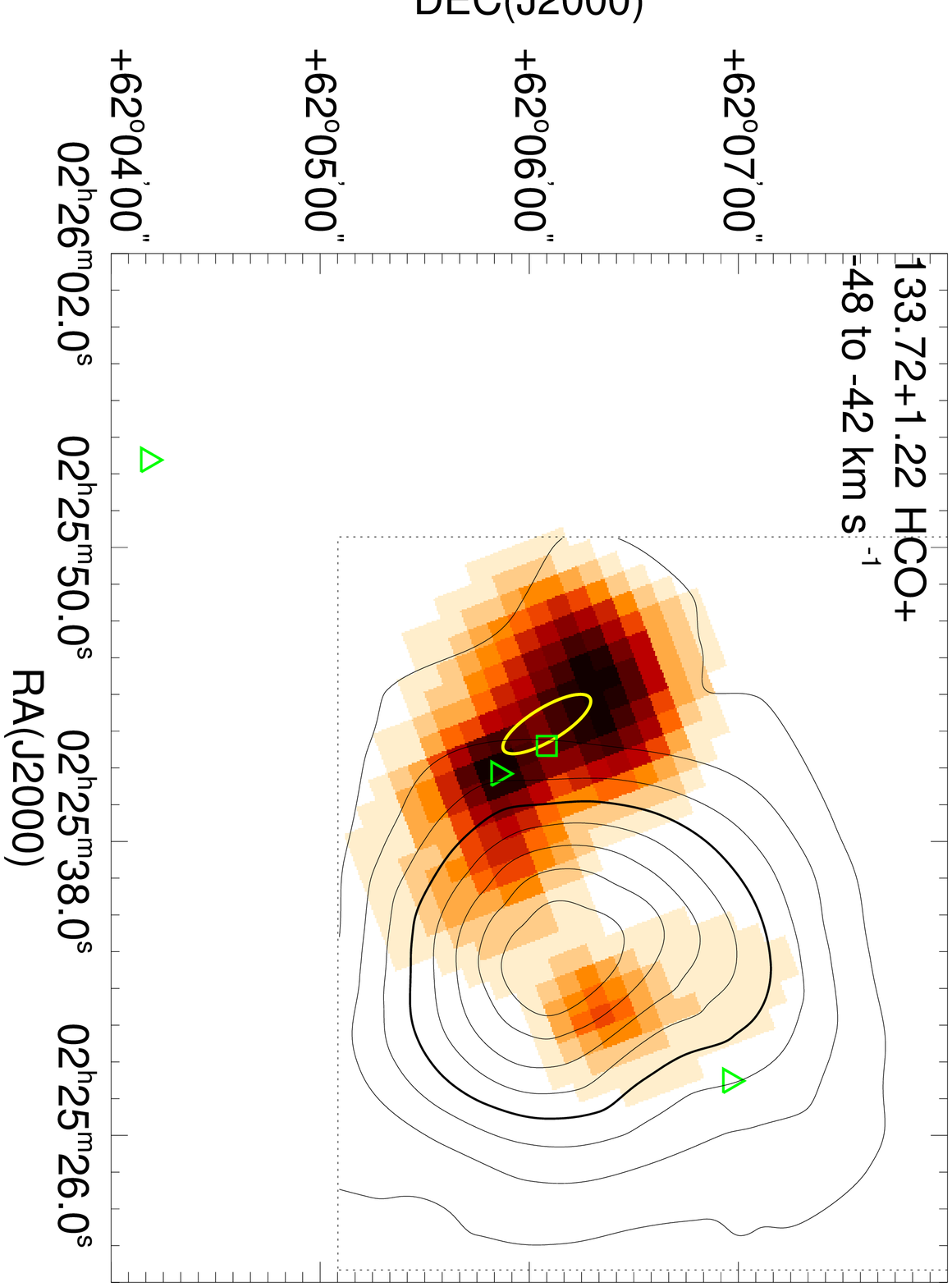}%
\hspace{-2mm}\includegraphics[height=45mm,angle=90]{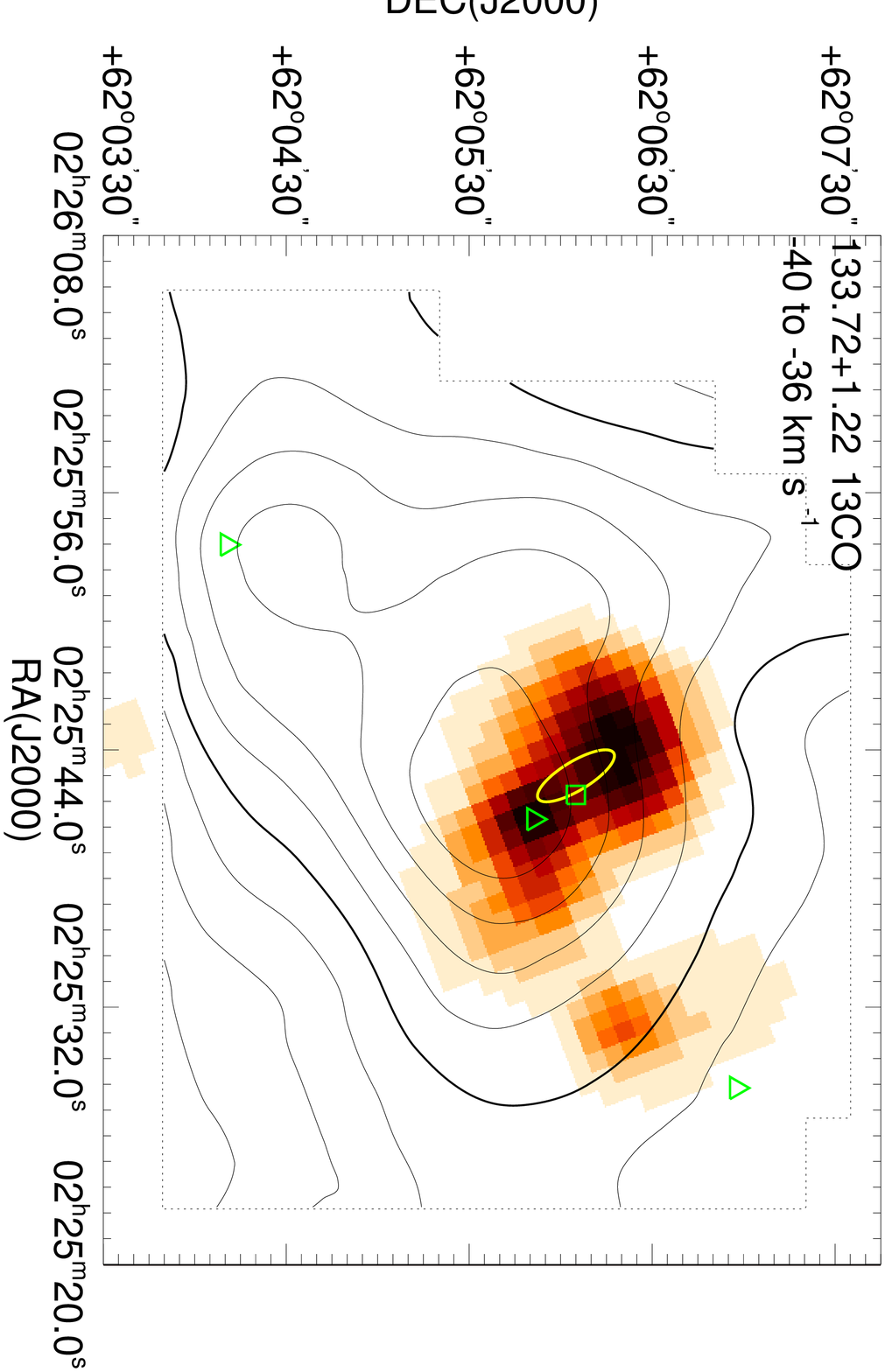}


\caption{\baselineskip 3.6mm $^{13}$CO(1--0), C$^{18}$O(1--0) and
HCO$^+$(1--0) maps of faint maser regions. The grey scale images are
MSX 21\,$\mu$m images except for 106.80+5.31, where the grey scale
image is{ the} MIPS 24\,$\mu$m image. The central blank pixels in the
24\,$\mu$m image {are} due to saturation. The squares,
triangles{,} cross{es} and ellipses denote the CH$_3$OH masers, H$_2$O
masers, OH masers and error ellipses of the IRAS point sources in
the fields, respectively. The contours are chosen to highlight the
most prominent features in each source, usually between 20\% and
90\% (steps of 10\%) of the peak integrated intensity. The thicker
lines denote the 50\% levels of the peak integrated intensity which
are used to determine the core sizes.}\label{Fig:maps}

\pagebreak
\end{figure}
\begin{figure}[h!!!]\setcounter{figure}{1}

\centering

\includegraphics[height=53mm,angle=90]{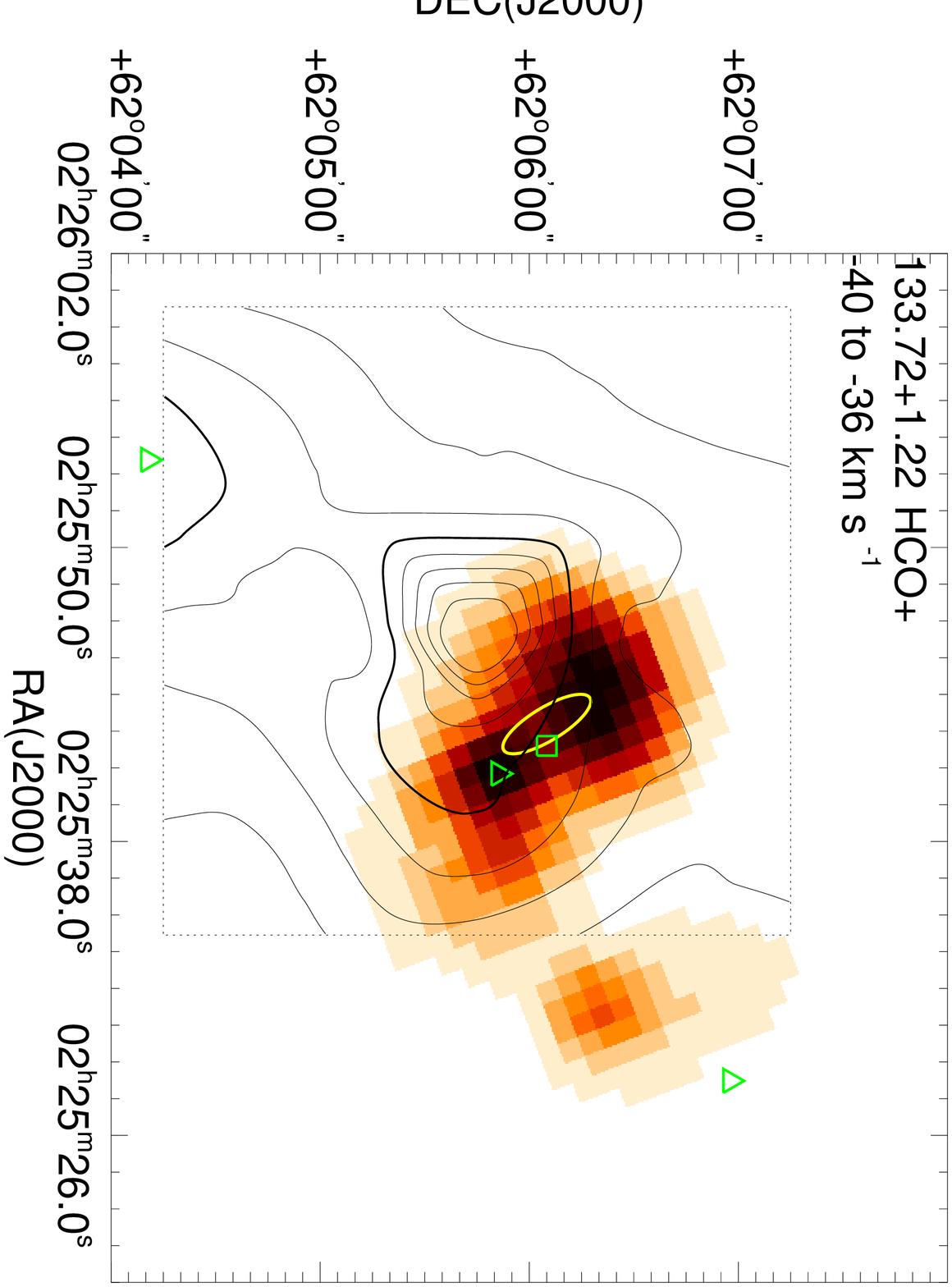}%
\hspace{-6mm}
\includegraphics[height=51mm,angle=90]{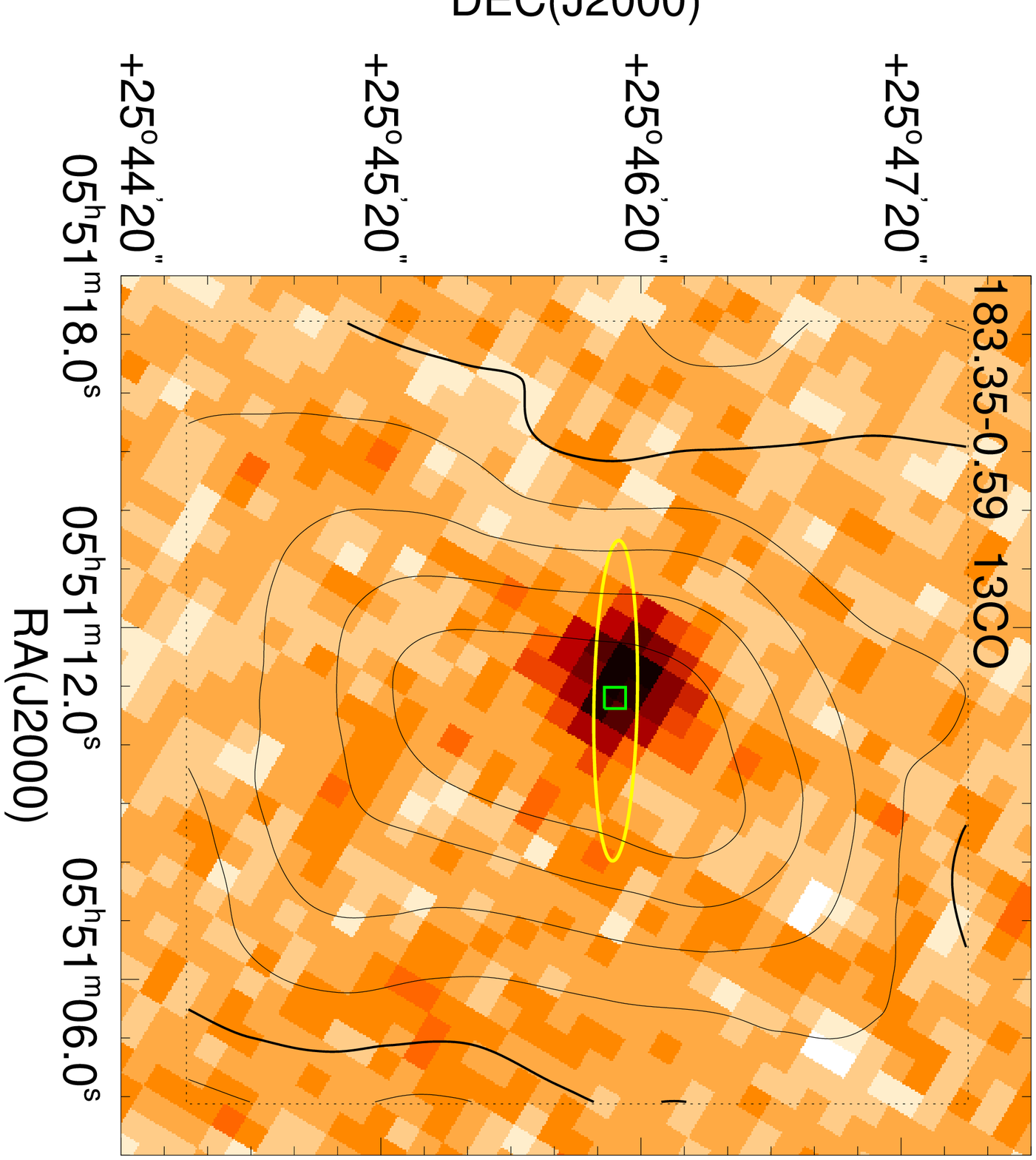}
\hspace{-12mm}\includegraphics[height=51mm,angle=90]{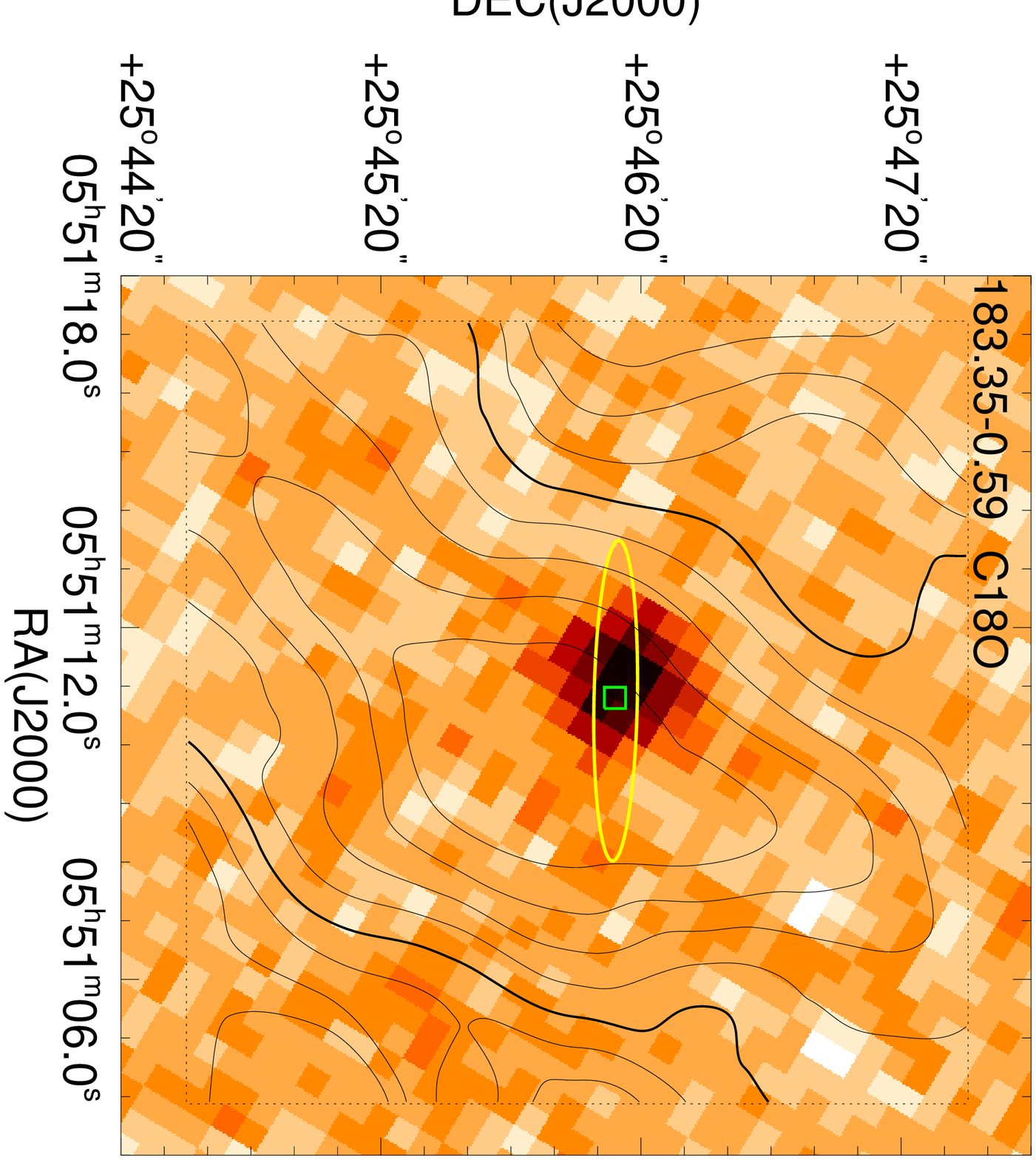}%

\includegraphics[height=55mm,angle=90]{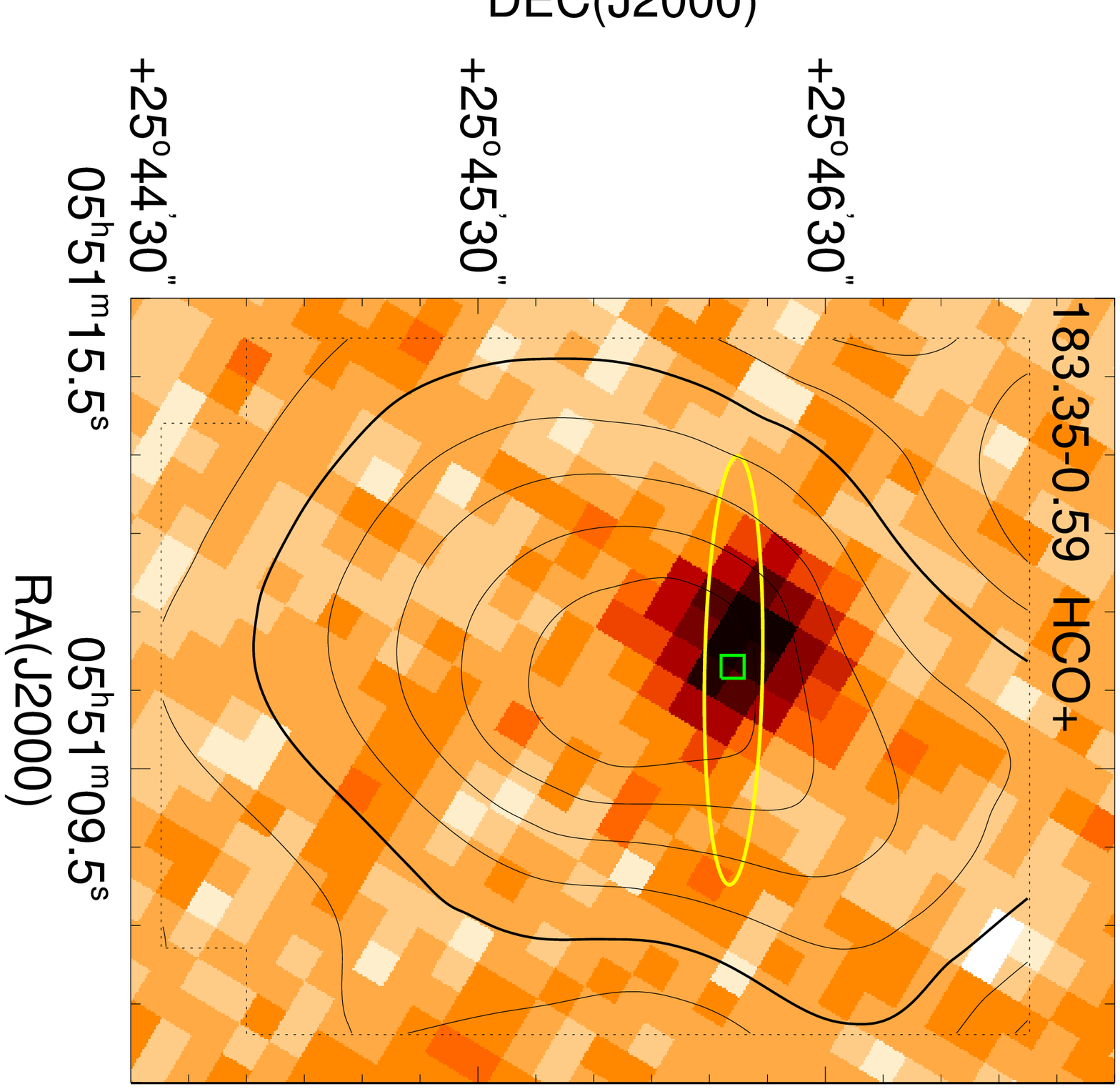}
\hspace{-12mm}\includegraphics[height=51mm,angle=90]{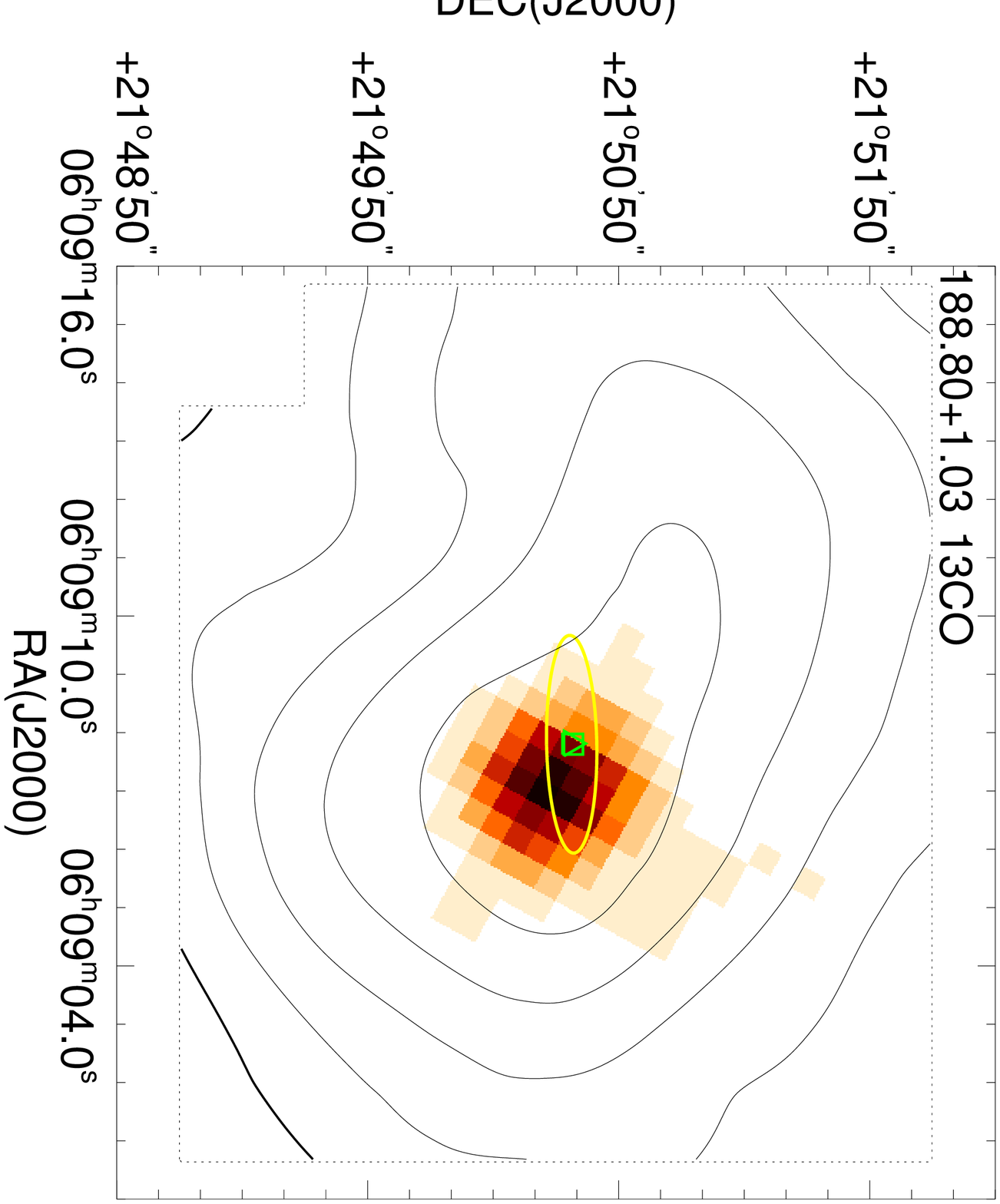}%
\hspace{-8mm}\includegraphics[height=51mm,angle=90]{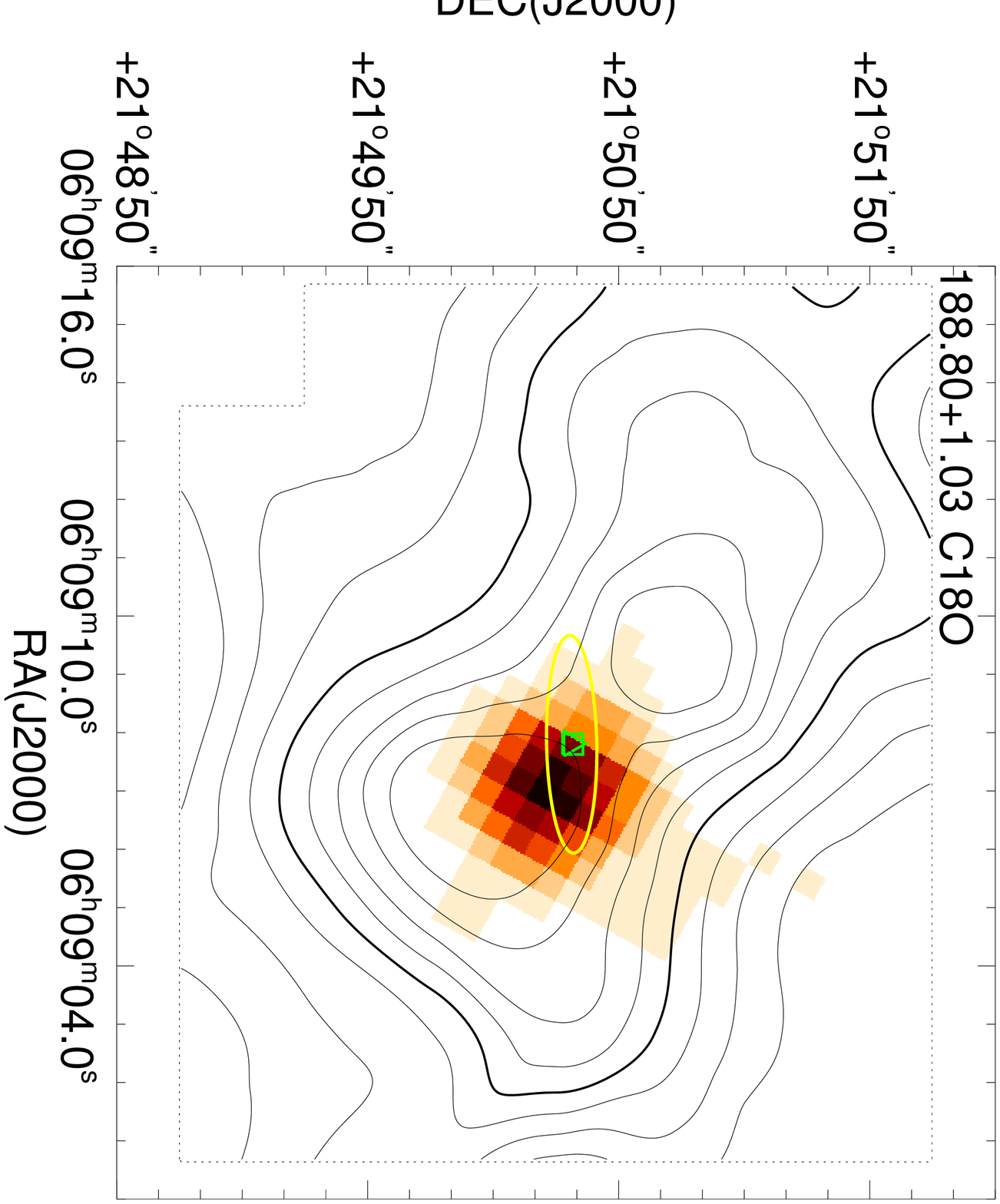}


\includegraphics[height=50mm,angle=90]{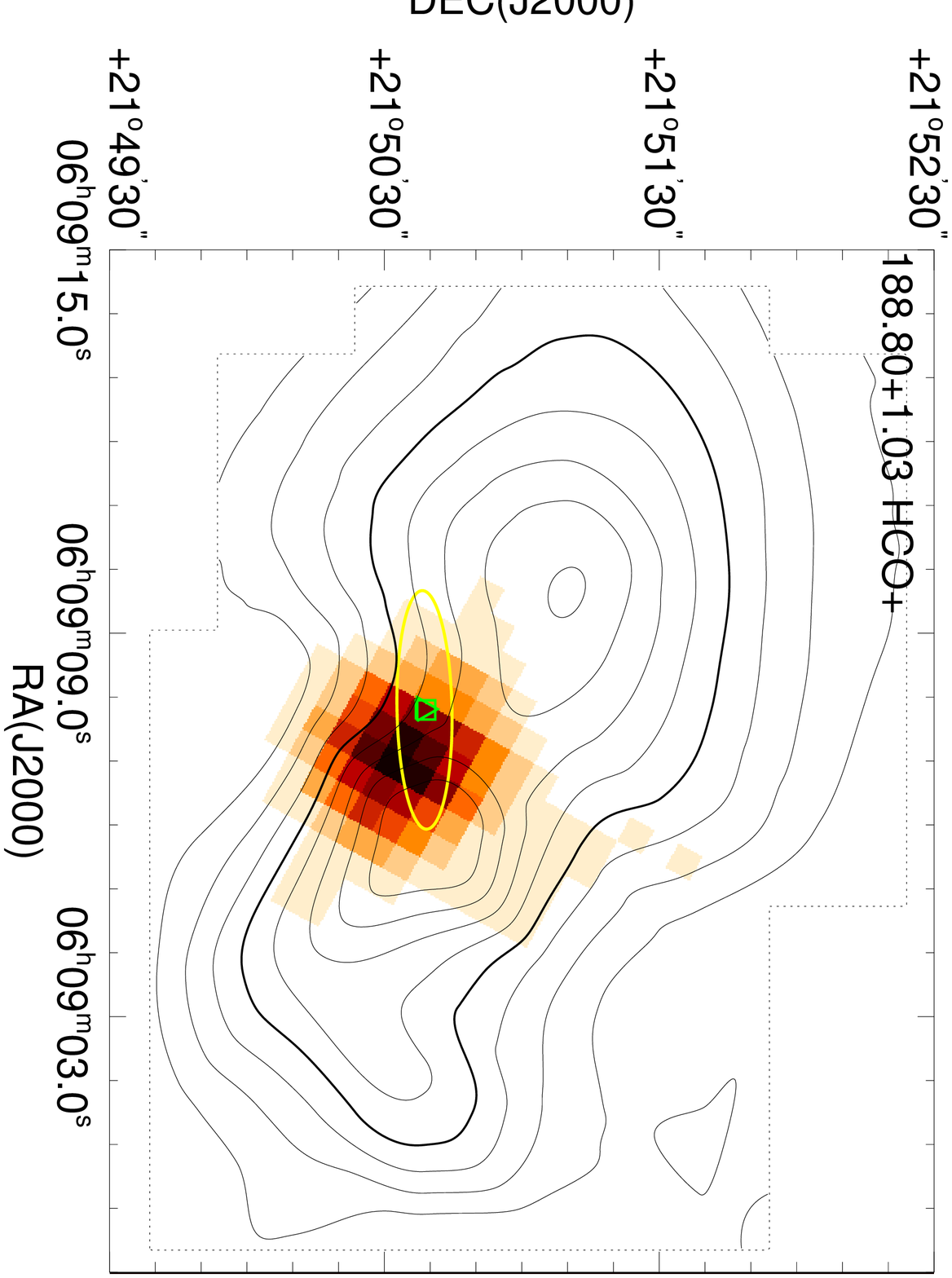}%
\hspace{-4mm}\includegraphics[height=50mm,angle=90]{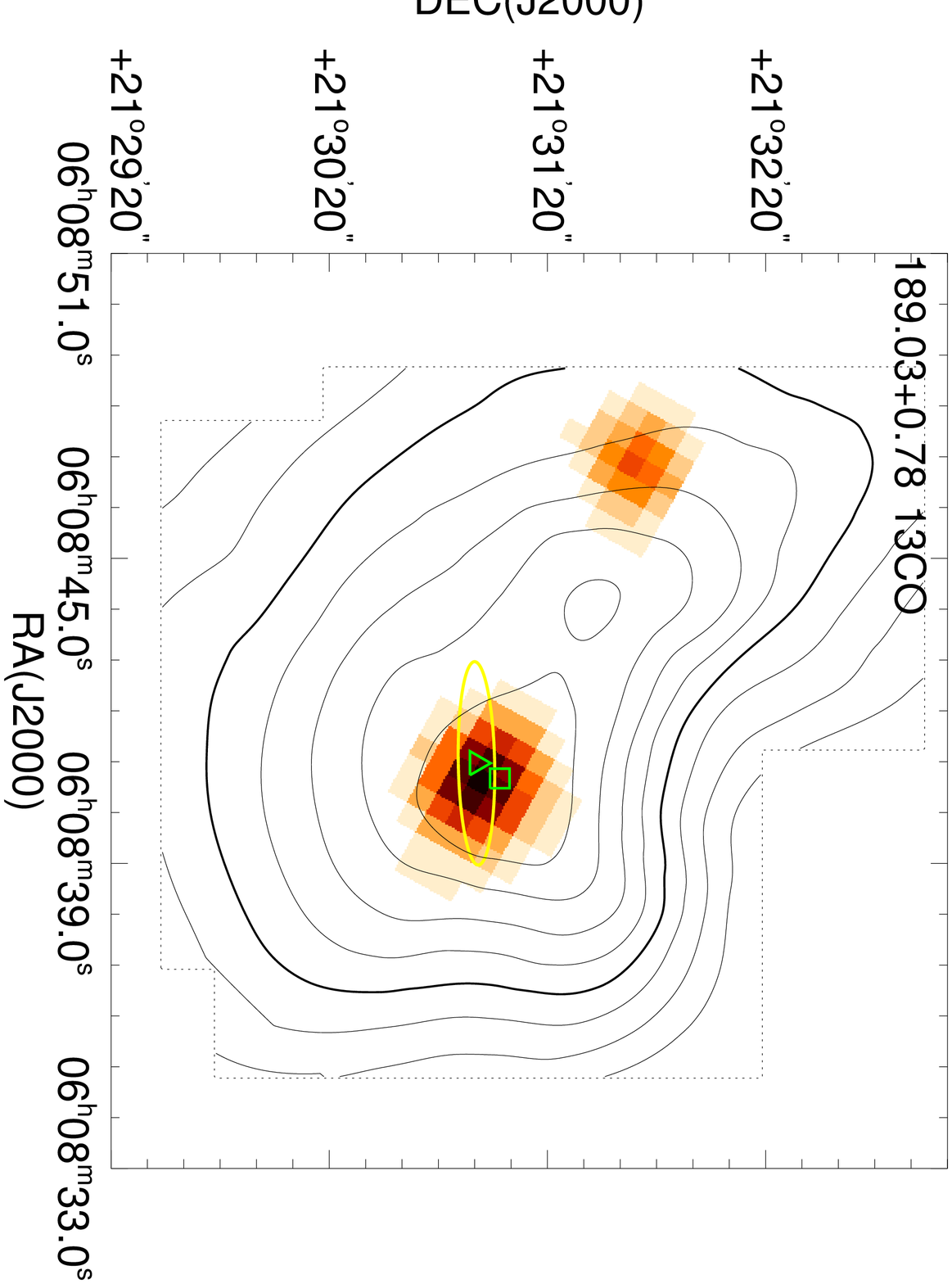}
\hspace{-9mm}\includegraphics[height=50mm,angle=90]{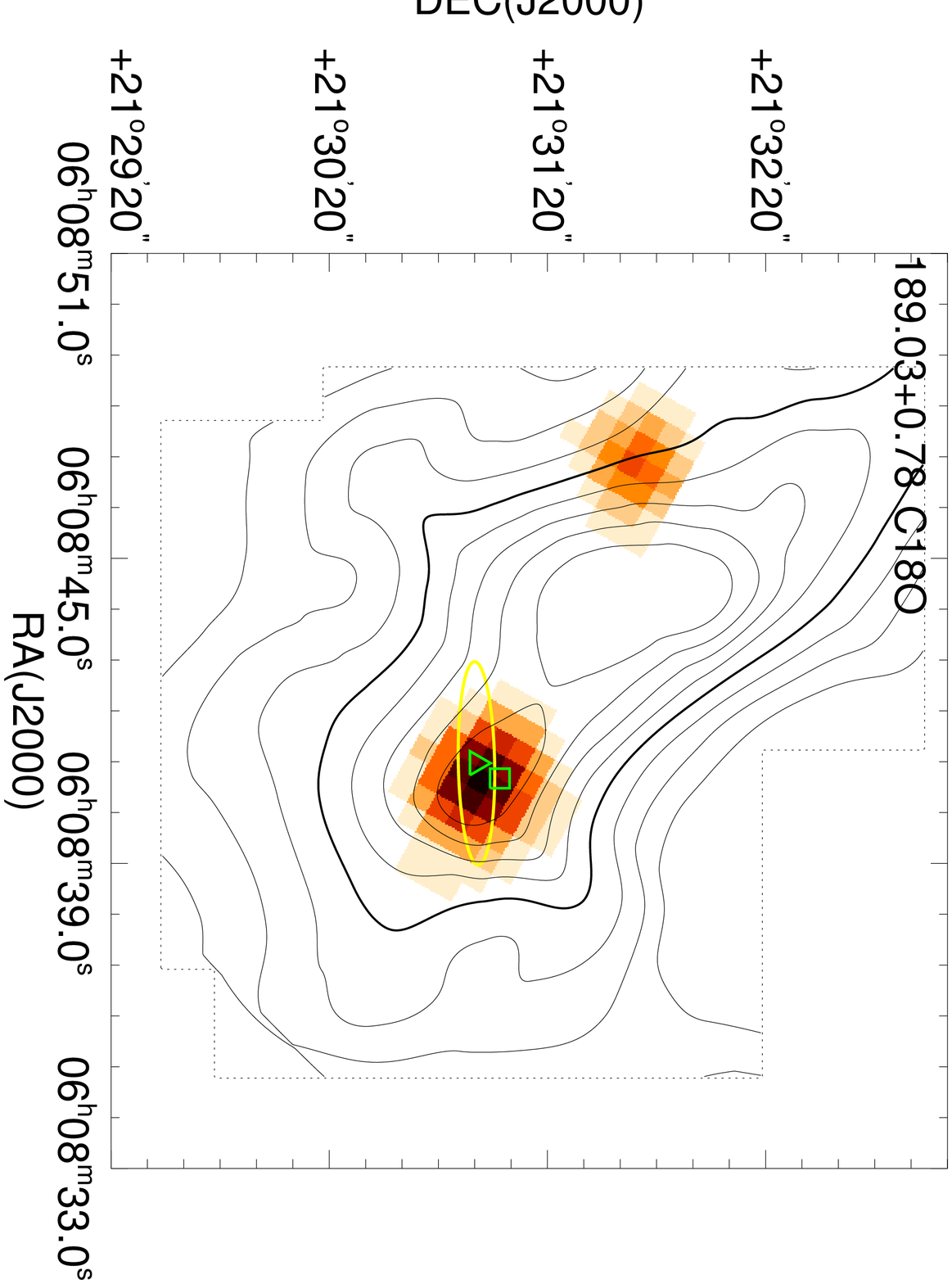}%

\includegraphics[height=51mm,angle=90]{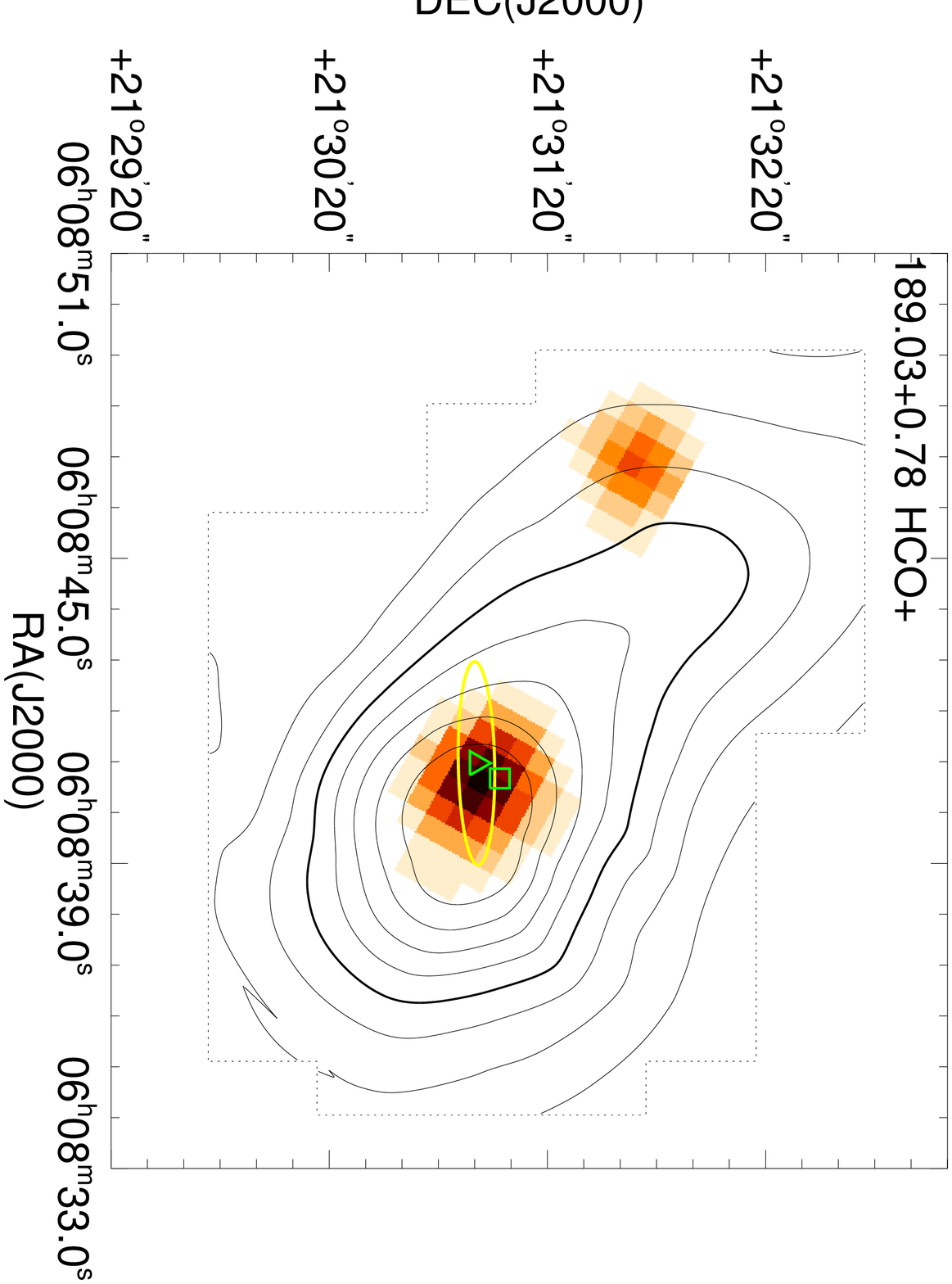}
\hspace{-7mm}\includegraphics[height=51mm,angle=90]{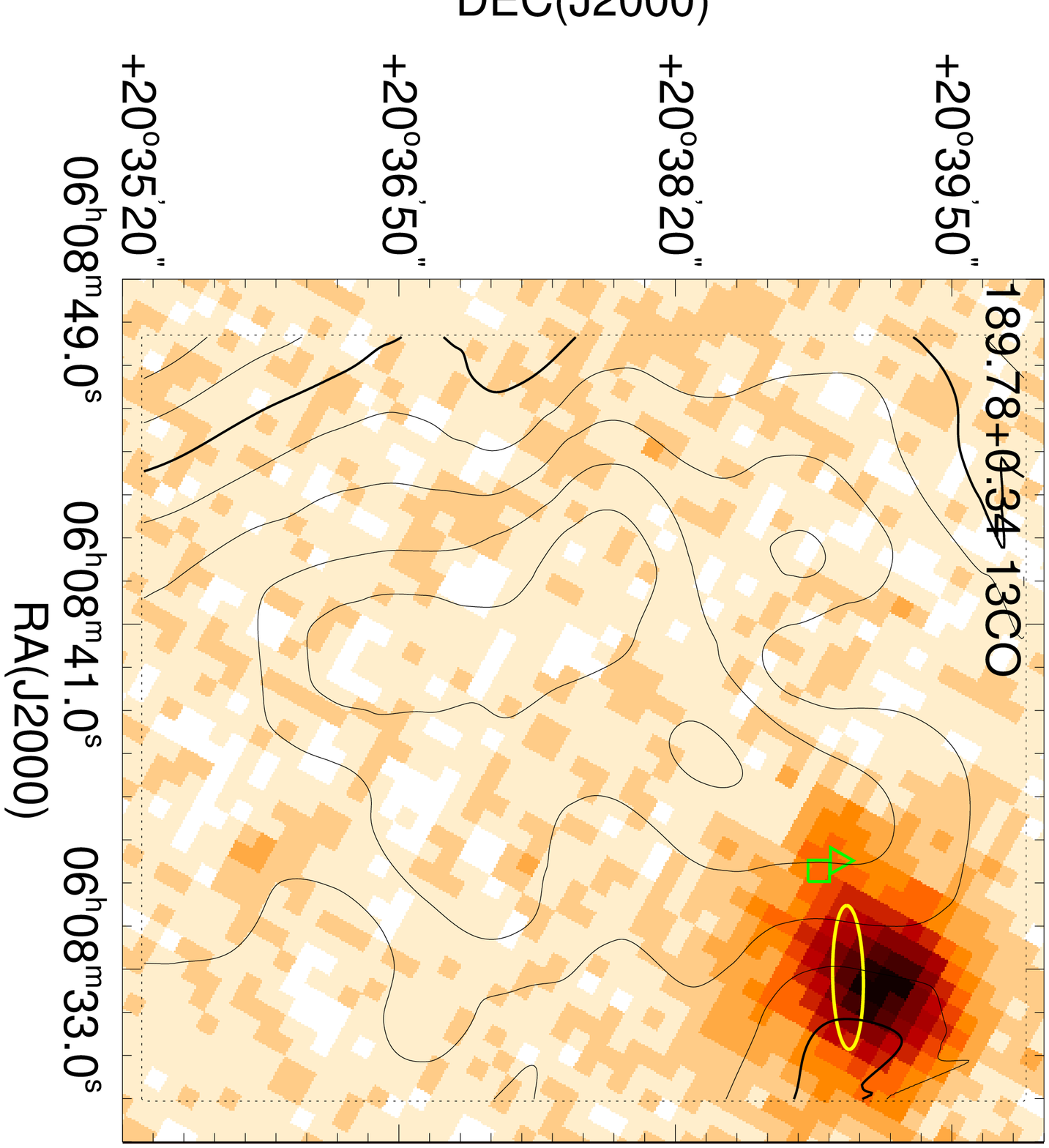}%
\hspace{-7mm}\includegraphics[height=51mm,angle=90]{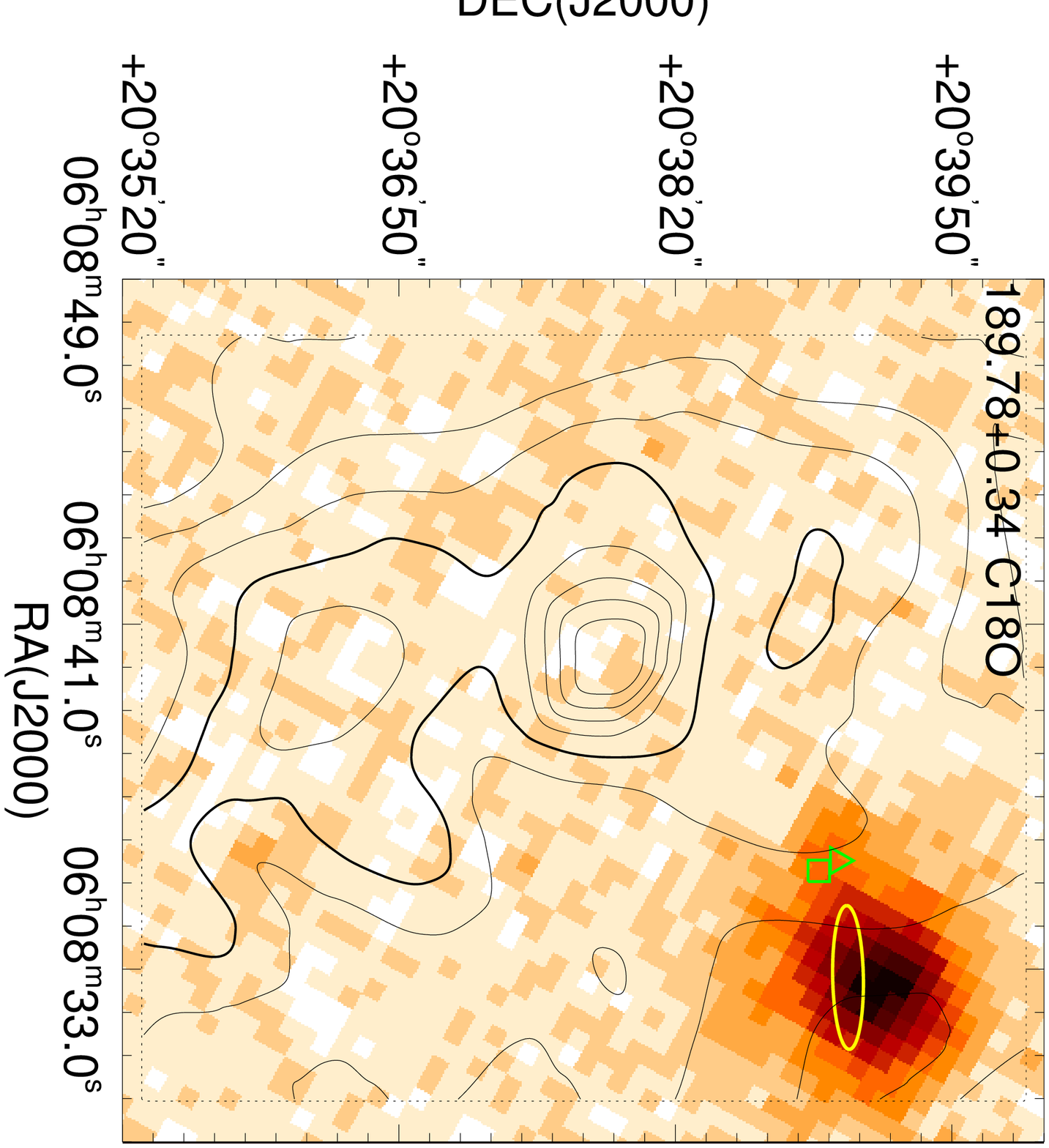}



\centering
\includegraphics[width=38mm,angle=90]{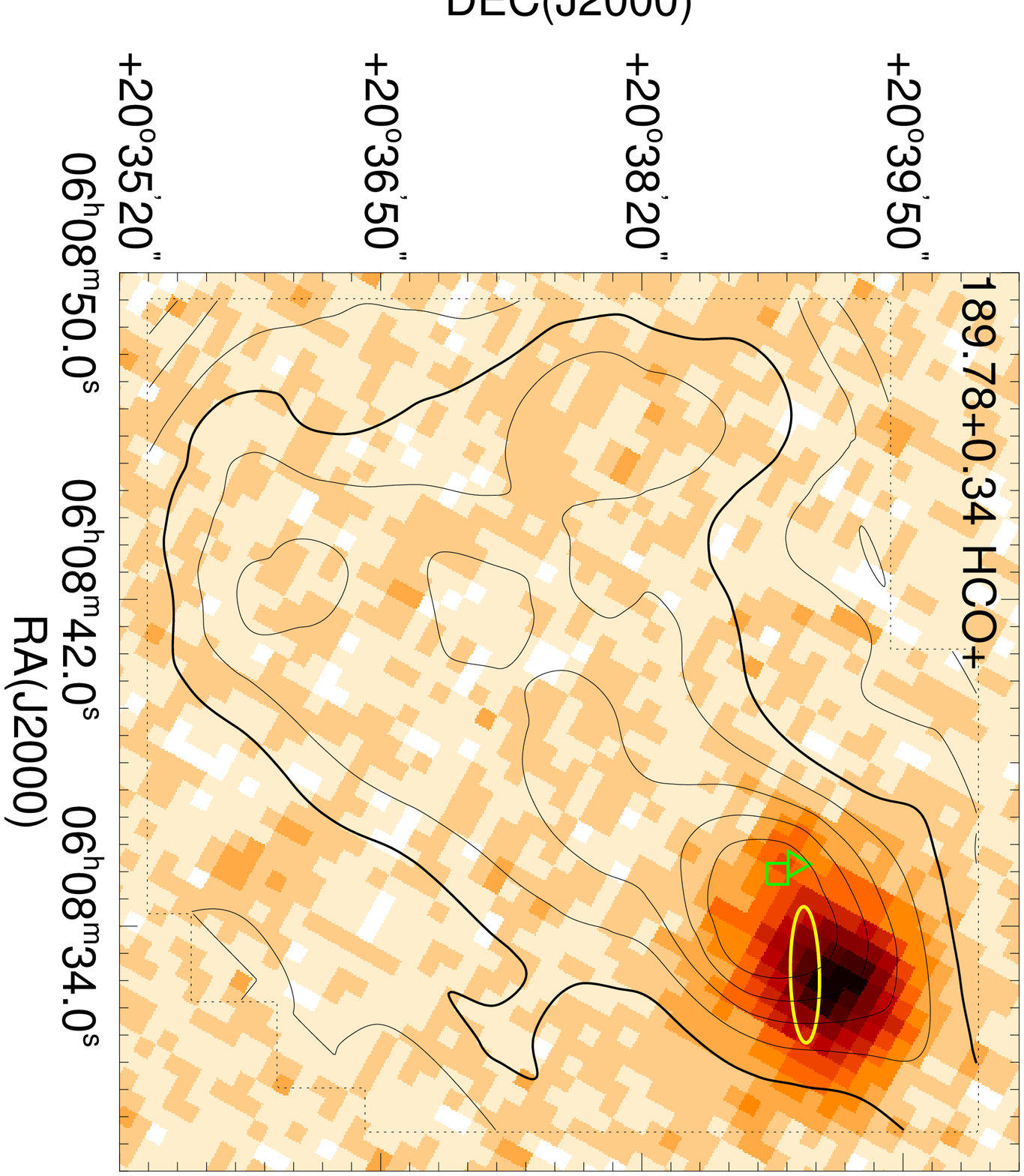}
\hspace{-14mm}
\includegraphics[width=55mm,angle=0]{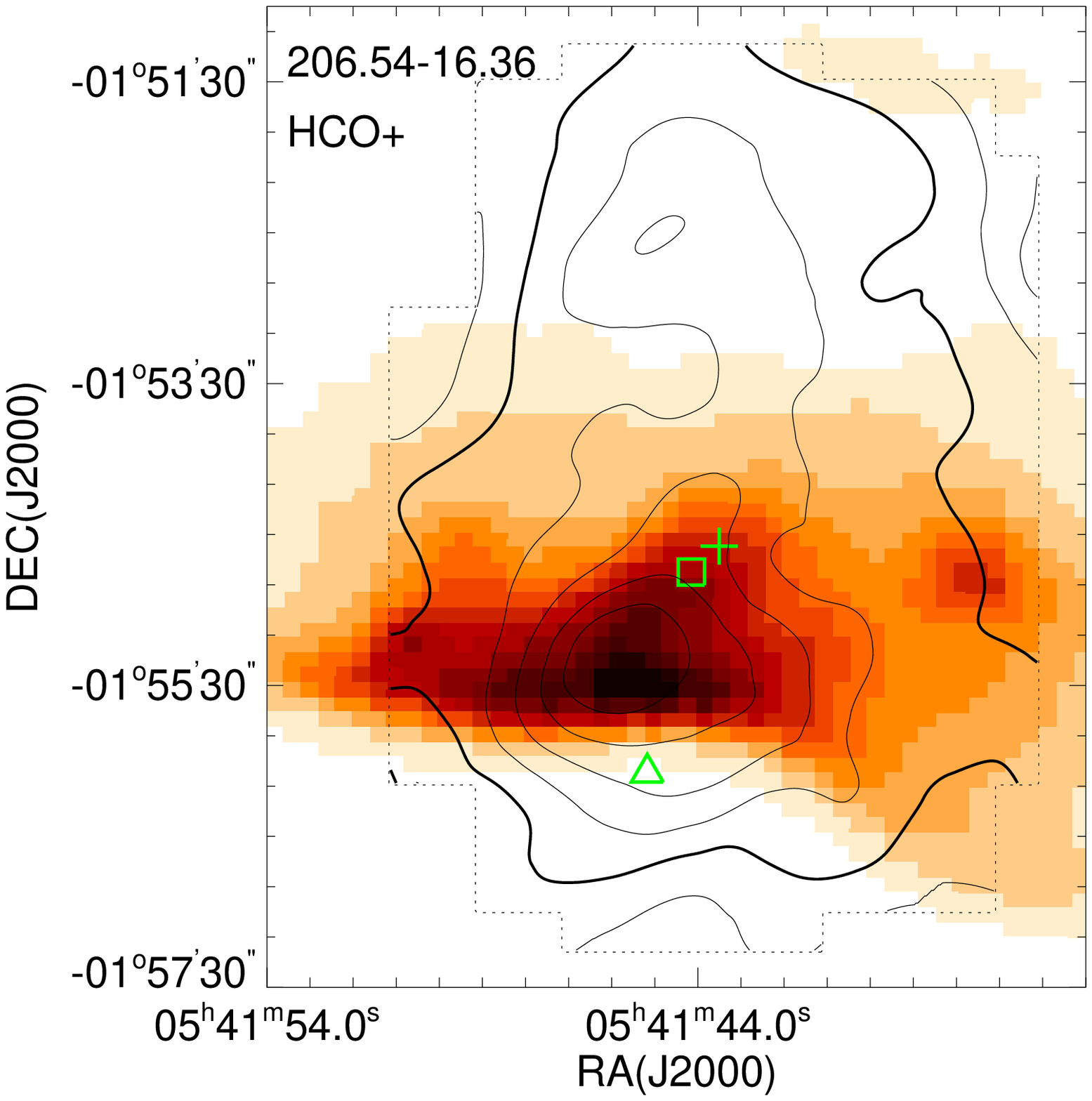}
\hspace{-14mm}\includegraphics[width=57mm]{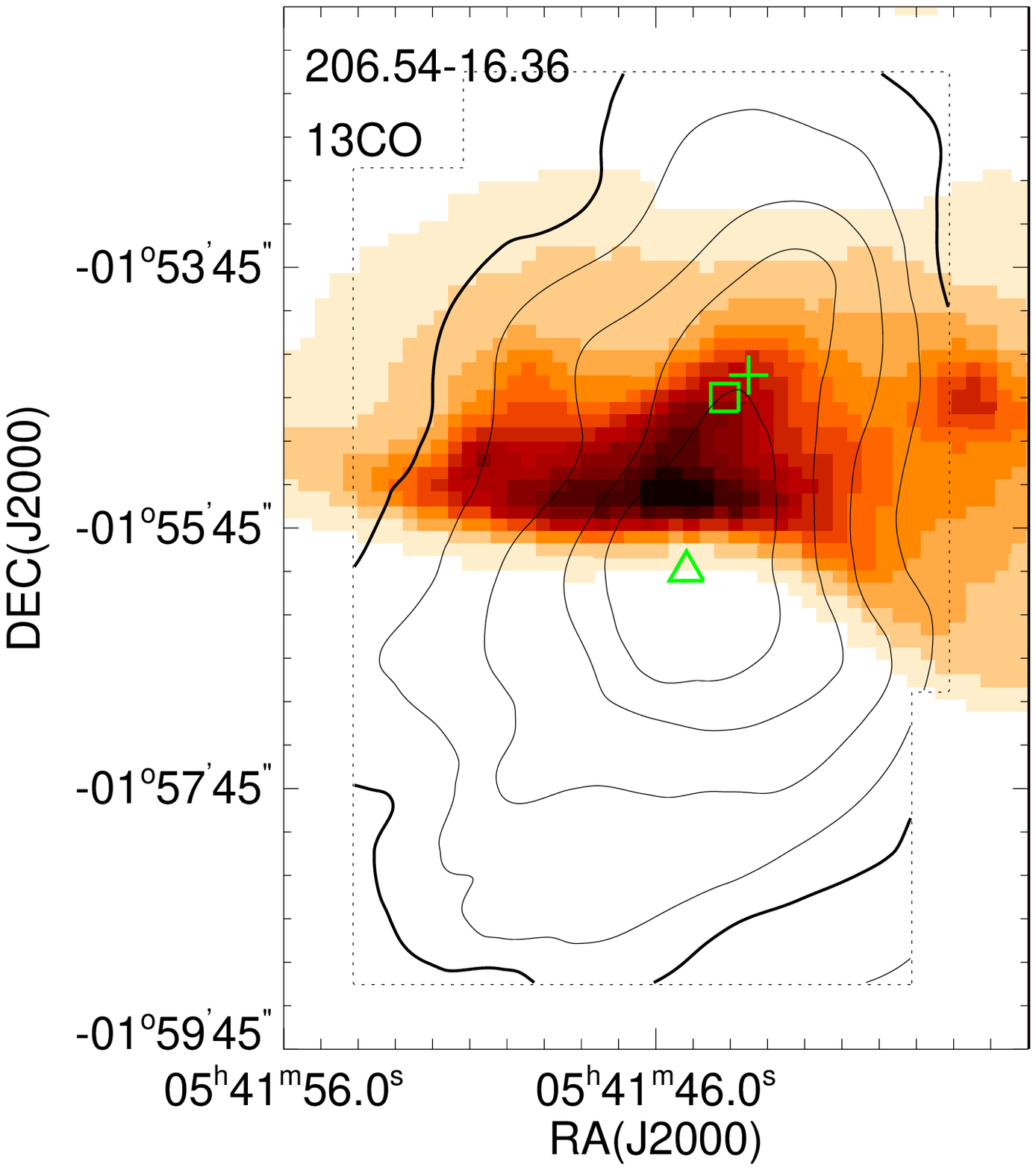}


\begin{minipage}[]{35mm}

\caption{ \it --- Continued.}\end{minipage}

\end{figure}

\begin{figure}\setcounter{figure}{1}

\centering
\hspace{10mm}\includegraphics[width=90mm,angle=0]{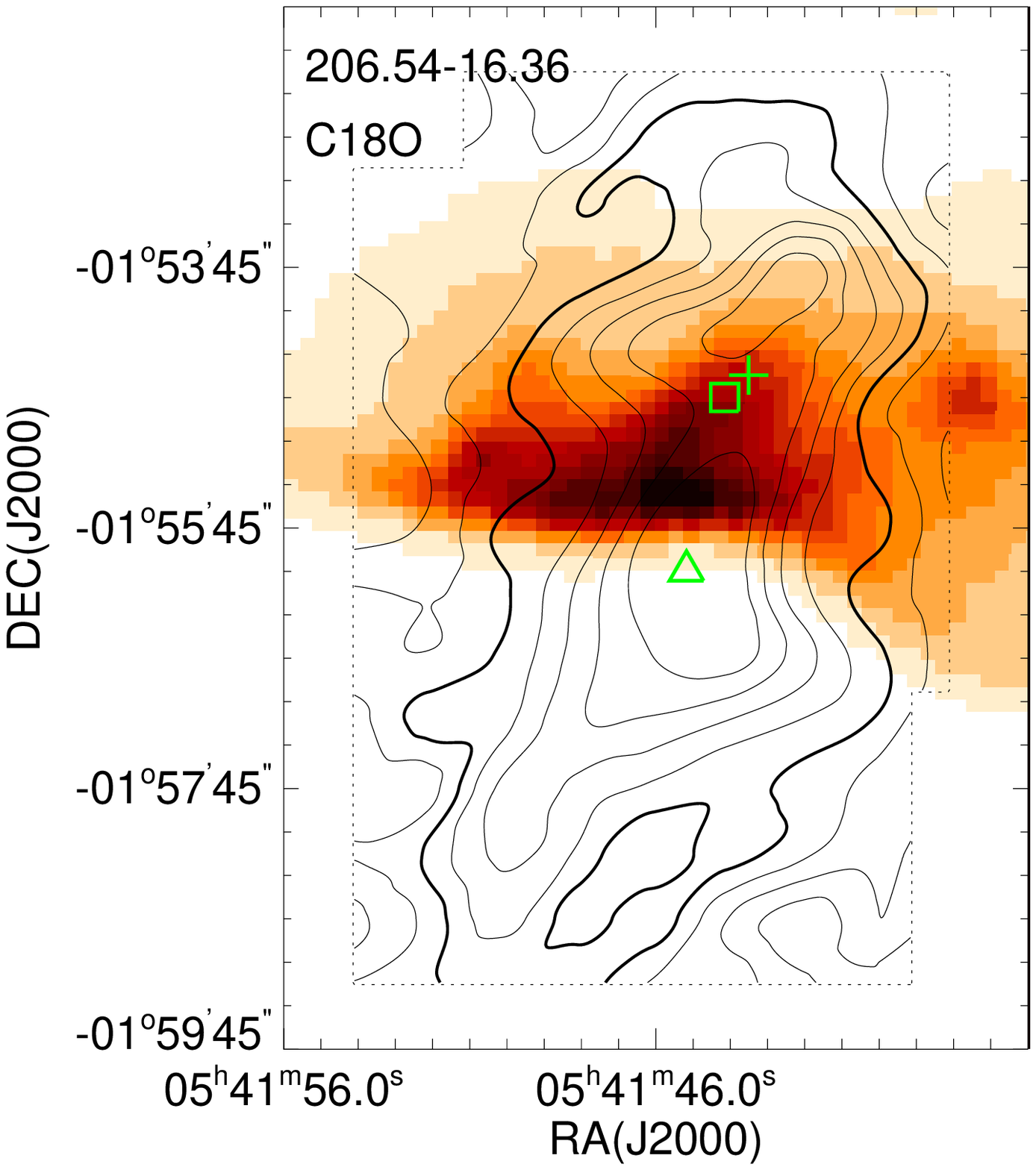}

\vspace{-4mm}
\begin{minipage}[]{35mm}

\caption{ \it --- Continued.}\end{minipage}
\end{figure}

\subsection{Comments on Individual Sources}

\begin{itemize}
\item[--] 
106.80+5.31 --- This source is located {in} the central part of the
S140 molecular complex. Both{ the} $^{13}$CO(1--0) and C$^{18}$O(1--0)
maps show a filamented structure extending southwest to northeast
and peak at the central infrared source. {The }HCO$^+$ map
reveals a single core surrounding the central infrared
source. All the three maser species, i.e., CH$_3$OH, H$_2$O and OH
masers{,} were found in this region.

\item[--] 
111.25--0.77 --- There is a single core peak towards the IRAS
23139+5939. {The }$^{13}$CO maps reveal a{n} extended emission
elongating towards the northern direction. Both H$_2$O and CH$_3$OH
masers were found in this region.

\item[--] 
121.24--0.34 --- There is a single core {which }elongates northwest to
southeast and peaks at IRAS 00338+6312. Both H$_2$O and CH$_3$OH
masers were found in this region.

\item[--] 
133.72+1.22 --- This is an intensively studied region of active star
formation, the W3 complex. $^{13}$CO(1--0) and HCO$^+$(1--0) spectra
show two velocity components, with velocity ranges of
$-48$$\sim$$-42$~km~s$^{-1}$ and $-40$$\sim$$-36$~km~s$^{-1}$,
respectively. In the $^{13}$CO integrated intensity map with{ a}
velocity range of --48$\sim$--42~km~s$^{-1}$, one can see a two-core
structure, with the eastern core (Core 1) peaking towards the
CH$_3$OH maser and the western core (Core 2) associated with the
western mid-infrared dust core. In the $^{13}$CO integrated
intensity map with{ a} velocity range of --40$\sim$--36~km~s$^{-1}${,}
there is a filament consisting of two cores, with the northwestern
core (Core 3) departing $\sim$ 30$^\prime$$^\prime$ south{ of} IRAS
02219+6152 and the southeastern core (Core 4) departing $\sim$
2$^\prime$ southeast {of} the IRAS source. Emission of HCO$^+$
show{s a} similar distribution. Three H$_2$O and one CH$_3$OH
masers are found in this {field}.

\item[--]
 183.35--0.59 --- A single core{ is} locate{d} southwest
and with{ an} offset of $\sim$ 15$^\prime$$^\prime$ from IRAS
05480+2545. Only {the }CH$_3$OH maser is found in this region.

\item[--] 
188.80+1.03 --- Both C$^{18}$O(1--0) and HCO$^+$ maps reveal a
filament consisting of two cores {which }extend southwest to northeast
through the field. Core 1 (the southeast core) is offset by $\sim$
30$^\prime$$^\prime$ southeast towards IRAS source 06061+2151 and
Core 2 (the northwest core) is offset by $\sim$
30$^\prime$$^\prime$ northeast towards the IRAS source. H$_2$O and
CH$_3$OH masers are found in this region.

\item[--] 
189.03+0.78 --- The mid-infrared dust emission of this region shows
two heat sources. Emission of CO and HCO$^+$ reveals a filament
extending from the southwest infrared source towards the northeast
one. The filament in{ the} C$^{18}$O map consist{s} of two cores. The
southwestern core (Core 1) is associated with IRAS 06056+2131. The
northeastern core (Core 2) {is }locate{d} in the middle of the two
mid-infrared cores departing $\sim$ 1$^\prime$ northeast towards
IRAS 06056+2131. Both H$_2$O and CH$_3$OH masers are found to{ be}
associated{ with} Core 1.

\item[--] 
 189.78+0.34 --- This source is part of the
S252A molecular complex. $^{13}$CO and C$^{18}$O maps reveal a
filamented structure consisting of two cores and extending from
north to south{, while the} HCO$^+$ map presents a filament extending
northwest to southeast and peaking towards the CH$_3$OH maser.

\item[--] 
206.54--16.36 ---It is located in the vicinity of{ the} NGC 2024 HII
region. A filament, called{ the} ``Molecular Ridge{,}'' {is }elongate{d} in{ the}
north-south direction \citep{chandler96}. The filament {is }segregate{d}
into {two} segments on{ the} C$^{18}$O maps. Along the ridge{,} there are seven
far infrared sources \citep{mezger88,mezger92}. The CH$_3$OH maser
is associated with FIR4 \citep{minier03}.

\end{itemize}

\begin{table}[b!!]

\centering
\renewcommand\arraystretch{1}

\begin{minipage}[]{65mm}

 \caption{Fitted
Profile Parameters for N$_2$H$^+$
}\label{Tab:ngauss}\end{minipage}

\vspace{-3mm}
\fns \tabcolsep 0.8mm
\begin{tabular}{lcccccccrcccccc}
\hline \noalign{\smallskip}
 & \multicolumn{2}{c}{7-component HFS fits $^{\alpha}$} 
 &
 \multicolumn{6}{c}{3-Gaussian fits $^{\gamma}$} && \\
 \cline{2-3} \cline{5-10} \noalign{\smallskip}
 & & & &
 \multicolumn{2}{c}{Group-1} &
  \multicolumn{2}{c}{Group-2} & \multicolumn{2}{c}{Group-3} &&\\
Source   & $V_{\rm LSR}^{\beta}$ & $\Delta V$ &
 & {\small $\int$} $T_R^{\ast} dV$ & $\Delta V$  & {\small $\int$}
 $T_{R}^{\ast}dV$
  &
$\Delta V$ & {\small $\int$} $ T_{R}^{\ast}dV$
 & $\Delta V$ \\
& (km s$^{-1}$) & (km s$^{-1}$) &
 & (K km s$^{-1}$) & (km
s$^{-1}$) & (K km s$^{-1}$) & (km s$^{-1}$)& (K km s$^{-1}$) & (km
s$^{-1}$) 
\\
\noalign{\smallskip}
\hline\noalign{\smallskip} 106.80+5.31   &--6.95(02)  &1.71(05) &
&1.20 &2.15(19) &4.82 &2.48(20) &2.87 &2.41(39)  \\
121.24--0.34   &--17.75(03) &1.56(07) &
&0.43 &1.84(24) &1.99 &2.39(19) &1.02 &1.96(24)  \\
183.35--0.59   &--9.70(06)  &1.60(11) &
&0.26 &1.41(30) &1.24 &2.39(20) &0.70 &2.11(34)  \\
189.03+0.78   &2.58(06)   &1.85(13) &
&0.40 &2.01(46) &1.79 &2.50(30) &1.27 &3.14(17)  \\
189.78+0.34   &7.44(06)   &1.55(15) &
&0.23 &0.94(34) &1.43 &2.47(28) &1.22 &2.93(74)  \\
206.54-16.36  &10.62(06)  &1.30(12) &
&---  &---      &1.07 &2.09(25) &0.63 &1.82(23)  \\
 \noalign{\smallskip}\hline\noalign{\smallskip}
\end{tabular}
\fns\parbox{130mm} {Notes:
 $\alpha$: fitted with {the }CLASS HFS routine considering{ the} 7-component
 structure; \\
$\beta$: 
 $V_{\rm LSR}$ is refer{ring} to the $F_1, F = 2, 3 
   \rightarrow 1, 2$
 transition, i.e. the center component of `Group-2';\\
$\gamma$: 
 Gaussian fits to the three groups.
 }
\end{table}

\subsection{Physical Quantities}
\subsubsection{Optical depth and excitation temperature}

In this section{,} we derive optical depths and excitation
temperatures of CO and N$_2$H$^+$ transitions. For $^{13}$CO and
C$^{18}$O, the optical depths and excitation temperatures are
estimated by assuming $^{12}$CO(1--0) lines are optically thick
with formulae 1--3 of \cite{wu09}. For N$_2$H$^+$, due to the
uncertainty introduced by the blend{ed} profiles, optical depths
of N$_2$H$^+$ {cannot} be directly fitted with the HFS routine,
but {rather can }be derived from the ratio of integrated
intensities, $\int T_{B}d\upsilon$, of the three{ blended} groups
with {a} method of \citet{purcell09}, using equation
\begin{equation}
\frac{\int T_{B,1}d\upsilon}{\int T_{B,2}d\upsilon}=\frac{1-{\rm
e}^{-\tau_{1}}} {1-{\rm e}^{-\tau_{2}}}=\frac{1-{\rm
e}^{-\tau_{1}}} {1-{\rm e}^{-a\tau_{1}}},\label{eq:optical}
\end{equation}
where `a' is the expected ratio of $\tau_{2}/\tau_{1}$, which should be 1: 5: 2 under
optically thin conditions. It is noted that the optical depth derived here is the
``group optical depth'' that contains contributions of (21$-$12), (23$-$12) and
(22$-$11) components.

Then we derived (23$-$12) excitation temperatures of
N$_2$H$^+$(1--0) using equation
\begin{equation}
T_{R}=T_{o}\left[\frac{1}{{\rm e}^{T_{o}/T_{\rm ex}^{*}}-1}-\frac{1}{{\rm
e}^{T_{o}/T_{\rm bg}}-1}\right]\left(1-{\rm e}^{-\tau}\right), \label{eq:N2HTex}
\end{equation}
where $T_{R}$ and $\tau$ are brightness temperature and optical
depth for the (23$-$12) component, $T_o=h\nu/k$, and $T{_{\rm bg}}
= 2.7$\,K.

The optical depths and excitation temperatures of $^{13}$CO,
C$^{18}$O and N$_2$H$^+$ are tabulated in
Table~\ref{Tab:faintabundance}. 
  The typical optical
depths are 0.7, 0.08 and 0.3 for $^{13}$CO, C$^{18}$O and
N$_2$H$^+$, respectively. CO excitation temperatures range from 15 K
to 48 K, with a mean value of 28 K. Excitation temperatures for
N$_2$H$^+$ range from 5{ K} to 12 K, with a mean value of 7 K.
{Since} the calculations are based on a uniform beam
filling factor, the excitation temperatures (especially for
N$_2$H$^+$) derived here {should} be the lower limits of the true
values.

\begin{table}[h!!]
\begin{center}

\begin{minipage}[]{85mm}

\caption{Optical Depths, Column Densities and
Abundances}\label{Tab:faintabundance}\end{minipage}

\vspace{-3mm} \fns \tabcolsep 0.4mm
\begin{tabular}{lccccccccccccccccc}
\hline\noalign{\smallskip}
 & & & \multicolumn{3}{c}{Optical Depth $\tau$ }& &\multicolumn{2}{c}
 {Excitation Temperature } & & \multicolumn{3}{c}{Column Density $N$}&
 & \multicolumn{3}{c}{Abundance $^\gamma$}  & \tabularnewline
\cline{3-6} \cline{8-9} \cline{11-13} \cline{15-16}\noalign{\smallskip} Region & Core
& & 
$^{13}$CO & 
C$^{18}$O  & 
N$_2$H$^+$ $^\alpha$ & & $T_{\rm ex}$ (CO) & $T_{\rm ex}$
(N$_2$H$^+$)$^\beta$ &
 & 
 $^{13}$CO  & 
 C$^{18}$O & 
 N$_2$H$^+$  & & X(C$^{18}$O) & X(\rm N$_2$H$^+$) & \\
&&&&&&&(K)&(K)&&(cm$^{-2}$)&(cm$^{-2}$)& $10^{12}$cm$^{-2}$
 & & $10^{-7}$
 &   & &\\
 \noalign{\smallskip}\hline\noalign{\smallskip} 106.80+5.31  & 1 & &0.62 &0.08  &0.30 &
&32.8(4.9)  &12.1(2.9)& &9.50E+16 &1.01E+16
&8.96
 &  &2.12 
   &1.89E--10  \\
111.25--0.77  & 1 & &0.85 &0.07  &---  & &23.8(3.6)  &---      &
&5.09E+16 &3.26E+15 &---
&  &1.28 
   &---       \\
121.24--0.34  & 1 & &0.80 &0.10  &0.27 & &22.0(3.3)  &7.4(2.7) & &3.63E+16 &3.85E+15 &2.48
 &  &2.12
   &1.37E--10  \\
133.72+1.22  & 1 & &0.31 &---   &---  & &35.0(5.7)  &---      & &6.01E+16 &---      &---      &  &---        &---       \\
133.72+1.22  & 2 & &0.24 &---   &---  & &38.3(5.7)  &---      & &6.46E+16 &---      &---      &  &---        &---       \\
133.72+1.22  & 3 & &0.33 &---   &---  & &32.3(4.8)  &---      & &8.60E+16 &---      &---      &  &---        &---       \\
133.72+1.22  & 4 & &0.49 &---   &---  & &27.7(4.5)  &---      & &5.14E+16 &---      &---      &  &---        &---       \\
183.35--0.59  & 1 & &1.02 &0.13  &0.32 & &21.2(3.2)  &5.1(3.1) & &3.32E+16 &2.85E+15 &1.23
 &  &1.72
   &7.42E-11  \\
188.80+1.03  & 1 & &0.69 &0.07  &---  & &20.7(3.1)  &---      & &3.15E+16 &3.27E+15 &---      &  &2.08
   &---       \\
188.80+1.03  & 2 & &1.23 &0.17  &---  & &14.6(2.2)  &---      & &3.12E+16 &2.68E+15 &---      &  &1.72
   &---       \\
189.03+0.78  & 1 & &0.72 &0.08  &0.25 & &23.1(3.5)  &7.3(6.3) & &4.56E+16 &4.69E+15 &2.19
 &  &2.06
   &9.62E--11  \\
189.03+0.78  & 2 & &0.89 &0.14  &---  & &17.2(2.6)  &---      & &3.45E+16 &3.89E+15 &---      &  &2.26
   &---       \\
189.78+0.34  & 1 & &0.72 &0.09  &0.37 & &33.1(4.9)  &5.0(2.3) & &9.76E+16 &8.58E+15 &1.44
 &  &1.76
   &2.94E--11  \\
189.78+0.34  & 2 & &0.82 &0.13  &---  & &29.1(4.4)  &---      & &1.08E+17 &1.16E+16 &---      &  &2.15
   &---       \\
189.78+0.34  & 3 & &0.76 &0.16  &---  & &27.1(4.1)  &---      & &9.17E+16 &7.82E+15 &---      &  &1.71
   &---       \\
206.54--16.36 & 1 & &0.94 &0.09  &0.33 & &36.0(5.4)  &5.0(4.3) & &1.60E+17 &1.44E+16 &1.05
 &  &1.80
   &1.31E--11  \\
206.54--16.36 & 2 & &1.54 &0.13  &---  & &36.7(5.5)  &---      & &2.42E+17 &1.90E+16 &---      &  &1.57
   &---       \\
\noalign{\smallskip}\hline\noalign{\smallskip}
mean$^\delta$         &---& &0.74 &0.09  &0.31 & &27.52      &6.97     & &6.78E+16 &6.37E+15 &2.89
 &  &1.87
   &8.98E--11  \\
median       &---& &0.72 &0.09  &0.31 & &23.80      &6.20     & &5.09E+16 &4.27E+15 &1.83
 &  &1.93
   &8.52E--11  \\
\noalign{\smallskip}\hline\noalign{\smallskip}
\end{tabular}
\parbox{135mm}
{\fns
$\alpha$: $\tau$(N$\rm _2H^+$) is  the ``group optical depth'' that contains
contributions of (21$-$12), (23$-$12) and (22$-$11)
components;\\
$\beta$: $T_{\rm ex}(\rm N_2H^+)$ is (23$-$12) excitation
temperature;\\
$\gamma$: the abundances were derived with the assumption of
X($^{13}$CO) $\sim 2\times10^{-6}$;\\
 $\delta$: mean
and median values are only for cores associated with masers, i.e.,
{Core 1}.}
\end{center}

\end{table}

\subsubsection{Column density and chemical abundance}

With optical depths and excitation temperatures, total column
densities of $^{13}$CO, C$^{18}$O and N$_2$H$^+$ can be obtained
from  (eq.~(A1), Scoville et al. 1986)
\begin{equation}
N=\frac{3k}{8\pi^{3}B\mu^{2}}\frac{\exp[hBJ_{l}(J_{l}+1)/kT_{\rm
ex}]}{(J_{l}+1)}\frac{T_{\rm ex}+hB/3k}{1-\exp[-h\upsilon/kT_{\rm ex}]}\int\tau
d\upsilon,\label{eq:density1}
\end{equation}
where {$B$} is the rotational constant of the molecule
and $\mu$ is the permanent dipole moment of the molecule. The values
of {$B$} and $\mu$ are taken to be 55.101 MHz and 0.1098
Debye for $^{13}$CO, 54.891 MHz and 0.1098 Debye for C$^{18}$O
\citep{lovas74},{ and} 46.587 MHz and 3.37 Debye for N$_2$H$^+$
\citep{botschwina1984}. {$J_{l}$} is the rotational
quantum number of the lower state in the observed transitions.
Column densities of $^{13}$CO, C$^{18}$O and N$_2$H$^+$ are
presented in Table~\ref{Tab:faintabundance}.
 Typical column densities of $^{13}$CO,
C$^{18}$O and N$_2$H$^+$ are 5$\times$10$^{16}$,
4$\times$10$^{15}$ and 1$\times$10$^{12}$~cm$^{-2}$, respectively.
For HCO$^+$, whose excitation temperature and optical depth
are lacking, we are unable to calculate reliable column densities.
It is noted that the column densities, especially for N$_2$H$^+$,
estimated here should be the lower limits, because of the
underestimate of excitation temperatures. The total N$_2$H$^+$
column density calculated with an excitation temperature of 20 K
should be {three} times larger than the value estimated with an
excitation temperature of 5\,K.

The relative abundance X between two species may be found directly
from the ratio of their volume densities. Assuming both molecules
occupy the same volume of space, the ratio of two species
X=$n_{1}/n_{2}$ $\thickapprox$ $N_{1}/N_{2}$. Recently, {the }COMPLETE
molecular survey gave an estimated value of [H$_2$/$^{13}$CO] in{ the}
range of 2.8 to 4.9 $\times$10$^5$ \citep{pineda08}. {The
c}hemical evolution model suggested a more stable CO
abundance relative to HCO$^+$ and N$_2$H$^+$ \citep{bergin97}. Thus,
assuming a moderate $^{13}$CO abundance of 3$\times$10$^{-6}$, we
derived the molecular abundances of C$^{18}$O and N$_2$H$^+$ for the
targeted low-luminosity 6.7-GHz maser regions. The results of column
densities and chemical abundances are presented in
Table~\ref{Tab:faintabundance}.

The abundances of C$^{18}$O range from 1.3 to 2.3$\times$10$^{-7}$, with a median
value of 1.9$\times$10$^{-7}${, but} the abundances of N$_2$H$^+$ range from
1.3$\times$10$^{-11}$  to 1.9$\times$10$^{-10}$, with a median value of
8.5$\times$10$^{-11}$. In contrast, {the }abundance fluctuation of N$_2$H$^+$ {is}
more dramatic than that of C$^{18}$O.

\subsubsection{Core size and mass}

The nominal core sizes, $l$, were determined from contours of{ the}
C$^{18}$O integral intensities by de-convolving the telescope beam,
using Equation~(\ref{eq:size}),
\begin{equation}
  l=D\left(\theta_{1/2}^2-\theta_{MB}^{2}\right)^{1/2} , \label{eq:size}
\end{equation}
where $D$ is the distance, $\theta_{1/2}$ is the half-power
angular size of the core, and $\theta_{MB}$ is the half-power beam
width of the telescope.

Core masses were computed by assuming a Gaussian column density
distribution with a full-width at half-maximum (FWHM) of $l$. If
$N(\rm H_{2}$) is the molecular hydrogen column density of {the }peak
position, $m_{\rm H_{2}}$ is the mass of {a} hydrogen molecule and
$\mu$ is the ratio of total gas mass to hydrogen mass (assumed to
be 1.36 based on \citealt{hild83}), the core mass is given by
\begin{equation}
  M_{\rm gas}\simeq\mu m_{\rm H_{2}}\int2\pi
  rN\left(H_{2}\right){\rm e}^{-4\ln2~(r/l)^{2}}dr.
  \label{eq:mass}
\end{equation}
With linewidths and core sizes, we also estimated the virial masses following the
approach by \citet{maclaren88}
\begin{equation}
  M_{\rm vir} = 126 r \Delta\upsilon^2\,,
  \label{eq:virialmass}
\end{equation}
where $r$ is the radius of the core in pc, $\Delta\upsilon$ is the
FWHM linewidth of C$^{18}$O in km~s$^{-1}$ and $M_{\rm vir}$ is the
virial mass in {$M_\odot$}.

The estimated nominal core sizes, gas masses and virial masses are
tabulated in Table~\ref{Tab:faintresult}. 

\begin{table}

\vs \centering

\begin{minipage}{60mm}
\caption{Physical Quantities of the Cores }\label{Tab:faintresult}\end{minipage}

\vspace{-3mm}

\fns \tabcolsep 2mm
\begin{tabular}{lcccccccc}

\hline \noalign{\smallskip} 
Region & Core & R.A. & Dec. 
& Angular & Size   & $N$(H$_2$)  & $M_{\rm gas}$ & $M_{\rm vir}$ \\
Name  &  &  (h~~m~~s) & ($^\circ$ $^\prime$ $^{\prime\prime}$)
&($^{\prime\prime}, ^{\prime\prime}$) & (pc)  &($\times
10^{22}$\,cm$^{-2}$)  &  ($M_\odot$)  &  ($M_\odot$)
\\
(1) & (2)& (3)& (4)& (5)& (6)& (7)& (8)& (9)\\
 \noalign{\smallskip}\hline\noalign{\smallskip}
106.80+5.31      & 1          &22:19:20.0 &+63:19:10   &(120, 90) &0.33  &4.75  &134   &88    \\
111.25--0.77      & 1          &23:16:10.0 &+59:55:30   &(70, 60)  &0.37  &2.55  &91    &49    \\
121.24--0.34      & 1          &00:36:47.8 &+63:28:57   &(110, 70) &0.23  &1.82  &25    &25    \\
133.72+1.22      & 1          &02:25:41.8 &+62:06:05   &(30, 30)  &0.29  &3.01  &66    &220   \\
133.72+1.22      & 2          &02:25:32.0 &+62:06:22   &(120, 90) &0.84  &3.23  &592   &638   \\
133.72+1.22      & 3          &02:25:41.8 &+62:05:40   &(150, 90) &0.98  &4.30  &1073  &685   \\
133.72+1.22      & 4          &02:25:53.5 &+62:04:35   &(90, 60)  &0.42  &2.57  &118   &105   \\
183.35--0.59      & 1          &05:51:10.5 &+25:46:05   &(120, 90) &0.76  &1.66  &249   &86    \\
188.80+1.03      & 1          &06:09:06.5 &+21:50:20   &(90, 70)  &0.45  &1.58  &83    &100   \\
188.80+1.03      & 2          &06:09:09.4 &+21:51:06   &(75, 70)  &0.35  &1.56  &50    &48    \\
189.03+0.78      & 1          &06:08:41.0 &+21:31:08   &(70, 70)  &0.23  &2.28  &31    &54    \\
189.03+0.78      & 2          &06:08:44.5 &+21:31:40   &(90, 60)  &0.27  &1.73  &33    &41    \\
189.78+0.34      & 1          &06:08:34.5 &+20:39:00   &(100, 80) &0.43  &4.88  &235   &68    \\
189.78+0.34      & 2          &06:08:40.5 &+20:38:00   &(90, 60)  &0.27  &5.40  &102   &78    \\
189.78+0.34      & 3          &06:08:40.5 &+20:36:30   &(90, 60)  &0.27  &4.59  &87    &33    \\
206.54--16.36     & 1          &05:41:43.5 &--01:54:05   &(90, 80)  &0.11  &8.00  &25    &13    \\
206.54--16.36     & 2          &05:41:45.0 &--01:56:05   &(120, 100)&0.16  &12.10 &81    &40    \\
\noalign{\smallskip}\hline\noalign{\smallskip}
Mean             &---         &---        &---         &---      &0.36  &3.39  &104   &78    \\
Median           &---         &---        &---         &---      &0.33  &2.55  &83    &68    \\
\noalign{\smallskip}\hline\noalign{\smallskip}
\end{tabular}
\parbox{125mm}{\baselineskip 3.6mm Cols.~(3) and (4) are
equatorial coordinates, Col.~(5) {is} angular extensions of the major and minor axes
of the core assuming a spherical geometry. Col.~(6) is the nominal core size.
Col.~(7) is molecular hydrogen column density, $N$(H$_2$). {G}as masses and virial
masses are listed in Cols.~(8)$-$(9), respectively. Mean and median values of these
quantities are listed in the last two rows.}
\end{table}

\section{Discussion}

\subsection{Line Profile{s}}

\begin{figure}
\centering

\vs

\includegraphics
[width=120mm]{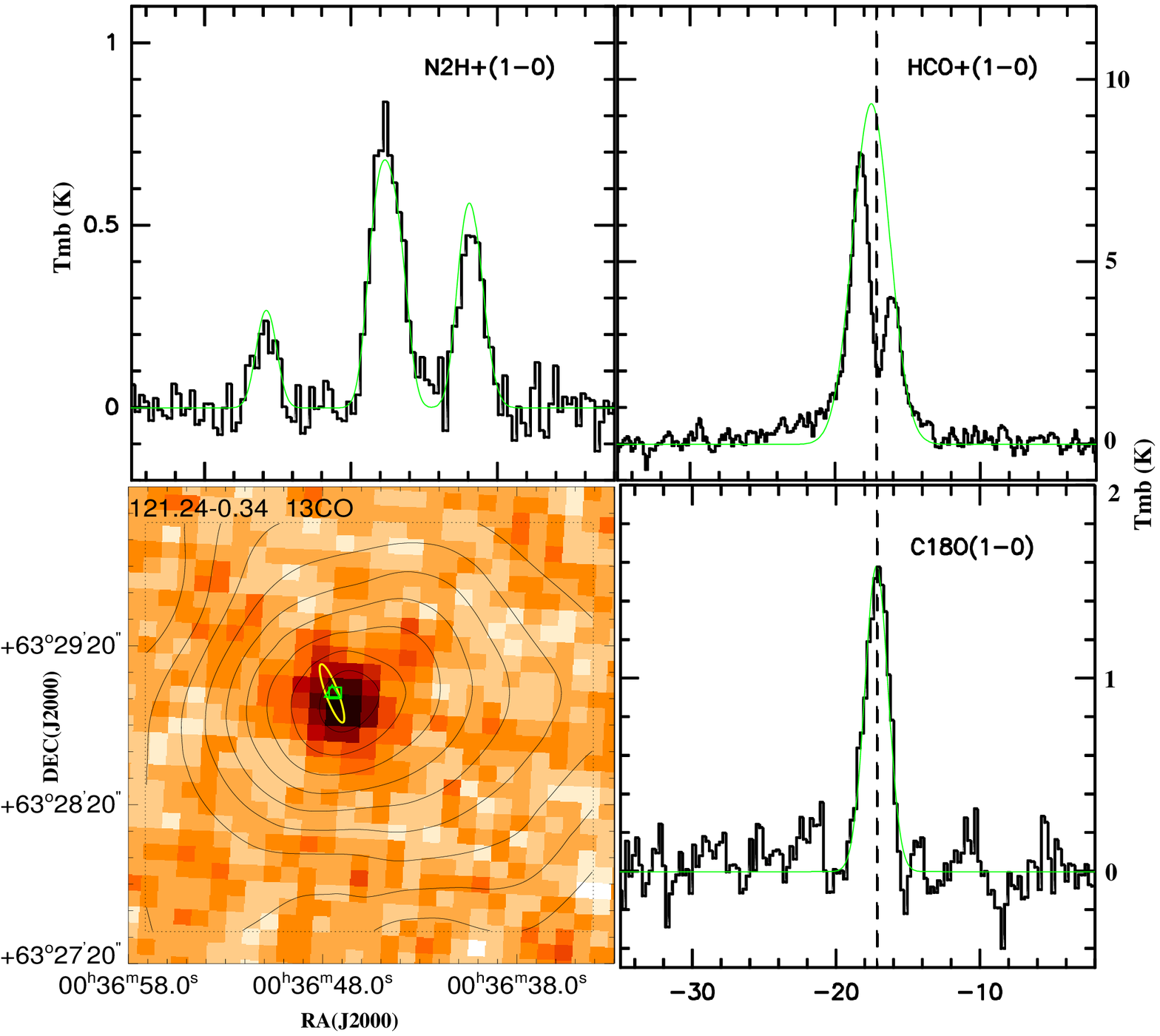}~~~~~~


 \caption{\baselineskip 3.6mm N$_2$H$^+$, HCO$^+$ and C$^{18}$O character{istic}
spectra for 121.24--0.34 alongside the $^{13}$CO(1-0) contours
overlaid {on the} 21 $\mu$m MSX image. On the image, the square
marks{ the} 6.7-GHz CH$_3$OH maser, the triangle marks the H$_2$O
maser and{ the} ellipse denotes{ the} error ellipse of the IRAS
point source. Green lines are hyperfine structure/Gaussian
fittings. The seven hyperfine components of N$_2$H$^+$ have
blended into three groups due to large linewidths. {The
}{s}pectrum of HCO$^+$ reveals{ a} self-absorbed line profile and
is fitted with {a }Gaussian by masking the self-absorption
dip.}\label{Fig:demospec}

\vs\vs \centering

\vspace{-4mm}
\includegraphics[height=85mm,angle=90]{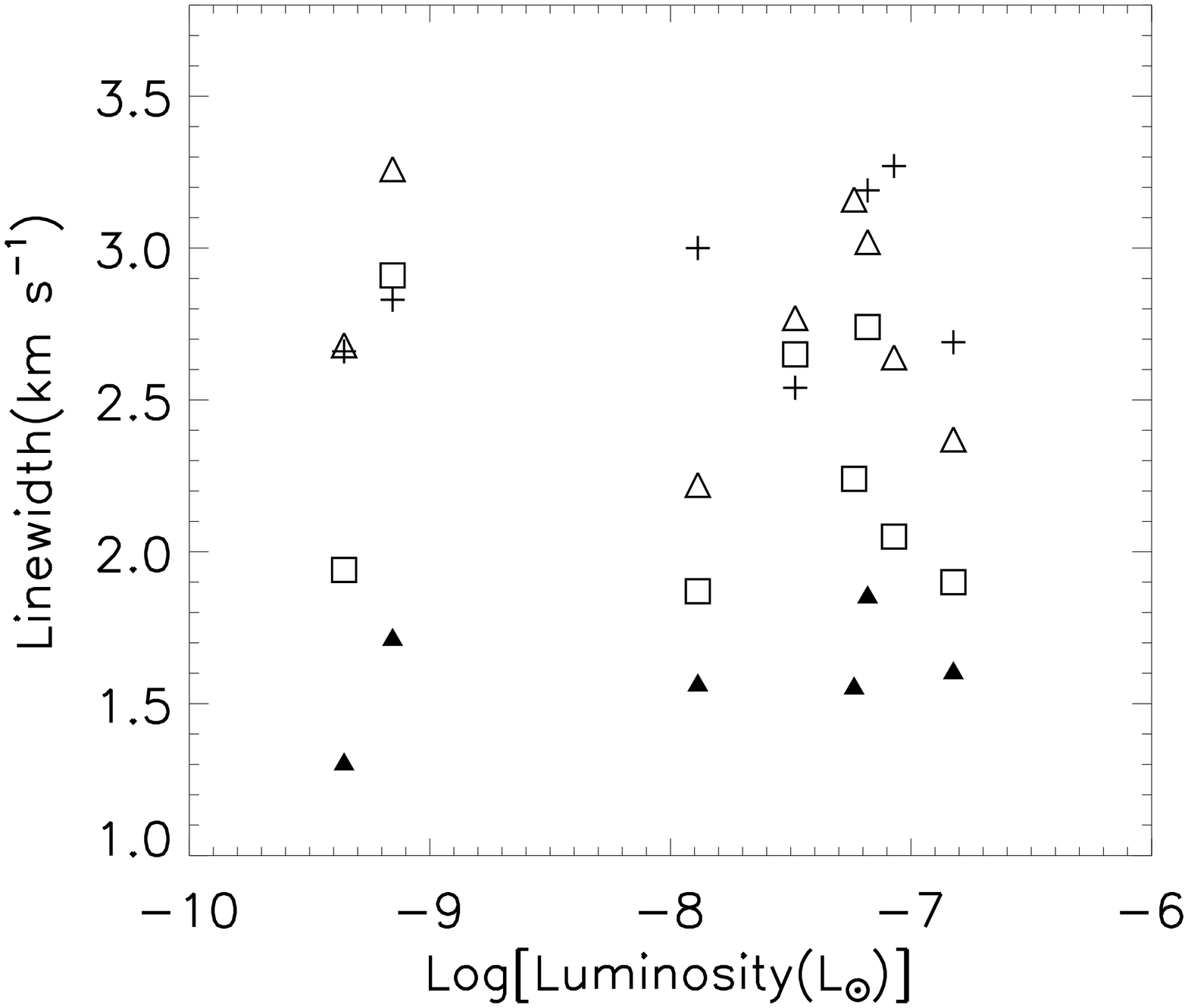}%

\vspace{-4mm} \caption{\baselineskip 3.6mm Line width as a
function of maser
   luminosity. Open triangles and squares indicate
   $^{13}$CO and C$^{18}$O{;} {pluses} and filled triangles indicate HCO$^+$ and N$_2$H$^+$.}
\label{Fig:linewidth}

\end{figure}

Line profiles of $^{13}$CO and C$^{18}$O are relatively simple and
can be well fitted with{ a} single Gaussian. In
Figure~\ref{Fig:demospec}
  we present example
spectra for the source 121.24--0.34 alongside the $^{13}$CO(1--0)
contour. {The }{s}pectrum of N$_2$H$^+$ is fitted with{ the} HFS routine in the
CLASS software to simultaneously fit the N$_2$H$^+$ profiles with
seven Gaussian{ distributions}. In case of {a }cold quiescent environment, e.g., {a }dark
cloud,{ the} spectrum of N$_2$H$^+$(1--0) can exhibit seven hyperfine
components \citep{womack92}. In our case, however, the seven
components of N$_2$H$^+$ are blended into three groups due to
large linewidths.

In Figure~\ref{Fig:demospec}, the spectrum of HCO$^+$ exhibits
{a }double-peak profile. {By contrast,} the spectrum of C$^{18}$O, an
optically thin line, shows {a }single peak locat{ed} at the dip
of HCO$^+$, which indicates that the double-peak profile of
HCO$^+$(1--0) is{ a} self-absorption feature. Therefore{,} we fit the
HCO$^+$ spectrum with{ a} single Gaussian by blanking the absorption
dip. This ``blue profile'' is considered as {a }signature of
inflow. \citet{wu07} conducted an HCO$^+$ survey towards high mass
star forming regions and obtained a high rate of{ incidence for} this kind of ``blue
profile'' (29\%). {However}, in our sample, this inflow signature is
only found towards 121.24--0.34, corresponding to a rate of 11\%.
{In addition}, {the }self-absorption feature is also found in{ the} HCO$^+$ spectrum
of 206.54--16.36, {but }in contrast to 121.24--0.34,{ the} HCO$^+$
spectrum of 206.54--16.36 shows an excess of {the }red profile which may
be evidence of expan{sion}. \citet{emper09} established a
sophisticated model based on {the }scenario of a PDR and the ``Blister
model'' to interpret the complex line shapes of{ the} multi{ple} CO transition
from 206.54--16.36.

Figure~\ref{Fig:linewidth} 
shows the linewidth of the four molecules as an function of maser luminosity. The
mean linewidths towards these low-luminosity maser regions are as follows: HCO$^+$:
3.3~km~s$^{-1}$, $^{13}$CO: 2.8~km~s$^{-1}$, C$^{18}$O: 2.3~km~s$^{-1}$ and
N$_2$H$^+$: 1.6~km~s$^{-1}$. In general{,} linewidths of HCO$^+$ are larger than
$^{13}$CO, C$^{18}$O and N$_2$H$^+$. \cite{purcell06,purcell09} surveyed 83 southern
methanol masers and obtained larger mean linewidths: HCO$^+$: 5.1~km~s$^{-1}$,
$^{13}$CO: 4.8~km~s$^{-1}$, and N$_2$H$^+$: 3.0~km~s$^{-1}$. It is obvious that
linewidths {in} our sample, including only low-luminosity masers, are smaller than
\cite{purcell06,purcell09}'s sample{,} which are composed of both high- and
low-luminosity masers. Previous observations of ammonia also indicate that
NH$_3$(1,1) (2,2) linewidths of low-luminosity maser regions are smaller than {those}
of high-luminosity maser regions (Paper {I}).

\subsection{Gas Mass vs Virial Mass}

\begin{figure}[h!!]

\centering 
\includegraphics[height=85mm,angle=90]
{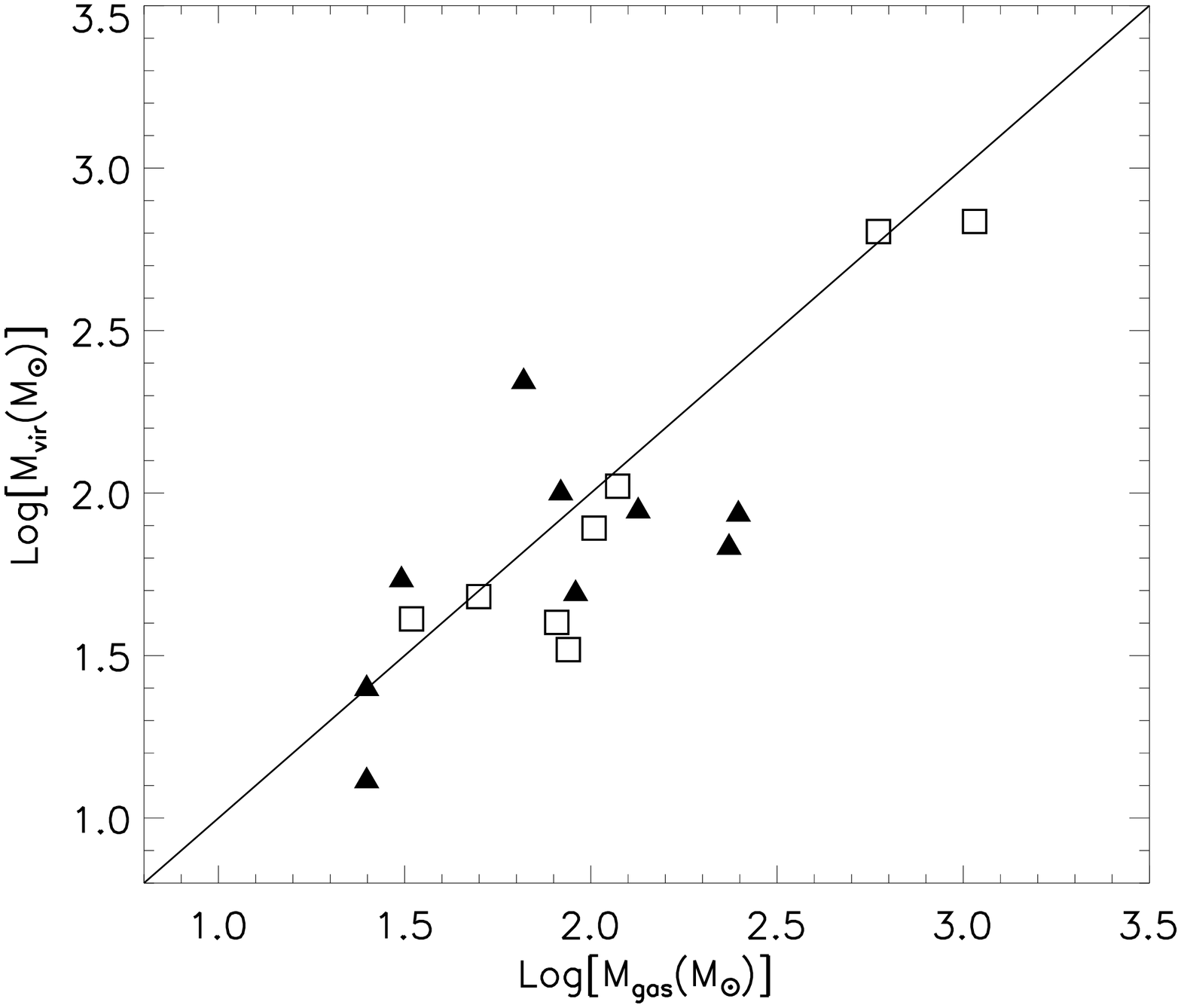}%

 \caption{\baselineskip 3.6mm Gas masses versus
virial masses. Filled triangles and open squares denote cores with
and without maser associations, respectively. The solid line is
the line {where} $M_{\rm gas}$ equal{s} $M_{\rm
vir}$.}\label{Fig:gasvsvir}
\end{figure}

In Figure~\ref{Fig:gasvsvir} 
  we plot the diagram of
gas mass versus virial mass. For the maser associated cores,
$M_{\rm gas}$ range{s} from 25 to 250 {$M_\odot$},
with a mean value of 104 {$M_\odot$} and $M_{\rm
vir}$ range{s} from 13 to 220 {$M_\odot$}, with a
mean value of 78 $M_\odot$. We can see that the value of $M_{\rm
vir}$ and $M_{\rm gas}$ are consistent with having the same
magnitude. As a general rule, clouds with $M_{\rm gas}/M_{\rm vir}>
1$ are considered {to be} gravitationally bound systems. Generally, we
propose that cores that contain low-luminosity 6.7-GHz methanol
masers should be gravitationally bound.

\subsection{Associations}

Apart from Class II methanol masers, outflows, {and }bright IRAS
sources, H$_2$O and OH masers are also signatures of star
formation. Since our targeted sources are all methanol masers, it
is meaningful to investigate the associations of methanol masers
with other star-forming phenomen{a}, such as outflows, bright IRAS
sources and two {other }kinds of masers. In
Table~\ref{Tab:assocation}, 
 we list these
associations. The association rates for{ the} IRAS source, outflow and
H$_2$O are all 7/9, {but} the association rate for{ the} OH maser is only
2/9. Positions of OH, H$_2$O and IRAS sources are also
marked in Figure~\ref{Fig:maps} 
 to show their spatial
coexistence. {W}e can see that these low-luminosity methanol masers
are highly {coincident} with bright IRAS sources, outflows and
H$_2$O masers. In contrast, associations of OH masers are
relatively weak. Though the number of {studied cases in }our sample is limited, {our sample}
clearly indicates the fact that{ the} Class II methanol maser phase {is}
more likely to be overlaid {on} outflow and{ the} H$_2$O maser phrase
than to be overlaid {on the} OH maser phrase
\citep{elling07a,breen10}.

\begin{table}
\centering

\vs\vs
\begin{minipage}[]{90mm}

\caption{ Associations of These Low-luminosity 6.7-GHz Masers}
\label{Tab:assocation}\end{minipage}

\vspace{-3mm}

\fns\tabcolsep 4mm

\begin{tabular}{ccccccccccccc}
\hline \noalign{\smallskip} 
Source Name     & IRAS Name      & Outflow & H$_2$O & OH \\
\noalign{\smallskip}\hline\noalign{\smallskip}
106.80+5.31 &22176+6303  &Y [1]    & Y [5]  &Y [7]   \\
111.25--0.77 &23139+5939  &Y [1]    & Y [5]  &N [8]    \\
121.24--0.34 &00338+6312  &Y [1]    & Y [5]  &...     \\
133.72+1.22 &02219+6152  &Y [1]    & Y [5]  &N [7]    \\
183.35--0.59 &05480+2545  &N [2]     &  N [5]  &...     \\
188.80+1.03 &06061+2151  &N [3]     & Y [5]  &N [7]    \\
189.03+0.78 &06056+2131  &Y [1]    & Y [5]  &N [7]    \\
189.78+0.34 &N           &Y [4]    & Y [6]  &...     \\
206.54--16.3 &N           &Y [1]    &  N [5]  &Y [9]  \\
\noalign{\smallskip}\hline\noalign{\smallskip}
{Association Rate}  &7/9  &7/9 &7/9  &2/9  \\
\noalign{\smallskip}\hline \noalign{\smallskip} 
\end{tabular}
\parbox{90mm}
{\baselineskip 3.8mm Notes: ``Y'' indicates association, ``N''
indicate{s no} association,
``...'' indicate{s} that no information is available; \\
References --- [1] \citet{wu04}; [2] \citet{Snell90}; [3]
\citet{kim06}; [4] \citet{xu06b}; {[5]} \citet{valde01}; [6]
\citet{lada81}; [7] \citet{baudry97}; [8] \citet{szym00b}; {[9]}
\citet{knowles76}. }

\vs\vs
\end{table}

\section{Conclusions}

We have performed multi-line observations, including
transitions of $^{13}$CO(1--0), C$^{18}$O(1--0), HCO$^+$(1--0) and
N$_2$H$^+$(1--0), towards {nine} low-luminosity 6.7-GHz
methanol masers. From integrated {intensity} emission maps, we identified 17 cores, among
which {nine} cores are closely associated with
low-luminosity masers and {eight} cores lack maser
associations. Physical quantities{ of these cores were derived}, including column densities, core
sizes, masses and molecular abundances.
Our major findings are as follows:

\begin{itemize}

\item[(1)] Linewidths of $^{13}$CO, HCO$^+$ and N$_2$H$^+$ of
these low-luminosity maser regions are smaller
 than \cite{purcell06,purcell09}'s sample.

\item[(2)] A ``blue profile'' of{ the} HCO$^+$(1--0) spectrum,
signature of inflow, is found towards one source,{ the}
 detection rate of which is {three} times {less} than \citet{wu07}.

\item[(3)] N$_2$H$^+$ abundances of these regions show larger
fluctuation{s} than {those} of CO.

\item[(4)] Virial masses and gas masses of the maser associated
cores are consistent with having the same
 magnitude, indicating that these cores that contains low-luminosity
  6.7-GHz masers are gravitationally bound systems.

\item[(5)] These low-luminosity masers are more inclined to
coexist with H$_2$O masers, outflows and bright IRAS sources
rather than to coexist with OH masers.
\end{itemize}

We however caution that the number of sources in our study is
limited. Consequently, the findings of this work ha{ve} to be
{confirmed by} a much larger
sample.

\vs
\begin{acknowledgements}
We wish to thank all the staff at Qinghai Station of Purple Mountain
Observatory for their assistance with our observations. This work
was supported by 
  the National Natural Science Foundation of China
(Grant Nos. 10673024, 10733030, 10703010 and 10621303) and the
National Basic Research Program of China-973 Program (2007CB815403).
\end{acknowledgements}

\end{document}